\documentclass[11pt]{article}
\pdfoutput=1
\usepackage{jheppub}
\usepackage{tabu}
\usepackage[vcentermath]{youngtab}
\usepackage[usenames,dvipsnames,table]{xcolor}
\usepackage{graphicx,amsmath,amssymb,amsthm,multirow,array,bm,bbm,esint}
\usepackage[mathscr]{eucal}
\usepackage[bbgreekl]{mathbbol}
\usepackage{epsf,amsfonts}
\usepackage{slashed}
\usepackage[numbers,sort&compress]{natbib}
\usepackage{minibox}
\usepackage{rotating}
\usepackage{pdflscape}
\usepackage{array,tikz-cd}
\usepackage{subcaption}
\usepackage{framed}

\newenvironment{boxeD}[1]
    {\begin{center}
    #1\\[1ex]
    
    \begin{tabular}{|@{\hspace{6mm}}m{0.8\textwidth}@{\hspace{6mm}}|}
    \hline
    }
    { 
    \\\hline
    \end{tabular}
    \end{center}
    }

\usepackage{slashed}

\include{UTqftTeXDefns}

\def \sgn{\mbox{sgn\,}}

\newcommand{\local}{J}
\newcommand{\hop}{J}
\newcommand{\gw}{J}
\newcommand{\effcoupling}{\mathcal{J}}
\newcommand{\effhop}{\mathcal{J}_{\text{hop}}}

\newcommand{\rank}{D}
\newcommand{\dipole}{\mathbb{D}}

\newcommand{\R}{r}
\newcommand{\G}{g}
\newcommand{\B}{b}

\newcommand{\ie}{\emph{i.e.}, }
\newcommand{\eg}{\emph{e.g.}, }

\newcommand{\etc}{\emph{etc}.}

\newcommand{\condition}{ULF property\ }
\newcommand{\col}{\mathfrak{c}}

\title{SYK-like Tensor Models on the Lattice}

\author{Prithvi Narayan}
\author{\!, Junggi Yoon}
\affiliation{International Centre for Theoretical Sciences (ICTS-TIFR), \\
Shivakote, Hesaraghatta Hobli, Bengaluru 560089, India.}
\emailAdd{prithvi.narayan@gmail.com}
\emailAdd{junggi.yoon@icts.res.in}

\abstract{
We study large $N$ tensor models on the lattice without disorder. We introduce techniques which can be applied to a wide class of models, and illustrate it by studying some specific rank-3 tensor models. In particular, we study Klebanov-Tarnopolsky model on lattice, Gurau-Witten model (by treating it as a tensor model on four sites) and also a new model which interpolates between these two models. In each model, we evaluate various four point functions at large $N$ and strong coupling, and discuss their spectrum and long time behaviors. We find similarities as well as differences from SYK model. We also generalize our analysis to rank-$\rank$ tensor models where we obtain analogous results as $\rank=3$ case for the four point functions which we computed. For $\rank>5$, we are able to compute the next-to-subleading ${1 \over N}$ corrections for a specific four point function.

}

\begin{document}
\maketitle


\section{Introduction}
\label{sec:introduction}

The Sachdev-Ye-Kitaev (SYK) model is a simple quantum mechanical model of $N$ fermions with disordered interacting fermions which turns out to be solvable at large $N$. The model was originally introduced by~\cite{Sachdev:1992fk} with motivations from condensed matter physics, but recently revived by~\cite{Sachdev:2010um,kitaevfirsttalk,KitaevTalks,Sachdev:2015efa} in the context of AdS/CFT correspondence. The systematic large $N$ expansion of the SYK model has been studied, and ``melonic'' diagram dominance makes the model solvable~\cite{KitaevTalks,Polchinski:2016xgd,Jevicki:2016bwu,Maldacena:2016hyu,Jevicki:2016ito}. Despite its simplicity, the model exhibits interesting features such as conformal invariance at low energies/strong coupling and the saturation of the chaos bound~\cite{Maldacena:2015waa,KitaevTalks,Maldacena:2016hyu}. All these features are also shared by Einstein gravity theories on AdS~\cite{Shenker:2013pqa,Roberts:2014isa}. This has given rise to the hope that these theories (or their cousins) might admit simple gravitational dual. Indeed the low energy action of SYK model has been shown to arise in dilaton gravity~\cite{Maldacena:2016upp,Jensen:2016pah} and in Liouville theories~\cite{Mandal:2017thl}. Moreover, it was also recently pointed out that the spectrum of SYK model suggests 3D scalar field coupled to gravity~\cite{Das:2017pif}. The SYK models have been generalized in various directions (\eg flavor, supersymmetry \etc)~\cite{Gross:2016kjj,Banerjee:2016ncu,Fu:2016vas,Nishinaka:2016nxg,Li:2017hdt,Gurau:2017xhf,Bonzom:2017pqs,Peng:2017kro}. The SYK model has also been realized on a lattice in higher dimensions~\cite{Gu:2016oyy,Berkooz:2016cvq,Turiaci:2017zwd,Berkooz:2017efq,Gu:2017ohj}.

It is interesting to ask whether the disorder is essential for having a SYK-like physics and in particular whether it is possible to have a more conventional vector/matrix model which can realize the same physics: the dominance of ``melonic'' and ``ladder'' diagrams in large $N$. This dominance was already observed in so-called ``tensor models'', and have been studied extensively in the literature~\cite{Gurau:2009tw,Gurau:2011aq,Gurau:2011xq,Bonzom:2011zz,Gurau:2011xp,Gurau:2011kk,Bonzom:2012hw,GurauSchaeffer,Carrozza:2015adg,Gurau:2016lzk,Bonzom:2016kqf}. 
Previously, except for~\cite{Gurau:2009tw}, only bosonic tensor models had been considered with various form of interactions. However, it was shown by Witten and Gurau~\cite{Witten:2016iux,Gurau:2016lzk} and by Klebanov and Tarnopolsky~\cite{Klebanov:2016xxf} that the fermionic tensor models with particular kind of interaction~(\emph{Tetrahedron} interaction) reproduce many of the features of SYK model at large $N$. However, the model is not entirely identical to SYK model and the physics seemingly differs at the level of $1/N$ corrections. It is now interesting to ask whether generalizations of the sort done in SYK models are possible in this class of tensor models. Some work has already been done in this regard. \eg a large $N$ supersymmetric tensor model was studied in~\cite{Peng:2016mxj}. Numerical analysis for finite $N$ tensor model has been done in~\cite{Krishnan:2017ztz,Krishnan:2016bvg}, and an abelian tensor model on the lattice will be analyzed in an adjoining paper~\cite{Chaudhuri:2017vrv}. As an aside, we mention here that tensor models have also been studied from the perspective of matrix models, where it arises in the limit of large number of matrices~\cite{Ferrari:2017ryl,Itoyama:2017emp,Itoyama:2017xid}. Also, for a discussion of how tensor models could possibly arise in string theories, see~\cite{Klebanov:1996mh,Beccaria:2017aqc}.

In this work, we will be interested in generalization of \emph{Klebanov-Tarnopolsky}(KT) model~\cite{Klebanov:2016xxf} by introducing lattice (Henceforth, we will call it \emph{KT chain model}). Although we mostly focus on a particular generalization closest in spirit to the lattice generalization of SYK model in~\cite{Gu:2016oyy}, the techniques we introduce can be used to analyze much wider class of lattice generalizations. As an example, we study \emph{Gurau-Witten}(GW) model~\cite{Witten:2016iux} as a particular case of KT model on a lattice. We also initiate a more systematic study of correlators in the tensor models. For example, there are many different gauge invariant four point functions possible in tensor models depending on the details of external gauge contractions, and we compute the four point function for many of them. We also compute the spectrum and the chaos exponent when possible for some of these other channels. Note that the special case of the translationally invariant modes of the KT chain model is just the same as KT model. Hence, KT model is a particular case of our results. 
We also consider rank-$\rank$ tensor models and compute some special class of four point functions.

The outline of the paper is as follows. In Section~\ref{sec:review}, we review the SYK model and the lattice generalization thereof~\cite{Gu:2016oyy}. We also give a short introduction to KT model~\cite{Klebanov:2016xxf}. In Section~\ref{sec:large N KT model on lattice}, we introduce the lattice generalization of KT model (\emph{KT chain} model), and discuss two point function. In Section~\ref{sec:four point function}, we introduce our general techniques, and illustrate it by working out various four point function channels in KT chain model. In Section~\ref{sec:models}, we introduce other models in particular the GW model and work out the various four point functions. In Section~\ref{sec:rank d tensor model}, we generalize our techniques for rank-$\rank$ tensor model. In Section~\ref{sec:conclusion}, we conclude with some future directions.

\section{Review}
\label{sec:review}

\subsection{Klebanov-Tarnopolsky Model}
\label{sec:single-site KT model}

We begin by the review of large $N$ results of Klebanov-Tarnopolsky (we call it \emph{KT} model from now on) model~\cite{Klebanov:2016xxf}. The KT model has a real fermion field $\psi_{ijk}(t)$ $(i,j,k=1,2,\cdots N)$ transforming in the tri-fundamental representation of $O(N)^3$ gauge symmetry. Since the fermion has three distinguishable indices, we call it rank-3 tensor field following the standard terminology. Note that $\psi$ has $N^3$ components. It is useful to introduce RGB color $\col$ to distinguish three different $O(N)$ groups. Henceforth, $\R$, $\G$ and $\B$ denotes the color of the first, second and third $O(N)$ group as well as the color of the index $i$ $j$ and $k$ of $\psi_{ijk}$, respectively. 

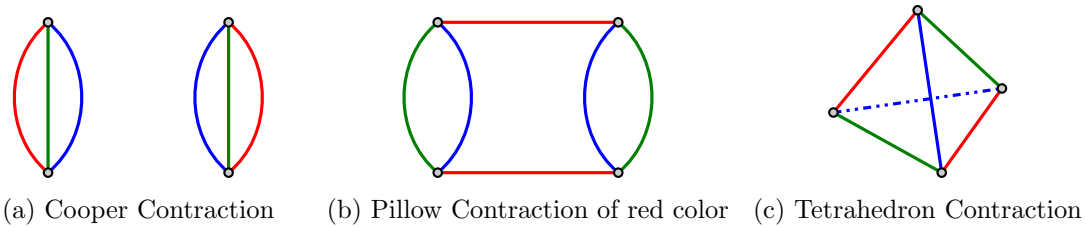
\begin{figure}
\centering
\begin{subfigure}[t]{0.3\linewidth}
\centering
\begin{tikzpicture}[scale=0.8]
\def\xa{0}  \def\ya{0}    
 
    \coordinate (a1) at (\xa,\ya);
    \coordinate (b1) at (\xa,\ya+2.5);
    \coordinate (c1) at (\xa+3,\ya+2.5);
    \coordinate (d1) at (\xa+3,\ya);
    
    \draw[bend left=50,very thick,color=red]  (a1) to   (b1);
    \draw[bend right=50,very thick,color=blue]  (a1) to   (b1);
    \draw[bend right=50,very thick,color=blue]  (c1) to   (d1);  
    \draw[bend left=50,very thick,color=red]  (c1) to   (d1);

      \draw[very thick,color=black!50!green]  (b1) -- (a1) ;
      \draw[very thick,color=black!50!green]   (d1) -- (c1);

	\fill[black!20, draw=black, thick] (a1) circle (2pt) ;
    \fill[black!20, draw=black, thick] (b1) circle (2pt);
    \fill[black!20, draw=black, thick] (c1) circle (2pt)  ;
    \fill[black!20, draw=black, thick] (d1) circle (2pt);      
   
\end{tikzpicture}
\caption{Cooper Contraction}
\label{fig:cooper contraction}
\end{subfigure}
\begin{subfigure}[t]{0.35\linewidth}
\centering
\begin{tikzpicture}[scale=0.8]
\def\xa{0}  \def\ya{0}    
 
    \coordinate (a1) at (\xa,\ya);
    \coordinate (b1) at (\xa,\ya+2.5);
    \coordinate (c1) at (\xa+3,\ya+2.5);
    \coordinate (d1) at (\xa+3,\ya);
    
    \draw[bend left=50,very thick,color=black!50!green]  (a1) to   (b1);
    \draw[bend right=50,very thick,color=blue]  (a1) to   (b1);
    \draw[bend right=50,very thick,color=blue]  (c1) to   (d1);  
    \draw[bend left=50,very thick,color=black!50!green]  (c1) to   (d1);

      \draw[very thick,color=red]  (d1) -- (a1) ;
      \draw[very thick,color=red]   (c1) -- (b1);

	\fill[black!20, draw=black, thick] (a1) circle (2pt) ;
    \fill[black!20, draw=black, thick] (b1) circle (2pt);
    \fill[black!20, draw=black, thick] (c1) circle (2pt)  ;
    \fill[black!20, draw=black, thick] (d1) circle (2pt);         
\end{tikzpicture}
\caption{Pillow Contraction of red color}
\label{fig:pillow contraction}
\end{subfigure}
\begin{subfigure}[t]{0.3\linewidth}
\centering
\begin{tikzpicture}[scale=0.8]
\def\xa{0}  \def\ya{0}   \def\xb{8}  \def\yb{0}    \def\xc{0}  \def\yc{6}    \def\xd{8}  \def\yd{6}

    \coordinate (a1) at (\xa+4.4,\ya+2.5);
    \coordinate (b1) at (\xa+3,\ya+0.8);
    \coordinate (c1) at (\xa+4.8,\ya-0.2);
    \coordinate (d1) at (\xa+5.8,\ya+1.2);
    
      \draw[very thick,color=red] (b1) -- (a1) -- cycle;
    \draw[very thick,color=black!50!green]  (a1) -- (d1) -- cycle;       
    \draw[very thick,color=red]  (d1) -- (c1) -- cycle;
    \draw[very thick,color=blue] (a1) -- (c1);
    \draw[very thick,color=black!50!green] (c1) -- (b1)  -- cycle;
    \draw[very thick,dash dot dot,color=blue] (b1) -- (d1);
    
	\fill[black!20, draw=black, thick] (a1) circle (2pt)  ;
    \fill[black!20, draw=black, thick] (b1) circle (2pt) ;
    \fill[black!20, draw=black, thick] (c1) circle (2pt) ;
    \fill[black!20, draw=black, thick] (d1) circle (2pt)  ;
   
\end{tikzpicture}
\caption{Tetrahedron Contraction}
\label{fig:tetrahedron contraction}
\end{subfigure}
\caption{Cooper, Pillow and Tetrahedron contractions. Each vertex represents a fermion, and the colored edge denotes the contraction of gauge index of the corresponding color in the two fermions.}
\label{fig:contractions}
\end{figure}

To write down a gauge invariant Hamiltonian for the tensor model, we need to classify the gauge invariant operators in the theory. Unlike vector model, the tensor model has various possible gauge contractions to generate gauge-invariants.  In particular if we want quartic interaction there are three different ways of gauge contractions of four rank-3 fermions: \emph{``Cooper''} contraction (\eg \eqref{def:cooper channel kt model}), \emph{``Pillow''} contraction and \emph{``Tetrahedron''} contraction (\eg \eqref{def: KT hamiltonian in large N}) (See Fig.~\ref{fig:contractions}). In Fig.~\ref{fig:contractions}, each vertex represents a fermion, and each colored edge denotes the gauge index of the corresponding color. The connection of two vertices means the gauge contraction of the corresponding color between the two fermions. It turned out that the quartic interaction with Tetrahedron contraction (we call it \emph{Tetrahedron} interaction) gives rise to melonic dominance similar to SYK model. The KT model is an example of such an interaction where the Hamiltonian is given by 
\begin{equation}
H_{\text{KT}}={1\over 4}\local N^{-{3\over 2}}    \psi_{i_1 j_1 k_1}\psi_{i_1 j_2 k_2}\psi_{i_2 j_1 k_2}\psi_{i_2 j_2 k_1}\label{def: KT hamiltonian in large N}
\end{equation}
where we scaled the coupling with $N$ so that the large $N$ limit~\cite{Gurau:2010ba,Bonzom:2011zz,Carrozza:2015adg} is taken to be
\begin{equation}\label{thooft like limit}
N\quad \longrightarrow \quad \infty \hspace{10mm} \mbox{ with fixed}\quad \local\ .
\end{equation}
%

\begin{figure}
\centering
\begin{subfigure}[b]{0.3\linewidth}
\centering
\begin{tikzpicture}[scale=0.85]
\draw[thick,color=red] (-1.5,0.5) -- (1.5,0.5);
\draw[thick,color=black!50!green] (-1.5,0) -- (1.5,0);
\draw[thick,color=blue] (-1.5,-0.5) -- (1.5,-0.5);
\draw[thick, color=black]
{
(0.0,-1.5) node [below] { \; }
};
\end{tikzpicture}
\caption{Free propagator}
\label{fig:free propagator in KT model}
\end{subfigure}
\quad  \quad
\begin{subfigure}[b]{0.3\linewidth}
\centering
\begin{tikzpicture}[scale=0.85]
\draw[thick,color=red] (-1.5,0.5) -- (-0.5,0.5) -- (-0.5,1.5);
\draw[thick,color=blue] (-1.5,-0.5) -- (-0.5,-0.5) -- (-0.5,-1.5);
\draw[thick,color=red] (1.5,-0.5) -- (0.5,-0.5) -- (0.5,-1.5);
\draw[thick,color=blue] (1.5,0.5) -- (0.5,0.5) -- (0.5,1.5);

\draw[thick,color=black!50!green] (-1.5,0.0) -- (1.5,0.0) ;
\draw[thick,color=black!50!green] (0.0,1.5) -- (0.0,0.2) ;
\draw[thick,color=black!50!green] (0.0,-1.5) -- (0.0,-0.2) ;
\draw[color=blue] (0.0,0.2) arc (90:-90:0.2);
\draw[thick, color=black]
{
(0.0,-1.5) node [below] {\small{$ \local N^{-{3 \over 2}}$}}
};
\end{tikzpicture}
\caption{Vertex}
\label{fig:vertex in KT model}
\end{subfigure} 
\caption{Free propagator and vertex in the KT model}
\end{figure}
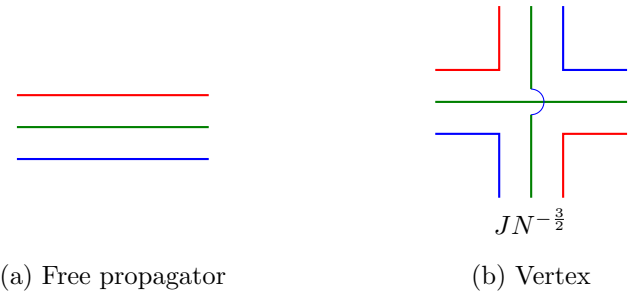
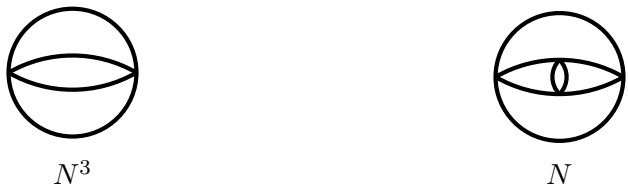
\begin{figure}
\centering
\begin{subfigure}[t]{0.4\linewidth}
\centering
\begin{tikzpicture}[scale=0.85]
\def\c{0} \def\s{1}   \def\r{2*\s} 
\draw[ultra thick,color=black] (\c,0) circle (\s);
\draw[ultra thick,color=black] (\c-\s,0) arc (90+30:90-30:\r);
\draw[ultra thick,color=black] (\c-\s,0) arc (-90-30:-90+30:\r);
\draw[thick, color=black]
{ 
(0,-\s-0.2) node [below] { $N^3$ }
};
\end{tikzpicture}
\caption{Melonic vacuum diagram}
\label{fig:example melonic diagram}
\end{subfigure}  \hspace{0cm}
\begin{subfigure}[t]{0.4\linewidth}
\centering
\begin{tikzpicture}[scale=0.85]
\def\c{0} \def\s{1}   \def\r{2*\s} 
\draw[ultra thick,color=black] (\c,0) circle (\s);
\draw[ultra thick,color=black] (\c-\s,0) arc (90+30:90-30:\r);
\draw[ultra thick,color=black] (\c-\s,0) arc (-90-30:-90+30:\r);
\draw[ultra thick,color=black] (\c,0.268) arc (45:-45:1.414*0.2682);
\draw[ultra thick,color=black] (\c,0.268) arc (180-45:180+45:1.414*0.2682);
\draw[thick, color=black]
{ 
(0,-\s-0.2) node [below] { $N$ }
};
\end{tikzpicture} 
\caption{Non-melonic vacuum diagram}
\label{fig:example non-melonic diagram}
\end{subfigure}	
\caption{Example : Melonic and non-melonic vacuum diagrams}
\label{fig:example melonic and non-melonic diagram}
\end{figure}

In Feynman diagrams of tensor models, a free propagator can be represented by strand of three colored lines corresponding to each of three indices of a tensor field (See Fig.~\ref{fig:free propagator in KT model}). For the Tetrahedron interaction, the vertex can be represented by Fig.~\ref{fig:vertex in KT model}. These strand Feynman diagrams are useful to determine the $N$ scaling of the diagrams. For some purposes (\eg Schwinger-Dyson equations) typically after determining the leading $N$ diagrams, it is more convenient to represent the strand by a single line (\eg See Fig.~\ref{fig:example melonic and non-melonic diagram}). 

%
%

Just as matrix models are dominated by planar diagrams at large $N$, ~\cite{Gurau:2010ba,Gurau:2011aq,Gurau:2011xq,Bonzom:2011zz,Gurau:2011kk,Carrozza:2015adg,Gurau:2011xp,Witten:2016iux,Gurau:2016lzk,Klebanov:2016xxf} showed that the tensor models in the large $N$ limit are dominated by the so called \emph{melonic} diagrams illustrated in Fig.~\ref{fig:example melonic diagram}. To define melonic diagrams we need two ingredients.
\begin{itemize}
\item \emph{Melonic operation} : This is defined as replacing any free propagator in a Feynman diagram by a \emph{melon}. In terms of diagrams, this is given by Fig.~\ref{fig:Melonic Operation}.  It is obvious from the figure that the melonic operation does not change the $N$ scaling of any Feynman diagram since it introduces three extra loops ($\sim N^3$) but at the cost of two vertices ($\sim N^{-3}$).

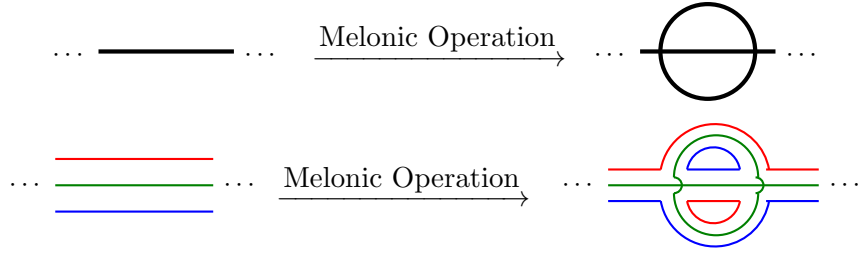
\begin{figure}[t!]
\centering
\begin{subfigure}[t]{\linewidth}
\centering 
\begin{tikzpicture}[scale=0.9]
\def\c{0}  \def\s{1}  \def\r{0.7}

\draw[ultra thick,color=black] (-\s+\c,0) --  (\s+\c,0);

\def\c{8} 
\draw[ultra thick,color=black] (-\s+\c,0) --  (\s+\c,0);
\draw[ultra thick,color=black] (\c,0) circle (\r); 

\draw[thick, color=black]
{
(-\s,-0.05) node [left] {\small{$\cdots$}}
(\s,-0.05) node [right] {\small{$\cdots$}}
(\s+1,0.1) node [right] {$\xrightarrow{\mbox{Melonic Operation}}$}
(8-\s,-0.05) node [left] {\small{$\cdots$}}
(8+\s,-0.05) node [right] {\small{$\cdots$}}
};
\end{tikzpicture} 
\end{subfigure}
\\
\vspace{0.2cm}
\begin{subfigure}[t]{\linewidth}
\centering
\begin{tikzpicture}[scale=0.7]
\draw[thick,color=red] (-1.5,0.5) -- (1.5,0.5);
\draw[thick,color=black!50!green] (-1.5,0) -- (1.5,0);
\draw[thick,color=blue] (-1.5,-0.5) -- (1.5,-0.5);
\draw[thick, color=black]
{
(0.0,-1.5) node [below] { \; }
(-1.5,0) node [left]  {\small{$\cdots$}}
(1.5,0) node [right]  {\small{$\cdots$}}
(2.5,0) node [right]  {$\xrightarrow{\mbox{Melonic Operation}}$}
};
\def\c{11} 
\draw[thick,color=red] (-2+\c,0.3) -- (-1+\c,0.3);
\draw[thick,color=red] (-1+\c,0.3) arc (170:10:1.04403);
\draw[thick,color=red] (2+\c,0.3) -- (1+\c,0.3); 

\draw[thick,color=blue] (-0.5+\c,0.3) -- (0.5+\c,0.3);
\draw[thick,color=blue] (-0.5+\c,0.3) arc (170:10:0.5095);

\draw[thick,color=black!50!green] (-2+\c,0) -- (2+\c,0);
\draw [thick,color=black!50!green] (-0.75+\c,0.15) arc (180:0:0.8);
\draw [thick,color=black!50!green] (-0.75+\c,-0.15) arc (-180:0:0.8);
\draw [thick,color=black!50!green] (-0.75+\c,0.15) arc (90:-90:0.15); 
\draw [thick,color=black!50!green] (0.8+\c,0.15) arc (90:-90:0.15);

\draw[thick,color=red] (-0.5+\c,-0.3) -- (0.5+\c,-0.3);
\draw[thick,color=red] (-0.5+\c,-0.3) arc (-170:-10:0.5095);

\draw[thick,color=blue] (-2+\c,-0.3) -- (-1+\c,-0.3);
\draw[thick,color=blue] (-1+\c,-0.3) arc (-170:-10:1.04403);
\draw[thick,color=blue] (2+\c,-0.3) -- (1+\c,-0.3);
\draw[thick, color=black]
{
(\c-2,0) node [left] {\small{$\cdots$}}
(\c+2,0) node [right] {\small{$\cdots$}}
};
\end{tikzpicture}
\end{subfigure}
\caption{Melonic operation in single line \& strand notation }
\label{fig:Melonic Operation}
\end{figure}

\item \emph{Elementary n-point diagrams}: These are some special diagrams for any given $n$-point function. For example, vacuum $0$-point diagram is just the vacuum bubble given in Fig.~\ref{fig:Elementary 0-point diagram}. For two point functions this is the propagator given in Fig.~\ref{fig:Elementary 2-point diagram}. For four point function, this is the ladder diagram given in Fig.~\ref{fig:Elementary 4-point diagram}. Since we will not require higher point functions in this work, we will not give the corresponding elementary $n$-point diagrams and refer reader to \cite{GurauSchaeffer,Gurau:2016lzk} for more details.
\begin{figure}
\centering
\begin{subfigure}[t]{0.4\linewidth}
\centering
\begin{tikzpicture}[scale=0.85]
\draw[ultra thick,color=black] (0,0) circle (1);
\draw[thick, color=black]
{ 
(0,-1) node [below] { $N^3$ }
};
\end{tikzpicture}
\caption{Elementary 0-point diagram}
\label{fig:Elementary 0-point diagram}
\end{subfigure}  \hspace{0cm}
\begin{subfigure}[t]{0.4\linewidth}
\centering
\begin{tikzpicture}[scale=0.85]
\draw[ultra thick,color=black] (-1,1) -- (1,1);
\draw[thick, color=black]
{ 
(0,0) node [below] { $N^0$ }
};
\end{tikzpicture}
\caption{Elementary 2-point diagram}
\label{fig:Elementary 2-point diagram}
\end{subfigure}  \hspace{0cm} 
\begin{subfigure}[t]{0.5\linewidth}
\centering
\vspace{0.5cm}
\begin{tikzpicture}[scale=0.85]
\def\l{0} \def\s{2.5}  \def\r{\s} \def\v{1.7}  
\draw[ultra thick,color=black] (-\r,0) -- (\r,0);
\draw[ultra thick,color=black] (-\r,-\v) --  (\r,-\v);
\draw[ultra thick,color=black] (-\r/3,0) arc (30:-30:1.7);
\draw[ultra thick,color=black] (-\r/3,0) arc (180-30:180+30:1.7);
\draw[ultra thick,color=black] (\r/3,0) arc (30:-30:1.7);
\draw[ultra thick,color=black] (\r/3,0) arc (180-30:180+30:1.7);
\draw[ultra thick,color=black] (0,0) arc (30:-30:1.7);
\draw[ultra thick,color=black] (0,0) arc (180-30:180+30:1.7);
\draw[thick, color=black] (\r/3+0.2,-\v/2) node [right] {\small{$  \cdots$}};
\draw[thick, color=black] (-\r/3-0.2,-\v/2) node [left] {\small{$  \cdots$}};
\draw[thick, color=black]
{ 
(0,-\v-0.1) node [below] { $N^{-2}$ }
};
\end{tikzpicture} 
\caption{Elementary 4-point diagram}
\label{fig:Elementary 4-point diagram}
\end{subfigure}	
\caption{Elementary $n$-point diagrams}
\label{fig:Elementary n-point diagrams}
\end{figure}
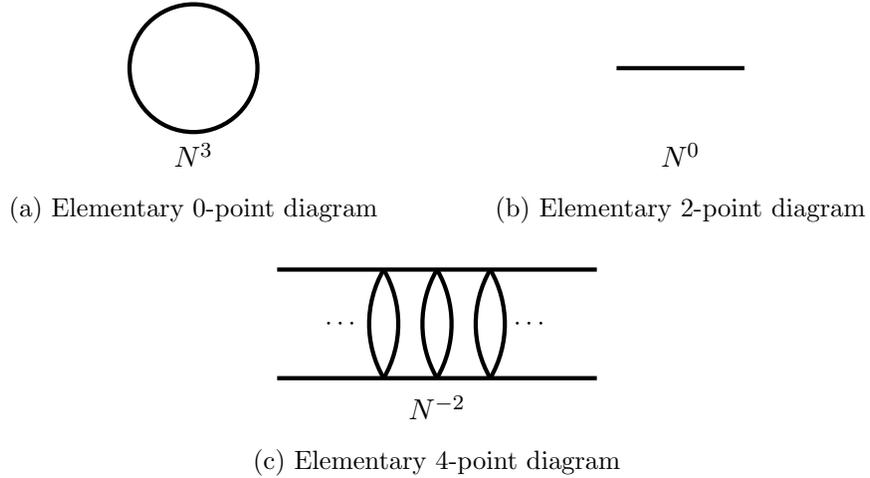
\end{itemize}
Given these ingredients, one can define the melonic $n$-point diagram as one which is obtained by recursively applying melonic operation on elementary $n$-point diagrams. \cite{GurauSchaeffer,Gurau:2016lzk} showed that the leading diagrams in large $N$ for any connected $n$-point function is a melonic $n$-point diagram. As an example, note that Fig \ref{fig:example melonic diagram} can be obtained from Fig \ref{fig:Elementary 0-point diagram} by a melonic operation, while Fig \ref{fig:example non-melonic diagram} cannot. 
 
%
%


\begin{figure}[t!]
\centering

\begin{tikzpicture}[scale=0.9]

\def\c{0}  \def\s{1} \def\r{0.7}

\draw[ultra thick,color=black] (-\s+\c,0) --  (\s+\c,0);

\def\c{3} 
\draw[ultra thick,color=black] (-\s+\c,0) --  (\s+\c,0);
\draw[ultra thick,color=black] (\c,0) circle (\r); 

\def\c{6} \def\v{0.8}
\draw[ultra thick,color=black] (-\s+\c,0) --  (\s+\c,0);
\draw[ultra thick,color=black] (\c,0) circle (\r); 
\draw[ultra thick,color=black] (\c,\v) circle (0.7*\r); 
\draw[ultra thick,color=black] (\s+\c,0) --  (3*\s+\c,0);
\draw[ultra thick,color=black] (2*\s + \c,0) circle (\r); 

\def\c{11} \def\v{0.8}
\draw[ultra thick,color=black] (-\s+\c,0) --  (\s+\c,0);
\draw[ultra thick,color=black] (\c,0) circle (\r); 
\draw[ultra thick,color=black] (\c,\v) circle (0.7*\r); 
\def\v{-0.8} 
\draw[ultra thick,color=black] (\c,\v) circle (0.7*\r); 
\draw[ultra thick,color=black] (\s+\c,0) --  (3*\s+\c,0);
\draw[ultra thick,color=black] (2*\s + \c,0) circle (\r);

\draw[thick, color=black]
{
(0+1+.7,0) node [left] {\small{$+$}}
(3+1+.7,0) node [left] {\small{$+$}}
(6+3+0.7,0) node [left] {\small{$+$}}
(12+3+0.7,0) node [left] {\small{$+ \cdots$}}
};
\end{tikzpicture} 
\caption{Contributions to two point function in single line notation}
\label{fig:Contributions to Two Point Function in Single Line notation}
\end{figure}
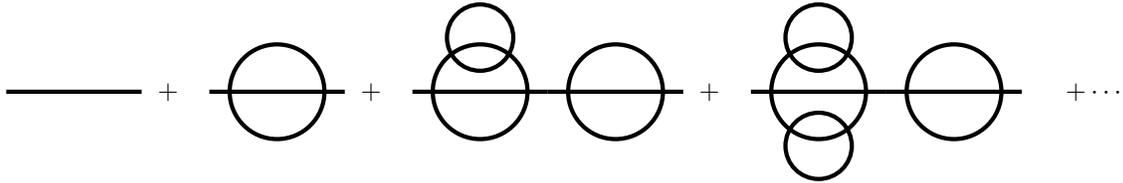

\begin{figure}[t!]
\centering
\begin{tikzpicture}[scale=1]
\draw[thick,color=red] (-1,0.1) -- (-0.5,0.1);
\draw[thick,color=red] (1,0.1) -- (0.5,0.1);
\draw[thick,color=red,dashed] (-1,-1.1) -- (-0.5,-1.1);
\draw[thick,color=red,dashed] (1,-1.1) -- (0.5,-1.1);
\draw[thick,color=red,dashed] (-1,0.1) arc (90:270:0.6);
\draw[thick,color=red,dashed] (1,0.1) arc (90:-90:0.6);
\draw[thick,color=red,dashed] (-0.5,-1.1) -- (0.5,-1.1);

\draw[thick,color=black!50!green] (-1,0) -- (-0.5,0);
\draw[thick,color=black!50!green] (1,0) -- (0.5,0);
\draw[thick,color=black!50!green,dashed] (-1,-1) -- (-0.5,-1);
\draw[thick,color=black!50!green,dashed] (1,-1) -- (0.5,-1);
\draw[thick,color=black!50!green,dashed] (-1,0) arc (90:270:0.5);
\draw[thick,color=black!50!green,dashed] (1,0) arc (90:-90:0.5);
\draw[thick,color=black!50!green,dashed] (-0.5,-1) -- (0.5,-1);

\draw[thick,color=blue] (-1,-0.1) -- (-0.5,-0.1);
\draw[thick,color=blue] (1,-0.1) -- (0.5,-0.1);
\draw[thick,color=blue,dashed] (-1,-0.9) -- (-0.5,-0.9);
\draw[thick,color=blue,dashed] (1,-0.9) -- (0.5,-0.9);
\draw[thick,color=blue,dashed] (-1,-0.1) arc (90:270:0.4);
\draw[thick,color=blue,dashed] (1,-0.1) arc (90:-90:0.4);
\draw[thick,color=blue,dashed] (-0.5,-0.9) -- (0.5,-0.9);

\draw[thick,color=black] (-0.5,-0.5) rectangle (0.5,0.5);

\draw[thick, color=black]
{
(2,-0.5) node [right] {\small{$=$}}
(-0.5,0) node [right] {\small{\ \ $G$}}
};
\end{tikzpicture} 
\begin{tikzpicture}[scale=1]
\draw[thick,color=red] (-1,0.1) --  (1,0.1);
\draw[thick,color=red,dashed] (-1,-1.1) -- (-0.5,-1.1);
\draw[thick,color=red,dashed] (1,-1.1) -- (0.5,-1.1);
\draw[thick,color=red,dashed] (-1,0.1) arc (90:270:0.6);
\draw[thick,color=red,dashed] (1,0.1) arc (90:-90:0.6);
\draw[thick,color=red,dashed] (-0.5,-1.1) -- (0.5,-1.1);

\draw[thick,color=black!50!green] (-1,0) --  (1,0) ;
\draw[thick,color=black!50!green,dashed] (-1,-1) -- (-0.5,-1);
\draw[thick,color=black!50!green,dashed] (1,-1) -- (0.5,-1);
\draw[thick,color=black!50!green,dashed] (-1,0) arc (90:270:0.5);
\draw[thick,color=black!50!green,dashed] (1,0) arc (90:-90:0.5);
\draw[thick,color=black!50!green,dashed] (-0.5,-1) -- (0.5,-1);

\draw[thick,color=blue] (-1,-0.1) --  (1,-0.1);
\draw[thick,color=blue,dashed] (-1,-0.9) -- (-0.5,-0.9);
\draw[thick,color=blue,dashed] (1,-0.9) -- (0.5,-0.9);
\draw[thick,color=blue,dashed] (-1,-0.1) arc (90:270:0.4);
\draw[thick,color=blue,dashed] (1,-0.1) arc (90:-90:0.4);
\draw[thick,color=blue,dashed] (-0.5,-0.9) -- (0.5,-0.9);

\draw[thick, color=black]
{
(2,-0.5) node [right] {\small{$+$}}
};
\end{tikzpicture} 
\def\ed{2}  \def\v1{1} \def\b{1.4} \def\gs{0.2} \def\lo{0.8} \def\gr{1.6}
\begin{tikzpicture}[scale=1]
\draw[thick,color=red] (-\ed,0.1) -- (-\v1,0.1);
\draw[thick,color=blue] (-\v1+0.3,0.1) -- (-\gs,0.1);
\draw[thick,color=blue] (\v1-0.3,0.1) -- (\gs,0.1);
\draw[thick,color=red] (\ed,0.1) -- (\gr+0.2,0.1);
\draw[thick,color=red] (\v1,0.1) -- (\gr-0.2,0.1);
\draw[thick,color=red,dashed] (-\ed,-\b -0.1) -- (-\v1,-\b-0.1);
\draw[thick,color=red,dashed] (\ed,-\b-0.1) -- (\v1,-\b-0.1);
\draw[thick,color=red,dashed] (-\ed,0.1) arc (90:270:\b/2+0.1);
\draw[thick,color=red,dashed] (\ed,0.1) arc (90:-90:\b/2+0.1);
\draw[thick,color=red,dashed] (-\v1,-\b-0.1) -- (\v1,-\b-0.1);

\draw[thick,color=black!50!green] (-\ed,0) --  (-\gs,0);
\draw[thick,color=black!50!green] (\v1,0) -- (\gs,0);
\draw[thick,color=black!50!green] (\ed,0) -- (\gr+0.2,0);
\draw[thick,color=black!50!green] (\gr-0.2,0) -- (\v1,0);
\draw[thick,color=black!50!green,dashed] (-\ed,-\b) -- (-\v1,-\b);
\draw[thick,color=black!50!green,dashed] (\ed,-\b) -- (\v1,-\b);
\draw[thick,color=black!50!green,dashed] (-\ed,0) arc (90:270:\b/2);
\draw[thick,color=black!50!green,dashed] (\ed,0) arc (90:-90:\b/2);
\draw[thick,color=black!50!green,dashed] (-\v1,-\b) -- (\v1,-\b);

\draw[thick,color=blue] (-\ed,-0.1) -- (-\v1,-0.1);
\draw[thick,color=red] (-\v1+0.3,-0.1) -- (-\gs,-0.1);
\draw[thick,color=red] (\v1-0.3,-0.1) -- (\gs,-0.1);
\draw[thick,color=blue] (\ed,-0.1) -- (\gr+0.2,-0.1);
\draw[thick,color=blue] (\v1,-0.1) -- (\gr-0.2,-0.1);
\draw[thick,color=blue,dashed] (-\ed,-\b+0.1) -- (-\v1,-\b+0.1);
\draw[thick,color=blue,dashed] (\ed,-\b+0.1) -- (\v1,-\b+0.1);
\draw[thick,color=blue,dashed] (-\ed,-0.1) arc (90:270:\b/2-0.1);
\draw[thick,color=blue,dashed] (\ed,-0.1) arc (90:-90:\b/2-0.1);
\draw[thick,color=blue,dashed] (-\v1,-\b+0.1) -- (\v1,-\b+0.1);

\draw[thick,color=red] (-\v1+0.3,-0.1) arc (180:260:\b-\v1+0.2);
\draw[thick,color=blue] (-\v1+0.3,0.1) arc (180:100:\b-\v1+0.2);
\draw[thick,color=red] (\v1-0.3,-0.1) arc (0:-80:\b-\v1+0.2);
\draw[thick,color=blue] (-\v1,-0.1) arc (180:270:\b-\v1+0.4);
\draw[thick,color=red] (-\v1,0.1) arc (180:90:\b-\v1+0.4);
\draw[thick,color=red] (\v1,0.1) arc (0:90:\b-\v1+0.4);
\draw[thick,color=blue] (\v1,-0.1) arc (0:-90:\b-\v1+0.4);
\draw[thick,color=red] (\v1,0.1) arc (0:90:\b-\v1+0.4);
\draw[thick,color=blue] (\v1-0.3,0.1) arc (0:80:\b-\v1+0.2);

\draw[thick,color=black] (-\gs,\lo-0.2) rectangle (\gs,\lo+0.2);
\draw[thick,color=black] (-\gs,-0.2) rectangle (\gs,0.2);
\draw[thick,color=black] (-\gs,-\lo-0.2) rectangle (\gs,-\lo+0.2);
\draw[thick,color=black] (\gr-0.2,-0.2) rectangle (\gr+0.2,0.2);

\draw[thick,color=black!50!green]  (\gs-\v1-0.05,0.1) arc (0:-83:\gs-\v1+0.08);
\draw[thick,color=black!50!green]  (\gs-\v1-0.05,-0.1) arc (0:83:\gs-\v1+0.08);
\draw[thick,color=black!50!green]  (\gs-\v1-0.05,-0.1) arc (-90:90:0.1);

\draw[thick,color=black!50!green]  (\v1-\gs+0.05,0.1) arc (-180:-180+83:\gs-\v1+0.08);
\draw[thick,color=black!50!green]  (\v1-\gs+0.05,-0.1) arc (180:180-83:\gs-\v1+0.08);
\draw[thick,color=black!50!green]  (\v1-\gs+0.05,-0.1) arc (-90:90:0.1);


\draw[thick, color=black]
{
(-0.5,0) node [right] {\small{\ \ $G$}}
(-0.5,\lo) node [right] {\small{\ \ $G$}}
(-0.5,-\lo) node [right] {\small{\ \ $G$}}
(\gr-0.5,0) node [right] {\small{\ \ $G$}}
};
\end{tikzpicture} \\ \vspace{0.4cm}


\begin{tikzpicture}[scale=1]
\def\h{0.92}
\draw[ultra thick,color=black] (-1,0+\h) -- (-0.5,0+\h);
\draw[ultra thick,color=black] (1,0+\h) -- (0.5,0+\h);

\draw[ultra thick,color=black] (2,0+\h) --  (3.5,0+\h) ; 

\draw[thick,color=black] (-0.5,-0.5+\h) rectangle (0.5,0.5+\h);

\draw[thick, color=black]
{
(1.2,0+\h) node [right] {\small{$=$}}
(-0.5,0+\h) node [right] {\small{\ \ $G$}}
(4.2,0+\h) node [right] {\small{$+$}}
(0,0) node [right] {\small{\ \ }}
};
\end{tikzpicture} 
\def\ed{2}  \def\v1{1} \def\b{1.4} \def\gs{0.2} \def\lo{0.8} \def\gr{1.6}
\begin{tikzpicture}[scale=1]

\draw[ultra thick,color=black] (-\ed,0) --  (-\gs,0);
\draw[ultra thick,color=black] (\v1,0) -- (\gs,0);
\draw[ultra thick,color=black] (\ed,0) -- (\gr+0.2,0);
\draw[ultra thick,color=black] (\gr-0.2,0) -- (\v1,0);

\draw[thick,color=black] (-\gs,\lo-0.2) rectangle (\gs,\lo+0.2);
\draw[thick,color=black] (-\gs,-0.2) rectangle (\gs,0.2);
\draw[thick,color=black] (-\gs,-\lo-0.2) rectangle (\gs,-\lo+0.2);
\draw[thick,color=black] (\gr-0.2,-0.2) rectangle (\gr+0.2,0.2);

\draw[ultra thick,color=black]  (\gs-\v1-0.05,0) arc (0:-85:\gs-\v1+0.08);
\draw[ultra thick,color=black]  (\gs-\v1-0.05,-0) arc (0:85:\gs-\v1+0.08);

\draw[ultra thick,color=black]  (\v1-\gs+0.05,0) arc (-180:-180+85:\gs-\v1+0.08);
\draw[ultra thick,color=black]  (\v1-\gs+0.05,-0) arc (180:180-85:\gs-\v1+0.08);


\draw[thick, color=black]
{
(-0.5,0) node [right] {\small{\ \ $G$}}
(-0.5,\lo) node [right] {\small{\ \ $G$}}
(-0.5,-\lo) node [right] {\small{\ \ $G$}}
(\gr-0.5,0) node [right] {\small{\ \ $G$}}
};
\end{tikzpicture}
\caption{Schiwinger-Dyson equation for two point function: strand \& single line}
\label{fig:SD eq for two point function strand/single line}
\end{figure}
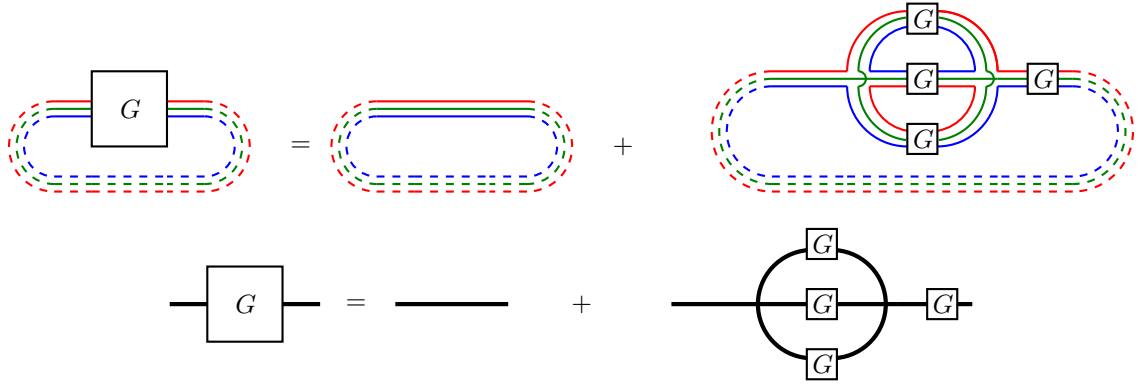

\paragraph{Two point Function:} To begin, let us consider the two point function. In large $N$ limit, full propagator
\begin{equation}
G ( \tau_1, \tau_2) \equiv {1 \over N^3}  \langle \psi_{ijk}(\tau_1) \psi_{ijk} (\tau_2) \rangle 
\end{equation}
is a sum of all possible melonic diagrams in Fig.~\ref{fig:Contributions to Two Point Function in Single Line notation} as we discussed above. This geometric series can be summed to produce a recursion relation as in Fig.~\ref{fig:SD eq for two point function strand/single line} which corresponds to Schwinger-Dyson equation for two point function in large $N$ limit :
\begin{equation}
G(\tau_1,\tau_2)=G_0(\tau_1,\tau_2)+J^2\int d\tau_3 d\tau_4 G_0(\tau_1,\tau_3)[G(\tau_3,\tau_4)]^3G(\tau_4,\tau_2)\label{eq:SD eq for two point function}
\end{equation}
In strong coupling limit $|J\tau_{12}|\gg 1$, a solution is given by~\cite{Sachdev:1992fk,kitaevfirsttalk, KitaevTalks,Polchinski:2016xgd,Jevicki:2016bwu,Maldacena:2016hyu} of \eqref{eq:SD eq for two point function}
\begin{equation}\label{eq: Two Point Function with effective coupling J KT model on lattice}
G(\tau_1,\tau_2)=b{\sgn(\tau_{12})\over |J\tau_{12}|^{1\over 2}}
\end{equation} 
where $b=-(4\pi)^{-{1\over 4}}$. Note that the two point function scales as a power law just like in scale invariant theory. In fact \cite{kitaevfirsttalk,KitaevTalks,Polchinski:2016xgd,Jevicki:2016bwu,Maldacena:2016hyu} showed that in the low energy (or, large $|J\tau|$) limit), the theory has a reparametrization invariance which is spontaneously broken to $SL(2,{\mathbb R} )$ by the above two point function~\eqref{eq: Two Point Function with effective coupling J KT model on lattice}. The reparametrization invariance can also be used to read off the finite temperature results from the zero temperature results. For example, the two point function at finite temperature $T = {1 \over \beta}$ is given by  $G(\tau_1, \tau_2) = b   {\sqrt  \pi \mbox{sgn}(\tau_{12})\over \sqrt{ \beta \sin({  \pi \tau_{12} \over \beta })}}$.


\begin{figure}[t!]
\centering

\begin{tikzpicture}[scale=1]

\def\l{0} \def\s{1.7} \def\r{\l + \s} \def\v{1.7}  \def\gap{0.5}
\draw[ultra thick,color=black] (\l,0) --  (\r,0);
\draw[ultra thick,color=black] (\l,-\v) --  (\r,-\v);

\draw[thick, color=black] (\r + \gap+0.2 ,-\v/2) node [left] {\small{$+$}};

\def\l{ 1.7 + 2 *\gap } \def\s{2} 
\draw[ultra thick,color=black] (\l,0) --  (\r,0);
\draw[ultra thick,color=black] (\l,-\v) --  (\r,-\v);
\def\ra{0.4}
\draw[ultra thick,color=black] (\l + \s/2,0) circle (\ra);
\draw[ultra thick,color=black] (\l + \s/2,-\v) circle (\ra);

\def\gap{0.65}
\draw[thick, color=black] (\r + \gap+0.3 ,-\v/2) node [left] {\small{$+ \cdots$}};
\def\gap{0.5}

\def\l{ 5.4+ 2 * \gap- 0 .4   } \def\s{1.7}  
\draw[ultra thick,color=black] (\l,0) -- (\r,0);
\draw[ultra thick,color=black] (\l,-\v) --  (\r,-\v);
\draw[ultra thick,color=black] (\l + \s/2,0) arc (30:-30:\s);
\draw[ultra thick,color=black] (\l + \s/2,0) arc (180-30:180+30:\s);

\draw[thick, color=black] (\r + \gap+0.2,-\v/2) node [left] {\small{$+$}};

\def\l{ 7.4+ 2 * \gap  } \def\s{1.7}  
\draw[ultra thick,color=black] (\l,0) -- (\r,0);
\draw[ultra thick,color=black] (\l,-\v) --  (\r,-\v);
\draw[ultra thick,color=black] (\l + \s/2,0) arc (30:-30:\s);
\draw[ultra thick,color=black] (\l + \s/2,0) arc (180-30:180+30:\s);
\draw[ultra thick,color=black] (\l+\s/4,0) circle (0.2); 

\draw[thick, color=black] (\r + \gap+0.6,-\v/2) node [left] {\small{$+\cdots $}};

\def\l{ 10.4+ 2 * \gap  } \def\s{2.5} 
\draw[ultra thick,color=black] (\l,0) -- (\r,0);
\draw[ultra thick,color=black] (\l,-\v) --  (\r,-\v);
\draw[ultra thick,color=black] (\l + \s/3,0) arc (30:-30:1.7);
\draw[ultra thick,color=black] (\l + \s/3,0) arc (180-30:180+30:1.7);
\draw[ultra thick,color=black] (\l + 2*\s/3,0) arc (30:-30:1.7);
\draw[ultra thick,color=black] (\l + 2*\s/3,0) arc (180-30:180+30:1.7);

\draw[thick, color=black] (\r + \gap+0.6,-\v/2) node [left] {\small{$+ \cdots$}};

\end{tikzpicture} 
\caption{Contributions to four point function in single line notation}
\label{fig:Contributions to Four Point Function in Single Line notation}
\end{figure}
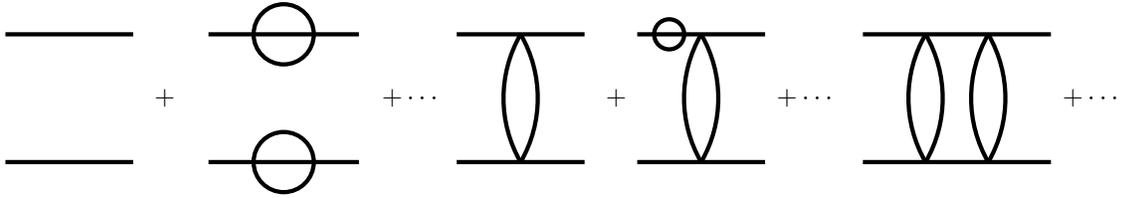

\paragraph{Four point Function:} Next consider the four point functions. The gauge invariant four point function that is the closest analogue of the maximally chaotic four point function in SYK model is the following.
\begin{equation}
F^C(\tau_1,\tau_2,\tau_3,\tau_4) = \langle \psi_{i_1j_1k_1}(\tau_1) \psi_{i_1j_1k_1}(\tau_2) \psi_{i_2j_2k_2}(\tau_3) \psi_{i_2j_2k_2} (\tau_4)\rangle\label{def:cooper channel kt model}
\end{equation}
The superscript $C$ on the four point function denotes that this is just one (what we call \emph{Cooper channel}) of the many four point function channels possible. We will consider the other channels in more detail in Section~\ref{sec:four point function}. For the connected four point function denoted by $\cal F^C$ (appropriately normalized. See Section~\ref{sec:four point function} for details), a similar technique of summing up the leading diagrams which are called ladder diagrams (See Fig.~\ref{fig:Contributions to Four Point Function in Single Line notation}) results in the following Schwinger-Dyson equations for four point functions  
\begin{equation}\label{4 point recursion}
{\cal F}^C(\tau_1,\tau_2,\tau_3,\tau_4)=\int dt dt' \; \mathcal{K}(\tau_1,\tau_2,\tau,\tau') {\cal F}^C (\tau,\tau',\tau_3,\tau_4)
\end{equation}
%
where the kernel $\mathcal{K}(\tau_1,\tau_2,\tau_3,\tau_4)$ is the same as that of SYK model and is given by
\begin{equation}
\mathcal{K}(\tau_1,\tau_2,\tau_3,\tau_4)\equiv -3 \local^2 G(\tau_{13})G(\tau_{24})[G(\tau_{34})]^2\label{eq:SYK kernel}
\end{equation}
It turns out that the four point function diverges if we use for $G(t)$ the two point function obtained in \eqref{eq: Two Point Function with effective coupling J KT model on lattice} in the strict large $|\local \tau |$ limit. Working at finite temperature, one can keep the leading ${1 \over \beta \effcoupling}$ corrections and \cite{Maldacena:2016hyu} were able to solve for the four point function by solving the recursion relation \eqref{4 point recursion}. For details we refer the reader to~\cite{Maldacena:2016hyu}.

\paragraph{Chaos in KT Model and the Lyapunov exponent:} An important property of the SYK, KT models and the generalizations thereof that we will discuss in this work is that they are maximally chaotic. This statement can be made more precise by using the results of \cite{Maldacena:2016hyu} which we will describe below. Consider an out-of-time-ordered four point function
\begin{equation}\label{def:Chaos Correlator Cooper}
F^C(t) \equiv \mbox{Tr} (  e^{-\beta H \over 4}    \psi_{i_1 j_1 k_1}(t) e^{-\beta H \over 4} \psi_{i_2 j_2 k_2}(0) e^{-\beta H \over 4}   \psi_{i_1 j_1 k_1}(t) e^{-\beta H \over 4} \psi_{i_2 j_2 k_2}(0) )
\end{equation}
\cite{Maldacena:2016hyu} showed that under reasonable assumptions, this four point function has the following behavior at large time $t$
\begin{equation}
F^C(t) = F^C_d -e^{\lambda_L ( t - t_*)} 
\end{equation}
and the exponent $\lambda_L$ (called Lyapunav exponent or chaos exponent) satisfies the chaos bound~\cite{Maldacena:2015waa} 
\begin{equation}
\lambda_L \le {2 \pi \over \beta}
\end{equation}
Here, $F^C_d$ is a disconnected piece, and $t_\ast$ is the scrambling time, and typically in a large $N$ theory $t_\ast \sim \log N^3$. At $t\sim t_\ast$, the second term starts to give a significant contribution to four point function. As we will see below and also later, the bound is saturated in KT models and its generalization. 

To compute the chaos exponent, one needs to look at the large time behavior of the connected part of the correlator in \eqref{def:Chaos Correlator Cooper}. This is obtained by taking an appropriate analytic continuation of the connected four point function ${\cal F}^C$obtained by solving~\eqref{4 point recursion}. This gives~\cite{Klebanov:2016xxf}
\begin{equation}
F(t) - F_d^C =  {1\over N^3} \left(c_1\  \beta J  e^{2 \pi t \over \beta} + {\cal O}( (\beta J)^0 ) \right) 
\end{equation} 
where $c_1$ is a number of order $\sim 1$.
From the exponential, one can read off the Lyapunov exponent to be
\begin{equation}
\lambda_L = {2 \pi \over \beta}
\end{equation}


\subsection{SYK model and Gu-Qi-Stanford Generalization Thereof} 
\label{sec:review of SYK}

In this work, we will be considering lattice  generalizations of KT models. A similar lattice generalization of SYK model has already been studied by \cite{Gu:2016oyy} (from now on, we will call it \emph{GQS SYK} model ) which we describe below. Consider $N_{\text{\tiny SYK}}$ fermions $\psi_i^a$ ($i=1,2,\cdots, N_{\text{\tiny SYK}}$) on each point of a $1$d lattice ($a=1,2,\cdots, L$). They interact via an SYK-like onsite interaction and also via a quartic nearest neighbor interaction which is also random. The Hamiltonian is given by
\begin{equation}
H=\sum_{a=1}^{L}\Big(\sum_{1\leq k<l<m<n\leq N}j_{klmn}\psi^{a}_{k} \psi^{a}_{l}\psi^{a}_{m}\psi^{a}_{n}+\sum_{1\leq k<l<m<n\leq N}j^{\prime}_{klmn}\psi^{a}_{k} \psi^{a}_{l}\psi^{a+1}_{m}\psi^{a+1}_{n}\Big)
\end{equation}
where $j_{klmn}$ and $j^{\prime}_{klmn}$ are random couplings drawn from gaussian ensemble such that $\overline{ j_{klmn}^2 } =\frac{J^{2} _{0} 3!}{N^3}$ and $\overline{ (j^\prime_{klmn})^2 }=\frac{J_1^2}{N^3}$, respectively. 

We will now briefly review the large $N_{\text{\tiny SYK}}$ results for this model. In this model, the two point function of fermions $G(\tau_1,\tau_2)$ defined as ${1 \over N_{\text{\tiny SYK}}} 	\sum_i \overline{\langle T[ \psi^{(a)}_i(\tau_1) \psi^{(b)}_i(\tau_2)  ]\rangle} = G(\tau_1,\tau_2) \delta^{ab}$ and satisfies the same Schwinger-Dyson equation as in \eqref{eq:SD eq for two point function} with the coupling $J$ replaced by an effective coupling
\begin{equation}
\effcoupling\equiv\sqrt{J_0^2 + J_1^2}
\end{equation}
Hence, the corresponding two point function becomes
\begin{equation}
G(\tau_1,\tau_2)=b{\sgn(\tau_{12})\over |\effcoupling \tau_{12}|^{1\over 2}}
\end{equation}
The four point function is defined in a analogous way as
\begin{equation}\label{GQS4PointFunction}
{\cal F}_{ab}(\tau_1,\tau_2,\tau_3,\tau_4) \equiv {1 \over N_{\text{\tiny SYK}}}\left[ \sum_{i,j=1}^N \overline{\langle  T(\psi_i^{a}(\tau_1) \psi_i^{a}(\tau_2) \psi_j^{b}(\tau_3) \psi_j^{b}(\tau_4) ) \rangle} -  N_{\text{\tiny SYK}}^2 G(\tau_{12})G(\tau_{34})\right]
\end{equation}
By using lattice translational invariance, it is convenient to shift to momentum space $p$. Using effective action techniques, \cite{Gu:2016oyy} obtained the following expression for $\cal F$:
\begin{equation}
{\cal F}_p(\tau_1,\tau_2,\tau_3,\tau_4) = {1 \over 1 - s(p) {\cal K}} {\cal F}_0  \label{eq:GuQiStanford4Point}
\end{equation}
where the structure factor $s(p)\equiv 1 + {2 J_1^2 \over 3 \effcoupling }(\cos p-1)$ and the kernel $\cal K$ is given by 
\begin{equation}
  {\cal K}(\tau_1,\tau_2,\tau_3,\tau_4) = - 3 \effcoupling^2 G(\tau_{13}) G(\tau_{24}) [ G(\tau_{34}) ]^2  
\end{equation}
and ${\cal F}_0 = - G(\tau_{13}) G(\tau_{24})+  G(\tau_{14}) G(\tau_{23})$. These results are reminiscent of SYK/Tensor models except that the four point function now has the dependence on spatial momentum $p$.  By relating the above 4 point function to stress tensor correlation function, \cite{Gu:2016oyy} extracted the Diffusion constant to be 
\begin{equation}\label{GQSDiffusionConstant}
D = { 2 \pi J_1^2 \over 3 \sqrt{2}\effcoupling \alpha_K}
\end{equation}
where $\alpha_K \approx 2.852$ is a constant which was defined in~\cite{Maldacena:2016hyu,Jevicki:2016ito} which appears in low energy effective theory for SYK model. 

\paragraph{Chaos in Gu-Qi-Stanford generalization of SYK model:} Since now the theory is defined on a lattice, one can study the chaotic behavior in spatial direction in addition to the temporal direction. Consider the out-of-time-ordered correlator with operators at different spatial locations
\begin{equation}
{F}_{a}(t) \equiv   {1 \over N_{SYK}^2} \mbox{Tr} \left[ e^{-{\beta H \over 4}} \psi^a_i(t)e^{-{\beta H \over 4}}  \psi^0_j(0) e^{-{\beta H \over 4}} \psi^a_i(t)   e^{-{\beta H \over 4}} \psi^0_j(0) \right]
\end{equation}
The large time behavior of the above correlator is expected to be
\begin{equation}
{F}_{a}(t)  -{F}_{d}  \propto e^{\lambda_L (t - a/v_B)}
\end{equation}
where ${F}_{d}$ is a constant independent of $t,a$. $v_B$ is defined to be the butterfly velocity: it characterizes the rate at which chaos propagates in space. 

This out-of-time-ordered correlator can also be obtained as an appropriate analytic continuation of the Euclidean correlator obtained by solving \eqref{GQS4PointFunction}. \cite{Gu:2016oyy} obtained the Lyapunov exponent $\lambda_L$ and the butterfly velocity $v_B$ to be 
\begin{equation}
\lambda_L = {2 \pi \over \beta} \hspace{10mm} v_B^2= {2 \pi D \over \beta}
\end{equation}
where $\beta$ is the inverse temperature and $D$ is the diffusion constant given in \eqref{GQSDiffusionConstant}. As we will see later, we will find similar results for many of the correlators in the models we consider in this work. It is interesting to note that this saturates the bound proposed for incoherent metals in \cite{Hartnoll:2014lpa} and verified for some holographic duals in \cite{Blake:2016wvh,Blake:2016sud}. We also note here that there are other lattice generalizations of SYK model \cite{Berkooz:2016cvq,Jian:2017unn,Jian:2017jfl}. It will be interesting to find tensor model versions of these generalizations.

\section{Klebanov-Tarnopolsky Chain Model}
\label{sec:large N KT model on lattice}

We will be interested in studying lattice generalizations of KT model. To be explicit, we will first consider the simplest class of such models: KT model on a lattice (which we term as \emph{KT chain} model) of $L$ sites with the nearest neighbor interaction of Gu-Qi-Stanford~\cite{Gu:2016oyy} type. Although as we will see later in Section~\ref{sec:GW model} and~\ref{sec:generalized GW moel}, most of the techniques will generalize in a straightforward way to other models to be described in the Section~\ref{sec:models}. In this section, we will study large $N$ diagrammatics of two point function and four point functions of three channels which we will shortly define.

\begin{figure}[t!]
\centering
\begin{tikzpicture}[scale=0.85]
\draw[thick,color=red] (-1.5,0.5) -- (-0.5,0.5) -- (-0.5,1.5);
\draw[thick,color=blue] (-1.5,-0.5) -- (-0.5,-0.5) -- (-0.5,-1.5);
\draw[thick,color=red] (1.5,-0.5) -- (0.5,-0.5) -- (0.5,-1.5);
\draw[thick,color=blue] (1.5,0.5) -- (0.5,0.5) -- (0.5,1.5);

\draw[thick,color=black!50!green] (-1.5,0.0) -- (1.5,0.0) ;
\draw[thick,color=black!50!green] (0.0,1.5) -- (0.0,0.2) ;
\draw[thick,color=black!50!green] (0.0,-1.5) -- (0.0,-0.2) ;
\draw[color=blue] (0.0,0.2) arc (90:-90:0.2);
\draw[thick, color=black]
{
(0.0,-1.7) node [below] {\small{$ \local N^{-{3 \over 2}}$}}
(-1.5,0) node [left] {\small{$ a  $}}
(2.0,0) node [left] {\small{$ a  $}}
(0.3,1.7) node [left] {\small{$ a  $}}
(0.3,-1.7) node [left] {\small{$ a  $}}
};
\end{tikzpicture}\\
\begin{tikzpicture}[scale=0.85]
\draw[thick,color=red] (-1.5,0.5) -- (-0.5,0.5) -- (-0.5,1.5);
\draw[thick,color=blue] (-1.5,-0.5) -- (-0.5,-0.5) -- (-0.5,-1.5);
\draw[thick,color=red] (1.5,-0.5) -- (0.5,-0.5) -- (0.5,-1.5);
\draw[thick,color=blue] (1.5,0.5) -- (0.5,0.5) -- (0.5,1.5);

\draw[thick,color=black!50!green] (-1.5,0.0) -- (1.5,0.0) ;
\draw[thick,color=black!50!green] (0.0,1.5) -- (0.0,0.2) ;
\draw[thick,color=black!50!green] (0.0,-1.5) -- (0.0,-0.2) ;
\draw[color=blue] (0.0,0.2) arc (90:-90:0.2);
\draw[thick, color=black]
{
(0.0,-1.7) node [below] {\small{$ {1\over \sqrt{2}}\hop_\R N^{-{3 \over 2}}$}}
(-1.5,0) node [left] {\small{$ a  $}}
(2.65,0) node [left] {\small{$ a\pm1  $}}
(0.3,1.7) node [left] {\small{$ a  $}}
(0.7,-1.7) node [left] {\small{$ a\pm1  $}}
};
\end{tikzpicture} 
 \hspace{0.5cm}
\begin{tikzpicture}[scale=0.85]
\draw[thick,color=black!50!green] (-1.5,0.5) -- (-0.5,0.5) -- (-0.5,1.5);
\draw[thick,color=blue] (-1.5,-0.5) -- (-0.5,-0.5) -- (-0.5,-1.5);
\draw[thick,color=black!50!green] (1.5,-0.5) -- (0.5,-0.5) -- (0.5,-1.5);
\draw[thick,color=blue] (1.5,0.5) -- (0.5,0.5) -- (0.5,1.5);
\draw[thick,color=red] (-1.5,0.0) -- (1.5,0.0) ;
\draw[thick,color=red] (0.0,1.5) -- (0.0,0.2) ;
\draw[thick,color=red] (0.0,-1.5) -- (0.0,-0.2) ;
\draw[color=red] (0.0,0.2) arc (90:-90:0.2);
\draw[thick, color=black]
{
(0.0,-1.7) node [below] {\small{${1\over \sqrt{2}} \hop_\G N^{-{3 \over 2}}$}}
(-1.5,0) node [left] {\small{$ a  $}}
(2.65,0) node [left] {\small{$ a\pm1  $}}
(0.3,1.7) node [left] {\small{$ a  $}}
(0.7,-1.7) node [left] {\small{$ a\pm1  $}}
};
\end{tikzpicture}
 \hspace{0.5cm}
\begin{tikzpicture}[scale=0.85]
\draw[thick,color=blue] (-1.5,0.5) -- (-0.5,0.5) -- (-0.5,1.5);
\draw[thick,color=red] (-1.5,-0.5) -- (-0.5,-0.5) -- (-0.5,-1.5);
\draw[thick,color=blue] (1.5,-0.5) -- (0.5,-0.5) -- (0.5,-1.5);
\draw[thick,color=red] (1.5,0.5) -- (0.5,0.5) -- (0.5,1.5);
\draw[thick,color=black!50!green] (-1.5,0.0) -- (1.5,0.0) ;
\draw[thick,color=black!50!green] (0.0,1.5) -- (0.0,0.2) ;
\draw[thick,color=black!50!green] (0.0,-1.5) -- (0.0,-0.2) ;
\draw[color=black!50!green] (0.0,0.2) arc (90:-90:0.2);
\draw[thick, color=black]
{
(0.0,-1.7) node [below] {\small{${1\over \sqrt{2}} \hop_\B N^{-{3 \over 2}}$}}
(-1.5,0) node [left] {\small{$ a  $}}
(2.65,0) node [left] {\small{$ a\pm1  $}}
(0.3,1.7) node [left] {\small{$ a  $}}
(0.7,-1.7) node [left] {\small{$ a\pm1  $}}
};
\end{tikzpicture} 
\caption{The Vertices of KT chain model on the lattice with the nearest neighbor interaction (GQS interaction)}
\label{fig:vertex for KT model on lattice}
\end{figure}
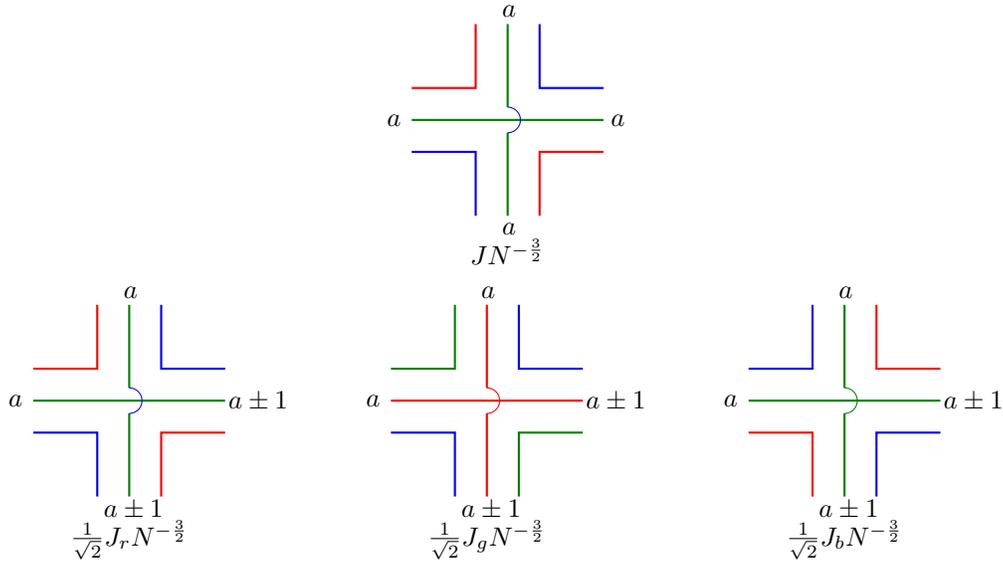

In KT chain model given by
\begin{align}
&H={\local N^{-{3\over 2}} \over 4} \sum_{a=1}^L  \psi^a_{i_1 j_1 k_1}\psi^a_{i_1 j_2 k_2}\psi^a_{i_2 j_1 k_2}\psi^a_{i_2 j_2 k_1}+{\hop_\R N^{-{3\over 2}} \over 2\sqrt 2} \sum_{a=1}^{L} \psi^{a}_{i_1 j_1 k_1}\psi^{a}_{i_1 j_2 k_2}\psi^{a+1}_{i_2 j_1 k_2}\psi^{a+1}_{i_2 j_2 k_1}\cr
&\hspace{0.5cm}+{\hop_\G N^{-{3\over 2}} \over 2\sqrt 2}  \sum_{a=1}^{L} \psi^{a}_{i_1 j_1 k_1}\psi^{a+1}_{i_1 j_2 k_2}\psi^{a}_{i_2 j_1 k_2}\psi^{a+1}_{i_2 j_2 k_1}+{\hop_\B N^{-{3\over 2}} \over 2\sqrt 2}  \sum_{a=1}^{L} \psi^{a}_{i_1 j_1 k_1}\psi^{a+1}_{i_1 j_2 k_2}\psi^{a+1}_{i_2 j_1 k_2}\psi^{a}_{i_2 j_2 k_1}\label{eq:hamiltonian of KT chain model}
\end{align}
there are four basic vertices: one on-site interaction and the three nearest-neighbor interactions shown in Fig.~\ref{fig:vertex for KT model on lattice}. The additional structure compared to the KT model is that the fermions now have a lattice index. But since number of lattice points $L \ll N$, this does not affect the large $N$ diagrammatics - the $n$-point functions are still dominated by $n$-point melonic diagrams.

%
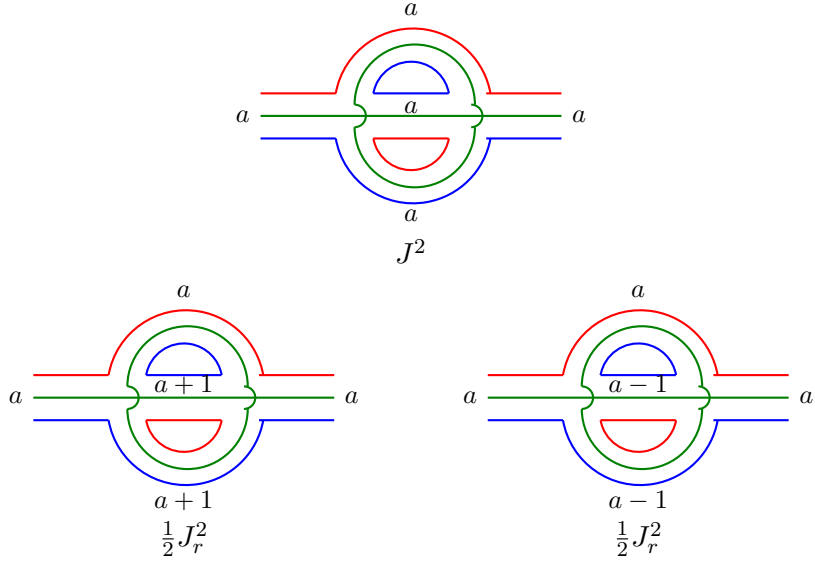
\begin{figure}[t!]
\centering
\begin{tikzpicture}[scale=1]
\draw[thick,color=red] (-2,0.3) -- (-1,0.3);
\draw[thick,color=red] (-1,0.3) arc (170:10:1.04403);
\draw[thick,color=red] (2,0.3) -- (1,0.3); 

\draw[thick,color=blue] (-0.5,0.3) -- (0.5,0.3);
\draw[thick,color=blue] (-0.5,0.3) arc (170:10:0.5095);

\draw[thick,color=black!50!green] (-2,0) -- (2,0);
\draw [thick,color=black!50!green] (-0.75,0.15) arc (180:0:0.8);
\draw [thick,color=black!50!green] (-0.75,-0.15) arc (-180:0:0.8);
\draw [thick,color=black!50!green] (-0.75,0.15) arc (90:-90:0.15); 
\draw [thick,color=black!50!green] (0.8,0.15) arc (90:-90:0.15);

\draw[thick,color=red] (-0.5,-0.3) -- (0.5,-0.3);
\draw[thick,color=red] (-0.5,-0.3) arc (-170:-10:0.5095);

\draw[thick,color=blue] (-2,-0.3) -- (-1,-0.3);
\draw[thick,color=blue] (-1,-0.3) arc (-170:-10:1.04403);
\draw[thick,color=blue] (2,-0.3) -- (1,-0.3);
\draw[thick, color=black]
{
(-2,0) node [left] {\small{$ a $}}
(2,0) node [right] {\small{$ a $}}
(0,-1.1) node [below] {\small{$ a $}}
(0,1.2) node [above] {\small{$ a $}}
(0,-0.1) node [above] {\small{$ a $}}
(0,-1.5) node [below] { {$ J^2 $}}
};
\end{tikzpicture}\\
\begin{tikzpicture}[scale=1]
\draw[thick,color=red] (-2,0.3) -- (-1,0.3);
\draw[thick,color=red] (-1,0.3) arc (170:10:1.04403);
\draw[thick,color=red] (2,0.3) -- (1,0.3); 

\draw[thick,color=blue] (-0.5,0.3) -- (0.5,0.3);
\draw[thick,color=blue] (-0.5,0.3) arc (170:10:0.5095);

\draw[thick,color=black!50!green] (-2,0) -- (2,0);
\draw [thick,color=black!50!green] (-0.75,0.15) arc (180:0:0.8);
\draw [thick,color=black!50!green] (-0.75,-0.15) arc (-180:0:0.8);
\draw [thick,color=black!50!green] (-0.75,0.15) arc (90:-90:0.15); 
\draw [thick,color=black!50!green] (0.8,0.15) arc (90:-90:0.15);

\draw[thick,color=red] (-0.5,-0.3) -- (0.5,-0.3);
\draw[thick,color=red] (-0.5,-0.3) arc (-170:-10:0.5095);

\draw[thick,color=blue] (-2,-0.3) -- (-1,-0.3);
\draw[thick,color=blue] (-1,-0.3) arc (-170:-10:1.04403);
\draw[thick,color=blue] (2,-0.3) -- (1,-0.3);
\draw[thick, color=black]
{
(-2,0) node [left] {\small{$ a $}}
(2,0) node [right] {\small{$ a $}}
(0,-1.1) node [below] {\small{$ a+1 $}}
(0,1.2) node [above] {\small{$ a $}}
(0,-0.1) node [above] {\small{$ a+1 $}}
(0,-1.5) node [below] { {$ {1\over 2} J_\R^2 $}}
};
\end{tikzpicture}\hspace{1cm}
\begin{tikzpicture}[scale=1]
\draw[thick,color=red] (-2,0.3) -- (-1,0.3);
\draw[thick,color=red] (-1,0.3) arc (170:10:1.04403);
\draw[thick,color=red] (2,0.3) -- (1,0.3); 

\draw[thick,color=blue] (-0.5,0.3) -- (0.5,0.3);
\draw[thick,color=blue] (-0.5,0.3) arc (170:10:0.5095);

\draw[thick,color=black!50!green] (-2,0) -- (2,0);
\draw [thick,color=black!50!green] (-0.75,0.15) arc (180:0:0.8);
\draw [thick,color=black!50!green] (-0.75,-0.15) arc (-180:0:0.8);
\draw [thick,color=black!50!green] (-0.75,0.15) arc (90:-90:0.15); 
\draw [thick,color=black!50!green] (0.8,0.15) arc (90:-90:0.15);

\draw[thick,color=red] (-0.5,-0.3) -- (0.5,-0.3);
\draw[thick,color=red] (-0.5,-0.3) arc (-170:-10:0.5095);

\draw[thick,color=blue] (-2,-0.3) -- (-1,-0.3);
\draw[thick,color=blue] (-1,-0.3) arc (-170:-10:1.04403);
\draw[thick,color=blue] (2,-0.3) -- (1,-0.3);
\draw[thick, color=black]
{
(-2,0) node [left] {\small{$ a $}}
(2,0) node [right] {\small{$ a $}}
(0,-1.1) node [below] {\small{$ a-1 $}}
(0,1.2) node [above] {\small{$ a $}}
(0,-0.1) node [above] {\small{$ a-1 $}}
(0,-1.5) node [below] { {$ {1\over 2}J_\R^2 $}}
};
\end{tikzpicture}
\caption{Contribution of interactions to melonic diagram. One can easily generate contributions from vertices of other colors by permuting lattice indices in the internal strand of melonic diagram.}
\label{fig:melonic diagram vertex contribution}
\end{figure}
%
\paragraph{Two point function:}We begin with an analysis of two point function. Imposing gauge invariance, there is only one possible contraction of external indices, namely
\begin{equation}\label{def:KT Lattice 2 point function}
\langle \psi^{a_1}_{ijk}(\tau_1) \psi^{a_2}_{ijk} (\tau_2)  \rangle
\end{equation}
From the structure of two point melonic diagrams (the simplest ones are given in Figure~\ref{fig:melonic diagram vertex contribution}), it is easy to deduce that these melonic diagrams always have the same lattice index on the two external legs. This result\footnote{As we will see later, this is also the feature of all models which obey \emph{Unique Last Fermion~(ULF) Property} defined in Section~\ref{sec:models}} also follows from the $\mathbb{Z}_2^L$ symmetry of KT chain model which corresponds to fermion parity conservation on each site as in the GQS SYK model~\cite{Gu:2016oyy}. Hence, to leading order in $N$, the two point function~\eqref{def:KT Lattice 2 point function} is diagonal in lattice indices. Thus, together with time translational invariance, it is enough to consider two point function   
%
\begin{equation}
G ( \tau_1, \tau_2) = G(\tau_1-\tau_2) \equiv {1 \over N^3}  \langle \psi^{a}_{ijk}(\tau_1) \psi^{a}_{ijk} (\tau_2) \rangle 
\end{equation}
where the indices $a$ are not summed. Since the RHS is independent of $a$, we have dropped the index $a$ on $G$ too. Also, since the $N$ scaling of the two point function is $N^3$ we have introduced a factor of ${1 \over N^3}$ in the definition, such that $G(\tau_1,\tau_2)$ does not scale with $N$ to leading order. 

Let us first consider the structure of the simplest two point melonic diagram. One can easily see that there are seven such diagrams depending on the details of the lattice index on the internal propagators (three of these diagrams are shown in Fig.~\ref{fig:melonic diagram vertex contribution}). One of the diagrams arises due to the on-site interaction, while the other six arise from the hopping interactions. Each two point melonic diagram gives the same contribution as that of the single-site KT model in Section~\ref{sec:single-site KT model}, but with different coupling constants. Note that we chose the numerical factor of hopping interaction in a way that the contribution of each melonic diagram in Fig.~\ref{fig:melonic diagram vertex contribution} with hopping interaction is ${1\over 2} J_{\mathfrak c}^2$ for any given color $\mathfrak c$ $(=\R, \G ,\B)$ while on-site interaction gives $\local^2$. Hence, the final contribution to two point function is the sum of these seven diagrams, which is effectively equivalent to KT model with effective coupling constant
\begin{equation}
\effcoupling^2 \equiv \local^2+\hop_\R^2+\hop_\G^2+\hop_\B^2\label{def:effective coupling constant KT model}
\end{equation}
One can easily extend these observations to any two point melonic diagram and this results in the same Schwinger-Dyson equation for $G(\tau_1,\tau_2)$ as in Figure \ref{fig:SD eq for two point function strand/single line}. We get
\begin{equation}
G(\tau_1,\tau_2)=G_0(\tau_1,\tau_2)+\effcoupling^2\int d\tau_3 d\tau_4 G_0(\tau_1,\tau_3)[G(\tau_3,\tau_4)]^3G(\tau_4,\tau_2)\label{eq:SD eq for two point function in KT model on lattice}
\end{equation}
As in KT model, this Schwinger-Dyson equation has reparametrization symmetry in the strong coupling limit $|\effcoupling \tau_{12}|\gg 1$, and one immediately has a solution~\cite{Sachdev:1992fk,kitaevfirsttalk, KitaevTalks,Polchinski:2016xgd,Jevicki:2016bwu,Maldacena:2016hyu} 
\begin{equation}\label{Two Point Function with effective coupling J}
G(\tau_1,\tau_2)=b{\sgn(\tau_{12})\over |\effcoupling \tau_{12}|^{1\over 2}}
\end{equation} 
where $b=-(4\pi)^{-{1\over 4}}$. Note that the theory has the same $SL(2,R)$ as single-site KT model.

\section{Four Point Function in Klebanov-Tarnopolsky Chain Model}

\label{sec:four point function}

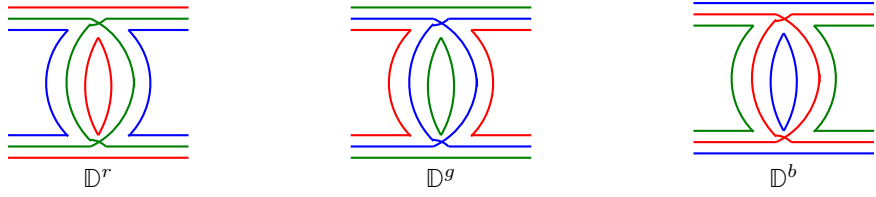
\begin{figure}[t!]
\centering
\begin{tikzpicture}[scale=1]
\def\xa{0}

\draw[thick,color=red] (\xa+-1.2,1) -- (\xa+1.2,1);

\draw[thick,color=blue]  (\xa+-1.2,0.7) -- (\xa+-0.4,0.7);
\draw[thick,color=blue]  (\xa+1.2,0.7) --  (\xa+0.4,0.7);
\draw[thick,color=blue]  (\xa+0.4,0.7) arc (45:-45:1);
\draw[thick,color=blue]  (\xa+-0.4,0.7) arc (135:225:1);

\draw[thick,color=black!50!green]  (\xa+-1.2,0.85)  -- (\xa+-0.1,0.85);
\draw[thick,color=black!50!green] (\xa+1.2,0.85)  -- (\xa+0.1,0.85);
\draw[thick,color=black!50!green] (\xa+-1.2,-0.85)  -- (\xa+-0.1,-0.85);
\draw[thick,color=black!50!green] (\xa+1.2,-0.85)  -- (\xa+0.1,-0.85);

\draw[thick,color=black!50!green] (\xa+-0.1,0.85) arc (65:0:1.0) ;
\draw[thick,color=black!50!green] (\xa+0.1,0.85) arc (-45:-45-45:0.3) ;
\draw[thick,color=black!50!green] (\xa+-0.1,0.78) arc (135:180+45:1.1) ;

\draw[thick,color=black!50!green] (\xa+-0.1,-0.85) arc (-65:0:1.0) ;
\draw[thick,color=black!50!green] (\xa+0.1,-0.85) arc (45:45+45:0.3) ;

\draw[thick,color=blue] (\xa+-1.2,-0.7) -- (\xa+-0.4,-0.7);
\draw[thick,color=blue] (\xa+1.2,-0.7) -- (\xa+0.4,-0.7);

\draw[thick,color=red] (\xa+0,0.6) arc (30:-30:1.3) ;
\draw[thick,color=red] (\xa+0,0.6) arc (180-30:180+30:1.3) ;
\draw[thick,color=red] (\xa+-1.2,-1) -- (\xa+1.2,-1);
\draw[thick, color=black]{
(\xa,-1) node [below] {\small{$ \dipole^\R$}}
};
\end{tikzpicture}\hspace{2cm}
\begin{tikzpicture}[scale=1]
\def\xa{0}

\draw[thick,color=black!50!green] (\xa+-1.2,1) -- (\xa+1.2,1);

\draw[thick,color=red]  (\xa+-1.2,0.7) -- (\xa+-0.4,0.7);
\draw[thick,color=red]  (\xa+1.2,0.7) --  (\xa+0.4,0.7);
\draw[thick,color=red]  (\xa+0.4,0.7) arc (45:-45:1);
\draw[thick,color=red]  (\xa+-0.4,0.7) arc (135:225:1);

\draw[thick,color=blue]  (\xa+-1.2,0.85)  -- (\xa+-0.1,0.85);
\draw[thick,color=blue] (\xa+1.2,0.85)  -- (\xa+0.1,0.85);
\draw[thick,color=blue] (\xa+-1.2,-0.85)  -- (\xa+-0.1,-0.85);
\draw[thick,color=blue] (\xa+1.2,-0.85)  -- (\xa+0.1,-0.85);

\draw[thick,color=blue] (\xa+-0.1,0.85) arc (65:0:1.0) ;
\draw[thick,color=blue] (\xa+0.1,0.85) arc (-45:-45-45:0.3) ;
\draw[thick,color=blue] (\xa+-0.1,0.78) arc (135:180+45:1.1) ;

\draw[thick,color=blue] (\xa+-0.1,-0.85) arc (-65:0:1.0) ;
\draw[thick,color=blue] (\xa+0.1,-0.85) arc (45:45+45:0.3) ;

\draw[thick,color=red] (\xa+-1.2,-0.7) -- (\xa+-0.4,-0.7);
\draw[thick,color=red] (\xa+1.2,-0.7) -- (\xa+0.4,-0.7);

\draw[thick,color=black!50!green] (\xa+0,0.6) arc (30:-30:1.3) ;
\draw[thick,color=black!50!green] (\xa+0,0.6) arc (180-30:180+30:1.3) ;
\draw[thick,color=black!50!green] (\xa+-1.2,-1) -- (\xa+1.2,-1);
\draw[thick, color=black]{
(\xa,-1) node [below] {\small{$ \dipole^\G$}}
};
\end{tikzpicture}\hspace{2cm}
\begin{tikzpicture}[scale=1]
\def\xa{0}

\draw[thick,color=blue] (\xa+-1.2,1) -- (\xa+1.2,1);

\draw[thick,color=black!50!green]  (\xa+-1.2,0.7) -- (\xa+-0.4,0.7);
\draw[thick,color=black!50!green]  (\xa+1.2,0.7) --  (\xa+0.4,0.7);
\draw[thick,color=black!50!green]  (\xa+0.4,0.7) arc (45:-45:1);
\draw[thick,color=black!50!green]  (\xa+-0.4,0.7) arc (135:225:1);

\draw[thick,color=red]  (\xa+-1.2,0.85)  -- (\xa+-0.1,0.85);
\draw[thick,color=red] (\xa+1.2,0.85)  -- (\xa+0.1,0.85);
\draw[thick,color=red] (\xa+-1.2,-0.85)  -- (\xa+-0.1,-0.85);
\draw[thick,color=red] (\xa+1.2,-0.85)  -- (\xa+0.1,-0.85);

\draw[thick,color=red] (\xa+-0.1,0.85) arc (65:0:1.0) ;
\draw[thick,color=red] (\xa+0.1,0.85) arc (-45:-45-45:0.3) ;
\draw[thick,color=red] (\xa+-0.1,0.78) arc (135:180+45:1.1) ;

\draw[thick,color=red] (\xa+-0.1,-0.85) arc (-65:0:1.0) ;
\draw[thick,color=red] (\xa+0.1,-0.85) arc (45:45+45:0.3) ;

\draw[thick,color=black!50!green] (\xa+-1.2,-0.7) -- (\xa+-0.4,-0.7);
\draw[thick,color=black!50!green] (\xa+1.2,-0.7) -- (\xa+0.4,-0.7);

\draw[thick,color=blue] (\xa+0,0.6) arc (30:-30:1.3) ;
\draw[thick,color=blue] (\xa+0,0.6) arc (180-30:180+30:1.3) ;
\draw[thick,color=blue] (\xa+-1.2,-1) -- (\xa+1.2,-1);
\draw[thick, color=black]{
(\xa,-1) node [below] {\small{$\dipole^\B$}}
};
\end{tikzpicture}
\caption{ Dipoles of three colors in KT Model}
\label{fig:dipole single site KT model}
\end{figure}

Before we discuss the large $N$ diagrammatics of four point functions, it is useful understand the structure of ladder diagrams. The standard terminology is to term the horizontal lines at the top and bottom as legs of a ladder and the lines connecting them as rungs of the ladder. It is also useful to define a basic building block for all the ladder diagrams: \emph{dipole} introduced in~\cite{Gurau:2010ba,Gurau:2011xq,Gurau:2011xp,Gurau:2011kk,Gurau:2016lzk}. A dipole is a four point function with two vertices shown in Fig.~\ref{fig:dipole single site KT model} for the case of single-site KT model. There are 3 types of dipoles of order $\mathcal{O}(N^{-2})$ depending on the color (\ie\ $\R, \G, \B$) which is transmitted along the ladder, and they will be denoted by $\dipole^{\mathfrak c}$ (${\mathfrak c}=\R, \G, \B$). It should be clear that stringing the dipoles together gives the full ladder diagram.

\begin{figure}[b!]
\centering
\begin{tikzpicture}[scale=1]
\def\xa{0}

\draw[thick,color=red] (\xa+-1.2,1) -- (\xa+1.2,1);

\draw[thick,color=blue]  (\xa+-1.2,0.7) -- (\xa+-0.4,0.7);
\draw[thick,color=blue]  (\xa+1.2,0.7) --  (\xa+0.4,0.7);
\draw[thick,color=blue]  (\xa+0.4,0.7) arc (45:-45:1);
\draw[thick,color=blue]  (\xa+-0.4,0.7) arc (135:225:1);

\draw[thick,color=black!50!green]  (\xa+-1.2,0.85)  -- (\xa+-0.1,0.85);
\draw[thick,color=black!50!green] (\xa+1.2,0.85)  -- (\xa+0.1,0.85);
\draw[thick,color=black!50!green] (\xa+-1.2,-0.85)  -- (\xa+-0.1,-0.85);
\draw[thick,color=black!50!green] (\xa+1.2,-0.85)  -- (\xa+0.1,-0.85);

\draw[thick,color=black!50!green] (\xa+-0.1,0.85) arc (65:0:1.0) ;
\draw[thick,color=black!50!green] (\xa+0.1,0.85) arc (-45:-45-45:0.3) ;
\draw[thick,color=black!50!green] (\xa+-0.1,0.78) arc (135:180+45:1.1) ;

\draw[thick,color=black!50!green] (\xa+-0.1,-0.85) arc (-65:0:1.0) ;
\draw[thick,color=black!50!green] (\xa+0.1,-0.85) arc (45:45+45:0.3) ;

\draw[thick,color=blue] (\xa+-1.2,-0.7) -- (\xa+-0.4,-0.7);
\draw[thick,color=blue] (\xa+1.2,-0.7) -- (\xa+0.4,-0.7);

\draw[thick,color=red] (\xa+0,0.6) arc (30:-30:1.3) ;
\draw[thick,color=red] (\xa+0,0.6) arc (180-30:180+30:1.3) ;
\draw[thick,color=red] (\xa+-1.2,-1) -- (\xa+1.2,-1);
\draw[thick, color=black]{
(\xa,-1) node [below] {\small{$ \dipole^\R$}}
};
\draw[thick, color=black]
{
(-1.2,0.85) node [left] {\small{$ a $}}
(1.2,0.85) node [right] {\small{$ b $}}
(-1.4, -0.65) node [below] {\small{$ a $}}
(1.4, -1.1) node [above] {\small{$ b $}}
};
\end{tikzpicture}\hspace{2cm}
\begin{tikzpicture}[scale=1]
\def\xa{0}

\draw[thick,color=black!50!green] (\xa+-1.2,1) -- (\xa+1.2,1);

\draw[thick,color=red]  (\xa+-1.2,0.7) -- (\xa+-0.4,0.7);
\draw[thick,color=red]  (\xa+1.2,0.7) --  (\xa+0.4,0.7);
\draw[thick,color=red]  (\xa+0.4,0.7) arc (45:-45:1);
\draw[thick,color=red]  (\xa+-0.4,0.7) arc (135:225:1);

\draw[thick,color=blue]  (\xa+-1.2,0.85)  -- (\xa+-0.1,0.85);
\draw[thick,color=blue] (\xa+1.2,0.85)  -- (\xa+0.1,0.85);
\draw[thick,color=blue] (\xa+-1.2,-0.85)  -- (\xa+-0.1,-0.85);
\draw[thick,color=blue] (\xa+1.2,-0.85)  -- (\xa+0.1,-0.85);

\draw[thick,color=blue] (\xa+-0.1,0.85) arc (65:0:1.0) ;
\draw[thick,color=blue] (\xa+0.1,0.85) arc (-45:-45-45:0.3) ;
\draw[thick,color=blue] (\xa+-0.1,0.78) arc (135:180+45:1.1) ;

\draw[thick,color=blue] (\xa+-0.1,-0.85) arc (-65:0:1.0) ;
\draw[thick,color=blue] (\xa+0.1,-0.85) arc (45:45+45:0.3) ;

\draw[thick,color=red] (\xa+-1.2,-0.7) -- (\xa+-0.4,-0.7);
\draw[thick,color=red] (\xa+1.2,-0.7) -- (\xa+0.4,-0.7);

\draw[thick,color=black!50!green] (\xa+0,0.6) arc (30:-30:1.3) ;
\draw[thick,color=black!50!green] (\xa+0,0.6) arc (180-30:180+30:1.3) ;
\draw[thick,color=black!50!green] (\xa+-1.2,-1) -- (\xa+1.2,-1);
\draw[thick, color=black]{
(\xa,-1) node [below] {\small{$ \dipole^\G$}}
};
\draw[thick, color=black]
{
(-1.2,0.85) node [left] {\small{$ a $}}
(1.2,0.85) node [right] {\small{$ b $}}
(-1.4, -0.65) node [below] {\small{$ a $}}
(1.4, -1.1) node [above] {\small{$ b $}}
};
\end{tikzpicture}\hspace{2cm}
\begin{tikzpicture}[scale=1]
\def\xa{0}

\draw[thick,color=blue] (\xa+-1.2,1) -- (\xa+1.2,1);

\draw[thick,color=black!50!green]  (\xa+-1.2,0.7) -- (\xa+-0.4,0.7);
\draw[thick,color=black!50!green]  (\xa+1.2,0.7) --  (\xa+0.4,0.7);
\draw[thick,color=black!50!green]  (\xa+0.4,0.7) arc (45:-45:1);
\draw[thick,color=black!50!green]  (\xa+-0.4,0.7) arc (135:225:1);

\draw[thick,color=red]  (\xa+-1.2,0.85)  -- (\xa+-0.1,0.85);
\draw[thick,color=red] (\xa+1.2,0.85)  -- (\xa+0.1,0.85);
\draw[thick,color=red] (\xa+-1.2,-0.85)  -- (\xa+-0.1,-0.85);
\draw[thick,color=red] (\xa+1.2,-0.85)  -- (\xa+0.1,-0.85);

\draw[thick,color=red] (\xa+-0.1,0.85) arc (65:0:1.0) ;
\draw[thick,color=red] (\xa+0.1,0.85) arc (-45:-45-45:0.3) ;
\draw[thick,color=red] (\xa+-0.1,0.78) arc (135:180+45:1.1) ;

\draw[thick,color=red] (\xa+-0.1,-0.85) arc (-65:0:1.0) ;
\draw[thick,color=red] (\xa+0.1,-0.85) arc (45:45+45:0.3) ;

\draw[thick,color=black!50!green] (\xa+-1.2,-0.7) -- (\xa+-0.4,-0.7);
\draw[thick,color=black!50!green] (\xa+1.2,-0.7) -- (\xa+0.4,-0.7);

\draw[thick,color=blue] (\xa+0,0.6) arc (30:-30:1.3) ;
\draw[thick,color=blue] (\xa+0,0.6) arc (180-30:180+30:1.3) ;
\draw[thick,color=blue] (\xa+-1.2,-1) -- (\xa+1.2,-1);
\draw[thick, color=black]{
(\xa,-1) node [below] {\small{$\dipole^\B$}}
};
\draw[thick, color=black]
{
(-1.2,0.85) node [left] {\small{$ a $}}
(1.2,0.85) node [right] {\small{$ b $}}
(-1.4, -0.65) node [below] {\small{$ a $}}
(1.4, -1.1) node [above] {\small{$ b $}}
};
\end{tikzpicture}
\caption{Dipoles $\dipole^{\col}_{ab}$ of three colors in KT chain model}
\label{fig:dipole}
\end{figure}
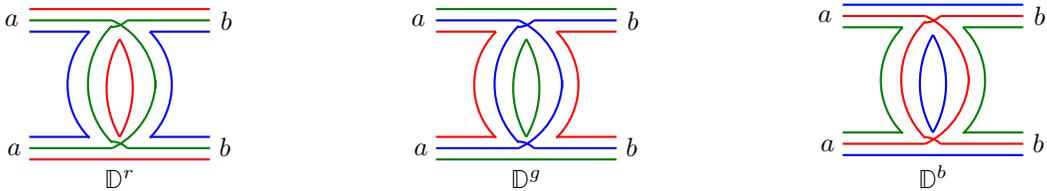

In KT chain model, each external leg of the dipole now carries a lattice index, which gives an extra structure for the dipoles. The $Z_2^L$ symmetry imposes that not all the external fermions can have different lattice index, but they must always come in pairs. Consequently, there are two classes of dipoles: The first class of dipoles is one with two identical lattice indices on the left (and, hence also on the right side) of the dipole - these are given in (Fig.~\ref{fig:dipole}) which are denoted by $\dipole^{\mathfrak c}_{ab}$ (${\mathfrak c}=\R, \G, \B$ and $a,b=1,2,\cdots, L$). The second class of dipoles is one which have two different lattice indices on the left side (and hence also on the right side) of dipole. Though one can easily evaluate all possible dipoles, we will, in this paper, focus on the dipole of the first class i.e $\dipole^{\mathfrak c}_{ab}$ given in Fig.~\ref{fig:dipole}. This is because we are interested in those four point functions which are analogous to those in GQS SYK model~\cite{Gu:2016oyy} and the second class of dipoles never appear in the leading ladder diagrams of the four point functions we study in this work.

\begin{figure}[t!]
\centering
\begin{tikzpicture}[scale=1]
\def\xa{0}

\draw[thick,color=red] (\xa+-1.2,1) -- (\xa+1.2,1);

\draw[thick,color=blue]  (\xa+-1.2,0.7) -- (\xa+-0.4,0.7);
\draw[thick,color=blue]  (\xa+1.2,0.7) --  (\xa+0.4,0.7);
\draw[thick,color=blue]  (\xa+0.4,0.7) arc (45:-45:1);
\draw[thick,color=blue]  (\xa+-0.4,0.7) arc (135:225:1);

\draw[thick,color=black!50!green]  (\xa+-1.2,0.85)  -- (\xa+-0.1,0.85);
\draw[thick,color=black!50!green] (\xa+1.2,0.85)  -- (\xa+0.1,0.85);
\draw[thick,color=black!50!green] (\xa+-1.2,-0.85)  -- (\xa+-0.1,-0.85);
\draw[thick,color=black!50!green] (\xa+1.2,-0.85)  -- (\xa+0.1,-0.85);

\draw[thick,color=black!50!green] (\xa+-0.1,0.85) arc (65:0:1.0) ;
\draw[thick,color=black!50!green] (\xa+0.1,0.85) arc (-45:-45-45:0.3) ;
\draw[thick,color=black!50!green] (\xa+-0.1,0.78) arc (135:180+45:1.1) ;

\draw[thick,color=black!50!green] (\xa+-0.1,-0.85) arc (-65:0:1.0) ;
\draw[thick,color=black!50!green] (\xa+0.1,-0.85) arc (45:45+45:0.3) ;

\draw[thick,color=blue] (\xa+-1.2,-0.7) -- (\xa+-0.4,-0.7);
\draw[thick,color=blue] (\xa+1.2,-0.7) -- (\xa+0.4,-0.7);

\draw[thick,color=red] (\xa+0,0.6) arc (30:-30:1.3) ;
\draw[thick,color=red] (\xa+0,0.6) arc (180-30:180+30:1.3) ;
\draw[thick,color=red] (\xa+-1.2,-1) -- (\xa+1.2,-1);
\draw[thick, color=black]{
(\xa,-1) node [below] {\small{$ \local^2 N^{-2}$}}
};
\draw[thick, color=black]
{
(-1.2,0.85) node [left] {\small{$ a $}}
(1.2,0.85) node [right] {\small{$ a $}}
(-1.4, -0.65) node [below] {\small{$ a $}}
(1.4, -1.05) node [above] {\small{$ a $}}
(-1.0,-0.3) node [above] {\small{$ a $}}
(0.95,-0.3) node [above] {\small{$ a $}}
};
\end{tikzpicture}\hspace{2cm}
\begin{tikzpicture}[scale=1]
\def\xa{0}

\draw[thick,color=red] (\xa+-1.2,1) -- (\xa+1.2,1);

\draw[thick,color=blue]  (\xa+-1.2,0.7) -- (\xa+-0.4,0.7);
\draw[thick,color=blue]  (\xa+1.2,0.7) --  (\xa+0.4,0.7);
\draw[thick,color=blue]  (\xa+0.4,0.7) arc (45:-45:1);
\draw[thick,color=blue]  (\xa+-0.4,0.7) arc (135:225:1);

\draw[thick,color=black!50!green]  (\xa+-1.2,0.85)  -- (\xa+-0.1,0.85);
\draw[thick,color=black!50!green] (\xa+1.2,0.85)  -- (\xa+0.1,0.85);
\draw[thick,color=black!50!green] (\xa+-1.2,-0.85)  -- (\xa+-0.1,-0.85);
\draw[thick,color=black!50!green] (\xa+1.2,-0.85)  -- (\xa+0.1,-0.85);

\draw[thick,color=black!50!green] (\xa+-0.1,0.85) arc (65:0:1.0) ;
\draw[thick,color=black!50!green] (\xa+0.1,0.85) arc (-45:-45-45:0.3) ;
\draw[thick,color=black!50!green] (\xa+-0.1,0.78) arc (135:180+45:1.1) ;

\draw[thick,color=black!50!green] (\xa+-0.1,-0.85) arc (-65:0:1.0) ;
\draw[thick,color=black!50!green] (\xa+0.1,-0.85) arc (45:45+45:0.3) ;

\draw[thick,color=blue] (\xa+-1.2,-0.7) -- (\xa+-0.4,-0.7);
\draw[thick,color=blue] (\xa+1.2,-0.7) -- (\xa+0.4,-0.7);

\draw[thick,color=red] (\xa+0,0.6) arc (30:-30:1.3) ;
\draw[thick,color=red] (\xa+0,0.6) arc (180-30:180+30:1.3) ;
\draw[thick,color=red] (\xa+-1.2,-1) -- (\xa+1.2,-1);
\draw[thick, color=black]{
(\xa,-1) node [below] {\small{${1\over 2} \hop_\R^2 N^{-2}$}}
};
\draw[thick, color=black]{
(-1.2,0.85) node [left] {\small{$ a $}}
(1.2,0.85) node [right] {\small{$ a $}}
(-1.4, -0.65) node [below] {\small{$ a $}}
(1.4, -1.05) node [above] {\small{$ a $}}
(-1.15,-0.3) node [above] {\small{$ a+1 $}}
(1.15,-0.3) node [above] {\small{$ a+1 $}}
};
\end{tikzpicture}\hspace{2cm}
\begin{tikzpicture}[scale=1]
\def\xa{0}

\draw[thick,color=red] (\xa+-1.2,1) -- (\xa+1.2,1);

\draw[thick,color=blue]  (\xa+-1.2,0.7) -- (\xa+-0.4,0.7);
\draw[thick,color=blue]  (\xa+1.2,0.7) --  (\xa+0.4,0.7);
\draw[thick,color=blue]  (\xa+0.4,0.7) arc (45:-45:1);
\draw[thick,color=blue]  (\xa+-0.4,0.7) arc (135:225:1);

\draw[thick,color=black!50!green]  (\xa+-1.2,0.85)  -- (\xa+-0.1,0.85);
\draw[thick,color=black!50!green] (\xa+1.2,0.85)  -- (\xa+0.1,0.85);
\draw[thick,color=black!50!green] (\xa+-1.2,-0.85)  -- (\xa+-0.1,-0.85);
\draw[thick,color=black!50!green] (\xa+1.2,-0.85)  -- (\xa+0.1,-0.85);

\draw[thick,color=black!50!green] (\xa+-0.1,0.85) arc (65:0:1.0) ;
\draw[thick,color=black!50!green] (\xa+0.1,0.85) arc (-45:-45-45:0.3) ;
\draw[thick,color=black!50!green] (\xa+-0.1,0.78) arc (135:180+45:1.1) ;

\draw[thick,color=black!50!green] (\xa+-0.1,-0.85) arc (-65:0:1.0) ;
\draw[thick,color=black!50!green] (\xa+0.1,-0.85) arc (45:45+45:0.3) ;

\draw[thick,color=blue] (\xa+-1.2,-0.7) -- (\xa+-0.4,-0.7);
\draw[thick,color=blue] (\xa+1.2,-0.7) -- (\xa+0.4,-0.7);

\draw[thick,color=red] (\xa+0,0.6) arc (30:-30:1.3) ;
\draw[thick,color=red] (\xa+0,0.6) arc (180-30:180+30:1.3) ;
\draw[thick,color=red] (\xa+-1.2,-1) -- (\xa+1.2,-1);
\draw[thick, color=black]{
(\xa,-1) node [below] {\small{${1\over 2} \hop_\R^2 N^{-2}$}}
};
\draw[thick, color=black]{
(-1.2,0.85) node [left] {\small{$ a $}}
(1.2,0.85) node [right] {\small{$ a $}}
(-1.4, -0.65) node [below] {\small{$ a $}}
(1.4, -1.05) node [above] {\small{$ a $}}
(-1.15,-0.3) node [above] {\small{$ a-1 $}}
(1.15,-0.3) node [above] {\small{$ a-1 $}}
};
\end{tikzpicture}\\
\begin{tikzpicture}[scale=1]
\def\xa{0}

\draw[thick,color=red] (\xa+-1.2,1) -- (\xa+1.2,1);

\draw[thick,color=blue]  (\xa+-1.2,0.7) -- (\xa+-0.4,0.7);
\draw[thick,color=blue]  (\xa+1.2,0.7) --  (\xa+0.4,0.7);
\draw[thick,color=blue]  (\xa+0.4,0.7) arc (45:-45:1);
\draw[thick,color=blue]  (\xa+-0.4,0.7) arc (135:225:1);

\draw[thick,color=black!50!green]  (\xa+-1.2,0.85)  -- (\xa+-0.1,0.85);
\draw[thick,color=black!50!green] (\xa+1.2,0.85)  -- (\xa+0.1,0.85);
\draw[thick,color=black!50!green] (\xa+-1.2,-0.85)  -- (\xa+-0.1,-0.85);
\draw[thick,color=black!50!green] (\xa+1.2,-0.85)  -- (\xa+0.1,-0.85);

\draw[thick,color=black!50!green] (\xa+-0.1,0.85) arc (65:0:1.0) ;
\draw[thick,color=black!50!green] (\xa+0.1,0.85) arc (-45:-45-45:0.3) ;
\draw[thick,color=black!50!green] (\xa+-0.1,0.78) arc (135:180+45:1.1) ;

\draw[thick,color=black!50!green] (\xa+-0.1,-0.85) arc (-65:0:1.0) ;
\draw[thick,color=black!50!green] (\xa+0.1,-0.85) arc (45:45+45:0.3) ;

\draw[thick,color=blue] (\xa+-1.2,-0.7) -- (\xa+-0.4,-0.7);
\draw[thick,color=blue] (\xa+1.2,-0.7) -- (\xa+0.4,-0.7);

\draw[thick,color=red] (\xa+0,0.6) arc (30:-30:1.3) ;
\draw[thick,color=red] (\xa+0,0.6) arc (180-30:180+30:1.3) ;
\draw[thick,color=red] (\xa+-1.2,-1) -- (\xa+1.2,-1);
\draw[thick, color=black]{
(\xa,-1) node [below] {\small{${1\over 2} \hop_\G^2 N^{-2}$}}
};
\draw[thick, color=black]{
(-1.2,0.85) node [left] {\small{$ a $}}
(1.2,0.85) node [right] {\small{$ a\pm1 $}}
(-1.4, -0.65) node [below] {\small{$ a $}}
(1.65, -1.1) node [above] {\small{$ a \pm 1  $}}
(-1.15,-0.3) node [above] {\small{$ a\pm 1 $}}
(0.95,-0.25) node [above] {\small{$ a $}}
};
\end{tikzpicture}\hspace{2cm}
\begin{tikzpicture}[scale=1]
\def\xa{0}

\draw[thick,color=red] (\xa+-1.2,1) -- (\xa+1.2,1);

\draw[thick,color=blue]  (\xa+-1.2,0.7) -- (\xa+-0.4,0.7);
\draw[thick,color=blue]  (\xa+1.2,0.7) --  (\xa+0.4,0.7);
\draw[thick,color=blue]  (\xa+0.4,0.7) arc (45:-45:1);
\draw[thick,color=blue]  (\xa+-0.4,0.7) arc (135:225:1);

\draw[thick,color=black!50!green]  (\xa+-1.2,0.85)  -- (\xa+-0.1,0.85);
\draw[thick,color=black!50!green] (\xa+1.2,0.85)  -- (\xa+0.1,0.85);
\draw[thick,color=black!50!green] (\xa+-1.2,-0.85)  -- (\xa+-0.1,-0.85);
\draw[thick,color=black!50!green] (\xa+1.2,-0.85)  -- (\xa+0.1,-0.85);

\draw[thick,color=black!50!green] (\xa+-0.1,0.85) arc (65:0:1.0) ;
\draw[thick,color=black!50!green] (\xa+0.1,0.85) arc (-45:-45-45:0.3) ;
\draw[thick,color=black!50!green] (\xa+-0.1,0.78) arc (135:180+45:1.1) ;

\draw[thick,color=black!50!green] (\xa+-0.1,-0.85) arc (-65:0:1.0) ;
\draw[thick,color=black!50!green] (\xa+0.1,-0.85) arc (45:45+45:0.3) ;

\draw[thick,color=blue] (\xa+-1.2,-0.7) -- (\xa+-0.4,-0.7);
\draw[thick,color=blue] (\xa+1.2,-0.7) -- (\xa+0.4,-0.7);

\draw[thick,color=red] (\xa+0,0.6) arc (30:-30:1.3) ;
\draw[thick,color=red] (\xa+0,0.6) arc (180-30:180+30:1.3) ;
\draw[thick,color=red] (\xa+-1.2,-1) -- (\xa+1.2,-1);
\draw[thick, color=black]{
(\xa,-1) node [below] {\small{${1\over 2} \hop_\B^2 N^{-2}$}}
};
\draw[thick, color=black]{
(-1.2,0.85) node [left] {\small{$ a $}}
(1.2,0.85) node [right] {\small{$ a \pm 1 $}}
(-1.4, -0.65) node [below] {\small{$ a $}}
(1.7, -1.1) node [above] {\small{$ a \pm 1 $}}
(-0.9,-0.25) node [above] {\small{$ a $}}
(1.15,-0.3) node [above] {\small{$ a \pm 1 $}}
};
\end{tikzpicture}
\caption{Contribution of interactions to dipoles}
\label{fig:dipole contribution from interaction}
\end{figure}
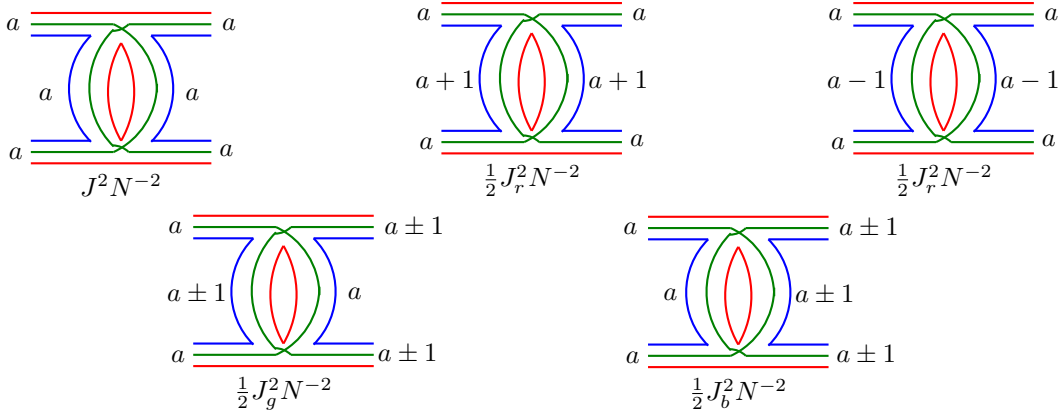
%

Dipoles denoted by $\dipole^{\mathfrak c}_{ab}$ (${\mathfrak c}=\R, \G, \B$ and $a,b=1,2,\cdots, L$) in Fig.~\ref{fig:dipole} can be treated as a matrix in the lattice space. Note that in KT chain model dipole $\dipole^{\mathfrak c}_{ab}$ vanishes unless $|a-b|\leqq 1$. To evaluate a dipole $\dipole^{\mathfrak c}_{ab}$, we need to find all possible Feynman diagrams consistent with external lattice indices and color of dipole (\eg See Fig.~\ref{fig:dipole contribution from interaction} for $\dipole^\R$). Moreover, one can easily see\footnote{In fact, this can be immediately deduced from \emph{Unique Last Fermion Property} which we will define in Section~\ref{sec:models}.} that two identical hopping vertices gives ${1\over 2} \hop_\col^2$ while two on-site vertex contribute $\local^2$, and therefore one can obtain, for example
\begin{equation}
N^{-2} \dipole^\R_{ab}(\tau_1,\tau_2,\tau_3,\tau_4) =  \begin{cases}
\hspace{0.1cm} -N^{-2}(\local^2+\hop_\R^2) G(\tau_{13}) G(\tau_{24}) [G(\tau_{34})]^2 & \hspace{0.5cm} \mbox{for}\;\; a=b\\
\hspace{0.1cm} -{1\over 2}N^{-2}(\hop_\G^2+\hop_\B^2) G(\tau_{13}) G(\tau_{24}) [G(\tau_{34})]^2 & \hspace{0.5cm} \mbox{for}\;\; |a-b|=1\\
\end{cases}\label{eq:dipole of red color}
\end{equation}
where we replaced free propagator with full propagators. In general, dipoles in this paper are symmetric matrices in lattice space because a horizontal reflection of the dipole generates another dipole with the same vertices. One can obtain a similar expression for other dipoles $\dipole^\G_{ab}$ and $\dipole^\B_{ab}$. Also, note that we make $\dipole^\R_{ab}$ of order $\mathcal{O}(N^0)$ by factoring out $N^{-2}$ in the LHS of~\eqref{eq:dipole of red color}.

Now that we have elucidated the building blocks of a ladder diagram, we next turn to an analysis of gauge invariant four point functions. One can then see that there are three different channels of the four point functions depending on contraction of gauge indices of external fermions. Specifically, we define
\begin{equation}\label{All4PointFunc}
\begin{split}
F^{C}_{a_1a_2} \equiv & \langle \psi^{a_1}_{i_1 j_1 k_1}(\tau_1) \psi^{a_1}_{i_1 j_1 k_1}(\tau_2) \psi^{a_2}_{i_2 j_2 k_2}(\tau_3) \psi^{a_2}_{i_2 j_2 k_2}(\tau_4) \rangle \\
F^{P,\R}_{a_1 a_2} \equiv &  \langle \psi^{a_1}_{i_1 j_1 k_1}(\tau_1) \psi^{a_1}_{i_2 j_1 k_1}(\tau_2) \psi^{a_2}_{i_1 j_2 k_2}(\tau_3) \psi^{a_2}_{i_2 j_2 k_2}(\tau_4) \rangle \\
F^{P,\G}_{a_1 a_2} \equiv &  \langle \psi^{a_1}_{i_1 j_1 k_1}(\tau_1) \psi^{a_1}_{i_1 j_2 k_1}(\tau_2) \psi^{a_2}_{i_2 j_1 k_2}(\tau_3) \psi^{a_2}_{i_2 j_2 k_2}(\tau_4) \rangle \\
F^{P,\B}_{a_1 a_2} \equiv &  \langle \psi^{a_1}_{i_1 j_1 k_1}(\tau_1) \psi^{a_1}_{i_1 j_1 k_2}(\tau_2) \psi^{a_2}_{i_2 j_2 k_1}(\tau_3) \psi^{a_2}_{i_2 j_2 k_2}(\tau_4) \rangle \\
F^T_{a_1 a_2 a_3 a_4}  \equiv &  \langle \psi^{a_1}_{i_1 j_1 k_1}(\tau_1) \psi^{a_2}_{i_1 j_2 k_2}(\tau_2) \psi^{a_3}_{i_2 j_1 k_2}(\tau_3) \psi^{a_4}_{i_2 j_2 k_1}(\tau_4) \rangle
\end{split}
\end{equation}
where ``$C$'' ``$P$''  and ``$T$'' represent ``Cooper'', ``Pillow'' and ``Tetrahedron'' channel, respectively. Since Pillow contraction is not symmetric in RGB color, there are three Pillow channels labelled by the three colors. Just like in the single-site KT model, the leading $N$ diagrams of KT chain model will be ladder diagrams.

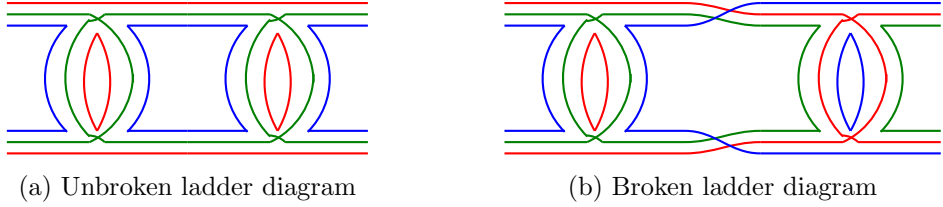
\begin{figure}[t!]
\centering
\begin{subfigure}[t]{0.4\linewidth}
\centering
\begin{tikzpicture}[scale=1]
\def\xa{0}

\draw[thick,color=red] (\xa+-1.2,1) -- (\xa+1.2,1);

\draw[thick,color=blue]  (\xa+-1.2,0.7) -- (\xa+-0.4,0.7);
\draw[thick,color=blue]  (\xa+1.2,0.7) --  (\xa+0.4,0.7);
\draw[thick,color=blue]  (\xa+0.4,0.7) arc (45:-45:1);
\draw[thick,color=blue]  (\xa+-0.4,0.7) arc (135:225:1);

\draw[thick,color=black!50!green]  (\xa+-1.2,0.85)  -- (\xa+-0.1,0.85);
\draw[thick,color=black!50!green] (\xa+1.2,0.85)  -- (\xa+0.1,0.85);
\draw[thick,color=black!50!green] (\xa+-1.2,-0.85)  -- (\xa+-0.1,-0.85);
\draw[thick,color=black!50!green] (\xa+1.2,-0.85)  -- (\xa+0.1,-0.85);

\draw[thick,color=black!50!green] (\xa+-0.1,0.85) arc (65:0:1.0) ;
\draw[thick,color=black!50!green] (\xa+0.1,0.85) arc (-45:-45-45:0.3) ;
\draw[thick,color=black!50!green] (\xa+-0.1,0.78) arc (135:180+45:1.1) ;

\draw[thick,color=black!50!green] (\xa+-0.1,-0.85) arc (-65:0:1.0) ;
\draw[thick,color=black!50!green] (\xa+0.1,-0.85) arc (45:45+45:0.3) ;

\draw[thick,color=blue] (\xa+-1.2,-0.7) -- (\xa+-0.4,-0.7);
\draw[thick,color=blue] (\xa+1.2,-0.7) -- (\xa+0.4,-0.7);

\draw[thick,color=red] (\xa+0,0.6) arc (30:-30:1.3) ;
\draw[thick,color=red] (\xa+0,0.6) arc (180-30:180+30:1.3) ;
\draw[thick,color=red] (\xa+-1.2,-1) -- (\xa+1.2,-1);

\def\xa{2.4}
\draw[thick,color=red] (\xa+-1.2,1) -- (\xa+1.2,1);

\draw[thick,color=blue]  (\xa+-1.2,0.7) -- (\xa+-0.4,0.7);
\draw[thick,color=blue]  (\xa+1.2,0.7) --  (\xa+0.4,0.7);
\draw[thick,color=blue]  (\xa+0.4,0.7) arc (45:-45:1);
\draw[thick,color=blue]  (\xa+-0.4,0.7) arc (135:225:1);

\draw[thick,color=black!50!green]  (\xa+-1.2,0.85)  -- (\xa+-0.1,0.85);
\draw[thick,color=black!50!green] (\xa+1.2,0.85)  -- (\xa+0.1,0.85);
\draw[thick,color=black!50!green] (\xa+-1.2,-0.85)  -- (\xa+-0.1,-0.85);
\draw[thick,color=black!50!green] (\xa+1.2,-0.85)  -- (\xa+0.1,-0.85);

\draw[thick,color=black!50!green] (\xa+-0.1,0.85) arc (65:0:1.0) ;
\draw[thick,color=black!50!green] (\xa+0.1,0.85) arc (-45:-45-45:0.3) ;
\draw[thick,color=black!50!green] (\xa+-0.1,0.78) arc (135:180+45:1.1) ;

\draw[thick,color=black!50!green] (\xa+-0.1,-0.85) arc (-65:0:1.0) ;
\draw[thick,color=black!50!green] (\xa+0.1,-0.85) arc (45:45+45:0.3) ;

\draw[thick,color=blue] (\xa+-1.2,-0.7) -- (\xa+-0.4,-0.7);
\draw[thick,color=blue] (\xa+1.2,-0.7) -- (\xa+0.4,-0.7);

\draw[thick,color=red] (\xa+0,0.6) arc (30:-30:1.3) ;
\draw[thick,color=red] (\xa+0,0.6) arc (180-30:180+30:1.3) ;
\draw[thick,color=red] (\xa+-1.2,-1) -- (\xa+1.2,-1);

\end{tikzpicture}
\caption{Unbroken ladder diagram }
\label{fig:unbroken ladder diagram}
\end{subfigure}
\quad\quad
\begin{subfigure}[t]{0.4\linewidth}
\centering
\begin{tikzpicture}[scale=1]
\def\xa{0}

\draw[thick,color=red] (\xa+-1.2,1) -- (\xa+1.2,1);

\draw[thick,color=blue]  (\xa+-1.2,0.7) -- (\xa+-0.4,0.7);
\draw[thick,color=blue]  (\xa+1.2,0.7) --  (\xa+0.4,0.7);
\draw[thick,color=blue]  (\xa+0.4,0.7) arc (45:-45:1);
\draw[thick,color=blue]  (\xa+-0.4,0.7) arc (135:225:1);

\draw[thick,color=black!50!green]  (\xa+-1.2,0.85)  -- (\xa+-0.1,0.85);
\draw[thick,color=black!50!green] (\xa+1.2,0.85)  -- (\xa+0.1,0.85);
\draw[thick,color=black!50!green] (\xa+-1.2,-0.85)  -- (\xa+-0.1,-0.85);
\draw[thick,color=black!50!green] (\xa+1.2,-0.85)  -- (\xa+0.1,-0.85);

\draw[thick,color=black!50!green] (\xa+-0.1,0.85) arc (65:0:1.0) ;
\draw[thick,color=black!50!green] (\xa+0.1,0.85) arc (-45:-45-45:0.3) ;
\draw[thick,color=black!50!green] (\xa+-0.1,0.78) arc (135:180+45:1.1) ;

\draw[thick,color=black!50!green] (\xa+-0.1,-0.85) arc (-65:0:1.0) ;
\draw[thick,color=black!50!green] (\xa+0.1,-0.85) arc (45:45+45:0.3) ;

\draw[thick,color=blue] (\xa+-1.2,-0.7) -- (\xa+-0.4,-0.7);
\draw[thick,color=blue] (\xa+1.2,-0.7) -- (\xa+0.4,-0.7);

\draw[thick,color=red] (\xa+0,0.6) arc (30:-30:1.3) ;
\draw[thick,color=red] (\xa+0,0.6) arc (180-30:180+30:1.3) ;
\draw[thick,color=red] (\xa+-1.2,-1) -- (\xa+1.2,-1);

\draw[thick,color=red] (\xa+1.2,1) to [out=0,in=180] (3.4-1.2,0.85);
\draw[thick,color=black!50!green] (\xa+1.2,0.85) to [out=0,in=180] (3.4-1.2,0.7);

\draw[thick,color=red] (\xa+1.2,-1) to [out=0,in=-180] (3.4-1.2,-0.85);
\draw[thick,color=black!50!green] (\xa+1.2,-0.85) to [out=0,in=-180] (3.4-1.2,-0.7);
\draw[thick,color=blue] (\xa+1.2,0.7) to [out=0,in=-180] (3.4-1.2,1);
\draw[thick,color=blue] (\xa+1.2,-0.7) to [out=0,in=180] (3.4-1.2,-1);

\def\xa{3.4}

\draw[thick,color=blue] (\xa+-1.2,1) -- (\xa+1.2,1);

\draw[thick,color=black!50!green]  (\xa+-1.2,0.7) -- (\xa+-0.4,0.7);
\draw[thick,color=black!50!green]  (\xa+1.2,0.7) --  (\xa+0.4,0.7);
\draw[thick,color=black!50!green]  (\xa+0.4,0.7) arc (45:-45:1);
\draw[thick,color=black!50!green]  (\xa+-0.4,0.7) arc (135:225:1);

\draw[thick,color=red]  (\xa+-1.2,0.85)  -- (\xa+-0.1,0.85);
\draw[thick,color=red] (\xa+1.2,0.85)  -- (\xa+0.1,0.85);
\draw[thick,color=red] (\xa+-1.2,-0.85)  -- (\xa+-0.1,-0.85);
\draw[thick,color=red] (\xa+1.2,-0.85)  -- (\xa+0.1,-0.85);

\draw[thick,color=red] (\xa+-0.1,0.85) arc (65:0:1.0) ;
\draw[thick,color=red] (\xa+0.1,0.85) arc (-45:-45-45:0.3) ;
\draw[thick,color=red] (\xa+-0.1,0.78) arc (135:180+45:1.1) ;

\draw[thick,color=red] (\xa+-0.1,-0.85) arc (-65:0:1.0) ;
\draw[thick,color=red] (\xa+0.1,-0.85) arc (45:45+45:0.3) ;

\draw[thick,color=black!50!green] (\xa+-1.2,-0.7) -- (\xa+-0.4,-0.7);
\draw[thick,color=black!50!green] (\xa+1.2,-0.7) -- (\xa+0.4,-0.7);

\draw[thick,color=blue] (\xa+0,0.6) arc (30:-30:1.3) ;
\draw[thick,color=blue] (\xa+0,0.6) arc (180-30:180+30:1.3) ;
\draw[thick,color=blue] (\xa+-1.2,-1) -- (\xa+1.2,-1);

\end{tikzpicture}
\caption{Broken ladder diagram}
\label{fig:broken ladder diagram}
\end{subfigure}
\caption{Unbroken and broken ladder diagrams. Unbroken ladder diagram is composed of dipoles of identical color (\eg $\dipole^\R\dipole^\R$) while broken ladder diagram contains at least one dipole of distinct color (\eg $\dipole^\R\dipole^\B$).}
\label{fig:unbroken and broken ladder diagrams}
\end{figure}

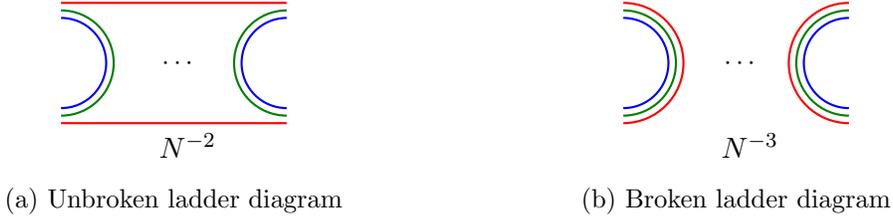
\begin{figure}
\centering
\begin{subfigure}[b]{0.4\linewidth}
\centering
\def\c{0} \def\s{1.5}  \def\v1{1/2} \def\b{1.4}  
\begin{tikzpicture}[scale=1]

\draw[thick,color=red] (-\s+\c,0.1) -- (\s+\c,0.1);
\draw[thick,color=red] (-\s+\c,-\b-0.1) -- (\s+\c,-\b-0.1);

\draw[thick,color=black!50!green] (-\s+\c,0) arc  (90:-90:\b/2);
\draw[thick,color=black!50!green] (\s+\c,0) arc (90:270:\b/2);

\draw[thick,color=blue] (-\s+\c,-0.1) arc (90:-90:\b/2-0.1);
\draw[thick,color=blue] (\s+\c,-0.1) arc (90:270:\b/2-0.1);

\draw[thick, color=black]
{
(\c-0.3,-\b/2) node [right] {\small{$\cdots$}} 
(\c+0.7,-\b-0.4) node [left] { $N^{-2}$ } 
};
\end{tikzpicture}
\caption{Unbroken ladder diagram}
\label{fig:general unbroken ladder diagram}
\end{subfigure} \hspace{1cm}
\begin{subfigure}[b]{0.4\linewidth}
\centering
\def\c{20} \def\s{1.5}  \def\v1{1/2} \def\b{1.4}  
\begin{tikzpicture}[scale=1]
\draw[thick,color=red] (-\s+\c,0.1) arc (90:-90:\b/2+0.1);
\draw[thick,color=red] (\s+\c,0.1) arc (90:270:\b/2+0.1);

\draw[thick,color=black!50!green] (-\s+\c,0) arc  (90:-90:\b/2);
\draw[thick,color=black!50!green] (\s+\c,0) arc (90:270:\b/2);

\draw[thick,color=blue] (-\s+\c,-0.1) arc (90:-90:\b/2-0.1);
\draw[thick,color=blue] (\s+\c,-0.1) arc (90:270:\b/2-0.1);

\draw[thick, color=black]
{
(\c-0.3,-\b/2) node [right] {\small{$\cdots$}} 
(\c+0.7,-\b-0.4) node [left] {$N^{-3}$ } 
};
\end{tikzpicture} 
\caption{Broken ladder diagram}
\label{fig:general broken ladder diagram}
\end{subfigure}
\caption{General structure of unbroken and broken ladder diagrams. In unbroken ladder diagram, one color (\eg red) is transmitted along the ladder. On the other hand, all three colors from the external strand turn back.}
\label{fig:general unbroken and broken ladder diagrams}
\end{figure}

\begin{figure}
\centering

\begin{subfigure}[b]{0.6\linewidth}
\centering
 \def\c{1} \def\s{1.2}  \def\v1{1/2} \def\b{1.4}  

\begin{tikzpicture}[scale=1]
\draw[thick,color=red] (-\s+\c,0.1) -- (\s+\c,0.1);
\draw[thick,color=red] (-\s+\c,-\b-0.1) -- (\s+\c,-\b-0.1);
\draw[thick,color=red,dashed] (-\s+\c,0.1) arc (90:270:\b/2+0.1);
\draw[thick,color=red,dashed] (\s+\c,0.1) arc (90:-90:\b/2+0.1);

\draw[thick,color=black!50!green,dashed] (-\s+\c,0) arc (90:270:\b/2);
\draw[thick,color=black!50!green] (-\s+\c,0) arc  (90:-90:\b/2);
\draw[thick,color=black!50!green,dashed] (\s+\c,0) arc (90:-90:\b/2);
\draw[thick,color=black!50!green] (\s+\c,0) arc (90:270:\b/2);

\draw[thick,color=blue,dashed] (-\s+\c,-0.1) arc (90:270:\b/2-0.1);
\draw[thick,color=blue] (-\s+\c,-0.1) arc (90:-90:\b/2-0.1);
\draw[thick,color=blue,dashed] (\s+\c,-0.1) arc (90:-90:\b/2-0.1);
\draw[thick,color=blue] (\s+\c,-0.1) arc (90:270:\b/2-0.1);

\draw[thick, color=black]
{
(-\s+\b/2+\c+0.2,-\b/2) node [right] {\small{$\cdots$}} 
};
\end{tikzpicture}
\caption{Unbroken ladder diagram in Cooper channel $\sim N^3$}
\label{fig:unbroken ladder diagram in cooper channel}
\end{subfigure} \hspace{10mm} \\
\vspace{0.5cm}
\begin{subfigure}[b]{0.6\linewidth}
\centering
\def\c{6} \def\s{1.2}  \def\v1{1/2} \def\b{1.4}  
\begin{tikzpicture}[scale=1]

\draw[thick,color=red] (-\s+\c,0.1) arc (90:-90:\b/2+0.1);
\draw[thick,color=red] (\s+\c,0.1) arc (90:270:\b/2+0.1);
\draw[thick,color=red,dashed] (-\s+\c,0.1) arc (90:270:\b/2+0.1);
\draw[thick,color=red,dashed] (\s+\c,0.1) arc (90:-90:\b/2+0.1);

\draw[thick,color=black!50!green,dashed] (-\s+\c,0) arc (90:270:\b/2);
\draw[thick,color=black!50!green] (-\s+\c,0) arc  (90:-90:\b/2);
\draw[thick,color=black!50!green,dashed] (\s+\c,0) arc (90:-90:\b/2);
\draw[thick,color=black!50!green] (\s+\c,0) arc (90:270:\b/2);

\draw[thick,color=blue,dashed] (-\s+\c,-0.1) arc (90:270:\b/2-0.1);
\draw[thick,color=blue] (-\s+\c,-0.1) arc (90:-90:\b/2-0.1);
\draw[thick,color=blue,dashed] (\s+\c,-0.1) arc (90:-90:\b/2-0.1);
\draw[thick,color=blue] (\s+\c,-0.1) arc (90:270:\b/2-0.1);

\draw[thick, color=black]
{
(-\s+\b/2+\c+0.2,-\b/2) node [right] {\small{$\cdots$}} 
};
\end{tikzpicture} 
\caption{Broken ladder diagram in Cooper channel $\sim N^3$}
\label{fig:broken ladder diagram in cooper channel}
\end{subfigure}
\caption{Unbroken and broken ladder diagram in Cooper channel. The dashed line represents contraction of external gauge indices}
\label{fig:unbroken and broken ladder diagram in cooper channel}
\end{figure}
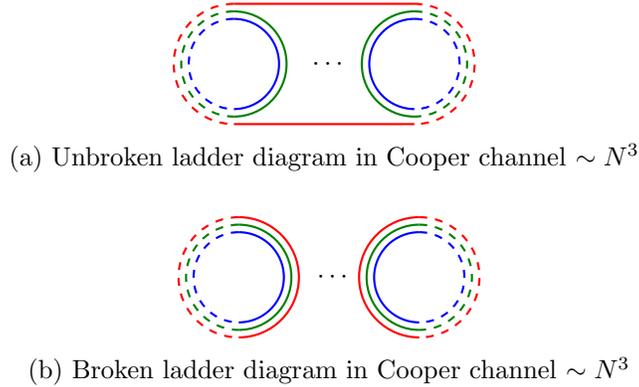

Various dipoles strung together horizontally form a ladder diagram with external color indices uncontracted (\eg Figure~\ref{fig:unbroken and broken ladder diagrams}). We will distinguish between two types of ladder diagrams : \textit{unbroken} and \textit{broken} ladder diagram~\cite{Gurau:2016lzk}. 
\begin{itemize}
\item \textbf{Unbroken ladder: }A ladder diagram is unbroken if it is made of dipoles of the same color (e.g. $\dipole^\R \dipole^\R$ in Fig.~\ref{fig:unbroken ladder diagram}) Note that one of external colors in an unbroken ladder diagram is transmitted along the ladder (See Fig.~\ref{fig:general broken ladder diagram}). Using this structure, one can easily show by induction that an unbroken ladder diagram (without contracting external indices) is of order $\mathcal{O}(N^{-2})$. For example, let us attach a dipole of red color to an unbroken ladder diagram of the same color (\eg See Fig.~\ref{fig:general unbroken ladder diagram}). This procedure produces two additional loops, and the corresponding $N^2$ contribution will be cancelled with $N^{-2}$ in the attached dipole. 

\item \textbf{Broken ladder: }A ladder diagram is broken if it has at least one different dipole (\ie $\dipole^\R \dipole^\B$ in Figure~\ref{fig:broken ladder diagram}, for example). Unlike unbroken ladder diagram, all three external colors return in the broken ladder diagrams (See Fig.~\ref{fig:general broken ladder diagram}). Also, one can show by induction that a broken ladder diagram is of order $\mathcal{O}(N^{-3})$. Namely, attaching any dipole to a broken ladder diagram in Fig.~\ref{fig:general broken ladder diagram} does not change order in $N$. However, when an unbroken ladder diagram is attached to a dipole of different color, it makes only one loop so that the unbroken ladder diagram of order $\mathcal{O}(N^{-2})$ becomes broken one of order $\mathcal{O}(N^{-3})$. 
\end{itemize}
Although the above argument seems to suggest that broken diagrams is suppressed at leading order in $N$, we will later see that depending on the contractions of external legs, they could give an important contribution even at leading order in $N$.

Now, we turn to analyzing each of the channels given in~\eqref{All4PointFunc}.

\subsection{Cooper Channel}
\label{sec:cooper channel}
 
We begin our analysis of four point function for the case of Cooper channel. This channel is interesting because the corresponding analogues in SYK~\cite{Maldacena:2016hyu} and the single-site KT model~\cite{Klebanov:2016xxf} exhibit maximal chaotic behavior. In the large $N$ limit, it can be written as
\begin{equation}
F^C_{ab} (\tau_1,\tau_2,\tau_3,\tau_4)= N^6 G(\tau_{12})G(\tau_{34}) + N^3\mathcal{F}^C_{ab} (\tau_1,\tau_2,\tau_3,\tau_4) \label{def:ladder diagram cooper channel}
\end{equation}
where the first term in the RHS of order $\mathcal{O}(N^6)$ corresponds to a disconnected diagram. Note that the disconnected piece is independent of lattice points.

The subleading piece $\mathcal{F}^C_{ab}$ consists of ladder diagrams with external gauge indices contracted. Recall that the unbroken and broken ladder diagram are of different order in $N$ without contraction of external gauge indices. But, when the external legs are contracted via the Cooper contraction both unbroken and broken ladder diagram become of order $\mathcal{O}(N^3)$. This is because the external contraction produces five additional loops for unbroken one while it produces six additional loops for broken one (See Fig.~\ref{fig:unbroken and broken ladder diagram in cooper channel}). Moreover, this also implies that the ladder diagram of Cooper channel $\mathcal{F}^C_{ab}$ is composed of arbitrary combination of three dipoles, and therefore, it can be easily written as a geometric series of the dipoles:
\begin{equation}
\mathcal{F}^C= \sum_{n=0}^\infty (\dipole^\R+\dipole^\G+\dipole^\B)^n \mathcal{F}_0\label{eq:geometric series for Cooper channel}
\end{equation}
where $\mathcal{F}_0$ is the first ladder diagram without rung
\begin{equation}
\mathcal{F}_0\equiv -G(\tau_{13})G(\tau_{24})+G(\tau_{14})G(\tau_{23}).
\end{equation}
Note that we have now introduced the matrix notation. For example, the product of two dipoles in \eqref{eq:geometric series for Cooper channel} must actually be understood as a matrix product in spatial coordinate as well as in bi-local time coordinate $(\tau_1,\tau_2)$, i.e 
\begin{equation}
(\dipole^{c_1} \dipole^{c_2})_{a_1a_2}(\tau_1,\tau_2;\tau_3,\tau_4)\equiv \sum_{a_3=1}^L\int d\tau_5 d\tau_6\; \dipole_{a_1a_3}^{c_1}(\tau_1,\tau_2;\tau_5,\tau_6)\dipole_{a_3 a_2}^{c_2}(\tau_5,\tau_6;\tau_3,\tau_4)\label{def:matrix product}
\end{equation}
and also ${\cal F}_0$ is understood to be identity in lattice space. Using \eqref{eq:dipole of red color}, the common ratio of the geometric series in \eqref{eq:geometric series for Cooper channel} can be written as
\begin{equation}
\begin{split}
&(\dipole^\R+\dipole^\G+\dipole^\B)_{a_1 a_2}(\tau_1,\tau_2,\tau_3,\tau_4) \\
=&  - \left[(3\local^2+\effhop^2)\delta_{a_1,a_2}+\effhop^2 \delta_{a_1, a_2\pm 1} \right]G(\tau_{13})G(\tau_{24})[G(\tau_{34})]^2 \equiv S^C_{a_1 a_2} \; \mathcal{K}(\tau_1,\tau_2,\tau_3,\tau_4) 
\end{split}\label{eq:common ratio for disconnected 4pt ftn}
\end{equation}
where we have defined 
\begin{equation}
\effcoupling_\text{hop}\equiv \sqrt{\hop_\R^2+\hop_\G^2+\hop_\B^2}
\end{equation}
and the $S^C_{a_1 a_2}$ is a hopping matrix defined by
\begin{equation}
S^C_{a_1 a_2} \equiv \delta_{a_1, a_2}+{\effhop^2 \over 3\effcoupling^2}(\delta_{a_1, a_2\pm 1} -2 \delta_{a_1 ,a_2} )\label{eq:hopping matrix for disconnected 4pt ftn}
\end{equation}
The kernel $\mathcal{K}(\tau_1,\tau_2,\tau_3,\tau_4)$ is the same as that of SYK model in \eqref{eq:SYK kernel} except that the coupling constant $J_{\text{\tiny SYK}}$ is replaced by the effective coupling constant $\effcoupling$ given in \eqref{def:effective coupling constant KT model}:
\begin{equation}
\mathcal{K}(\tau_1,\tau_2,\tau_3,\tau_4)\equiv -3 \effcoupling^2G(\tau_{13})G(\tau_{24})[G(\tau_{34})]^2\label{eq:SYK kernel 2}
\end{equation}
In a somewhat formal way, the geometric series in \eqref{eq:geometric series for Cooper channel} can be summed up to give
\begin{equation}
\mathcal{F}^C = {1 \over 1  - S^C \cal K} F_0
\end{equation}	
By utilizing the translational invariance, it is now convenient to to move to lattice momentum space to have the following expression for ${\cal F}_p$. 
\begin{equation}
{\cal F}_p(\tau_1,\tau_2,\tau_3,\tau_4) = {1 \over 1 - s(p) {\cal K}} {\cal F}_0  \label{eq:FpinKTLattice}
\end{equation}
where the function $s(p)\equiv 1 - {2 \effhop^2 \over 3 \effcoupling }(1 - \cos p)$. At this stage, one can recognize that this is exactly the same expression obtained in \cite{Gu:2016oyy} as we mentioned in \eqref{eq:GuQiStanford4Point}. Using the $SL(2,{\mathbb R} )$ invariance, one can diagonalize $\mathcal{K}(\tau_1,\tau_2,\tau_3,\tau_4)$ by hypergeometric function $\Psi_h(\chi)$ (for more details refer to \cite{Gu:2016oyy,Maldacena:2016hyu}), one has
\begin{align}
\mathcal{F}^C_p(\chi)\equiv{\mathcal{F}^C_p(\tau_1,\tau_2,\tau_3,\tau_4)\over G(\tau_{12})G(\tau_{34})}=&{4\pi \over 3}\left[\int_{-\infty}^\infty {ds\over 2\pi} \left.{h-{1\over 2}\over \pi \tan{\pi h\over 2}}{k_c(h)\over 1-s^C(p) k_c(h)}\Psi_h(\chi)\right|_{h={1\over 2}+is }\right.\cr
&\hspace{1cm}\left.+\left.\sum_{n=1}^\infty \left({2h-1\over \pi^2 }{k_c(h)\over 1- s^C(p) k_c(h)}\Psi_h(\chi) \right)\right|_{h=2n}\right]\label{eq:Cooper channel ladder diagram result}
\end{align}
%
%
where the cross ratio $\chi$ is given by $\chi={\tau_{12}\tau_{34}\over \tau_{13}\tau_{24} }$. Moreover, eigenvalue $k_c(h)$ of $\mathcal{K}$ is
\begin{equation}
k_c(h)=-{3\over 2} {\tan {\pi\over 2}(h-{1\over 2})\over h-{1\over 2}}
\end{equation}
%
The resulting four point function in \eqref{eq:Cooper channel ladder diagram result} is identical to the result in \cite{Gu:2016oyy}, and therefore, one can immediately conclude that Lypunov exponent is maximal. The only difference is that hopping constant $J_1$ and the effective coupling constant $\sqrt{J^2+J_1^2}$ in \cite{Gu:2016oyy} are replaced by $\effcoupling_\text{hop}=\sqrt{\hop_\R^2+\hop_\G^2+\hop_\B^2}$ and $\effcoupling=\sqrt{\local^2+\effcoupling_\text{hop}^2}$, respectively. One can then immediately conclude from their results that 
\begin{equation}
\lambda_L^C = {2 \pi \over \beta}
\end{equation}
In addition, diffusion constant and butterfly velocity are found to be
\begin{equation}
\dipole^C = { 2 \pi \effcoupling_\text{hop}^2 \over 3 \sqrt{2}\effcoupling \alpha_K}\;,\hspace{1cm} v^C_B = {2 \pi D \over \beta} 
\end{equation}
where $\alpha_K \approx 2.852$. At the scrambling time $t_\ast^C \sim \log N^3$, the ladder diagrams begin to be of the same order as the leading disconnected diagram (See in~\eqref{def:ladder diagram cooper channel}).
%

\begin{figure}
\centering
\begin{subfigure}[t]{\linewidth}
\centering
\def\c{-2} \def\s{1}  \def\gap{0.1}  \def\v{0.7} \def\ra{2}
\begin{tikzpicture}[scale=1]

\draw[thick,color=black!50!green] (-\s,-\v-0.05) -- (-\s,\v+0.05);
\draw[thick,color=blue] (-\s-\gap,-\v) -- (-\s-\gap,\v);
\draw[thick,color=red] (-\s+\gap,-\v) -- (-\s+\gap,\v);
\draw[thick,color=black!50!green] (\s,-\v-0.05) -- (\s,\v+0.05);
\draw[thick,color=red] (\s-\gap,-\v) -- (\s-\gap,\v);
\draw[thick,color=blue] (\s+\gap,-\v) -- (\s+\gap,\v);

\draw[thick,color=red,dashed] (-\s+\gap,\v) arc (90+30:90-30:\ra-2*\gap);
\draw[thick,color=red,dashed] (-\s+\gap,-\v) arc (270-30:270+30:\ra-2*\gap);
\draw[thick,color=blue,dashed] (-\s-\gap,\v) arc (90:270:\v);
\draw[thick,color=blue,dashed] (\s+\gap,\v) arc (90:-90:\v);
\draw[thick,color=black!50!green,dashed] (-\s,\v+0.05) arc (90-10:270+10:\v+0.08);
\draw[thick,color=black!50!green,dashed] (\s,\v+0.05) arc (90+10:-90-10:\v+0.08);

\end{tikzpicture}
\caption{Disconnected diagram in Pillow channel  $\sim N^5$}
\label{fig:disconnected diagram in pillow channel}
\end{subfigure}\\
\vspace{0.5cm}


\begin{subfigure}[t]{\linewidth}
\centering
\def\c{0} \def\s{1.2}  \def\v1{1/2} \def\b{1.4}  

\begin{tikzpicture}[scale=1]
\draw[thick,color=red] (-\s+\c,0.1) -- (\s+\c,0.1);
\draw[thick,color=red] (-\s+\c,-\b-0.1) -- (\s+\c,-\b-0.1);

\draw[thick,color=red,dashed] (\s+\c,0.1-\b-0.2) arc (90-20:-90+20:0.212);
\draw[thick,color=red,dashed] (-\s+\c,0.1-\b-0.2) arc (90+20:270-20:0.212);
\draw[thick,color=red,dashed] (-\s+\c,-\b-0.1-0.4) arc (-90-10:-90+10:7);
\draw[thick,color=red,dashed] (-\s+\c,0.1) arc (270-20:95:0.212);
\draw[thick,color=red,dashed] (\s+\c,0.1) arc (-90+20:90-5:0.212);
\draw[thick,color=red,dashed] (-\s+\c,0.5) arc (90+10:90-10:7);

\draw[thick,color=black!50!green,dashed] (-\s+\c,0) arc (90:270:\b/2);
\draw[thick,color=black!50!green] (-\s+\c,0) arc  (90:-90:\b/2);
\draw[thick,color=black!50!green,dashed] (\s+\c,0) arc (90:-90:\b/2);
\draw[thick,color=black!50!green] (\s+\c,0) arc (90:270:\b/2);

\draw[thick,color=blue,dashed] (-\s+\c,-0.1) arc (90:270:\b/2-0.1);
\draw[thick,color=blue] (-\s+\c,-0.1) arc (90:-90:\b/2-0.1);
\draw[thick,color=blue,dashed] (\s+\c,-0.1) arc (90:-90:\b/2-0.1);
\draw[thick,color=blue] (\s+\c,-0.1) arc (90:270:\b/2-0.1);

\draw[thick, color=black]
{
(\c-0.3,-\b/2) node [right] {\small{$\cdots$}} 
};
\end{tikzpicture}
\caption{Unbroken ladder diagram of the same color in Pillow channel $\sim N^4$}
\label{fig:unbroken ladder diagram of the same color in pillow channel}
\end{subfigure}
\caption{Leading and sub-leading diagrams in Pillow channel. The dashed line represents contraction of external gauge indices.}
\label{fig:leading and sub-leading diagrams in pillow channel}
\end{figure}
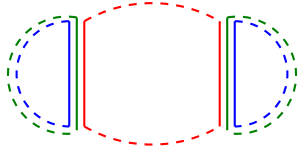
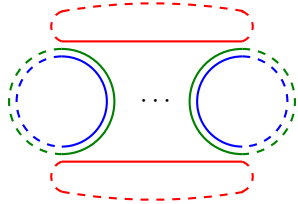

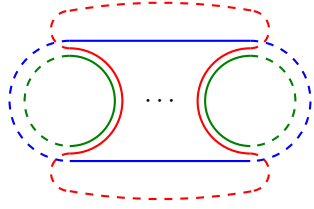
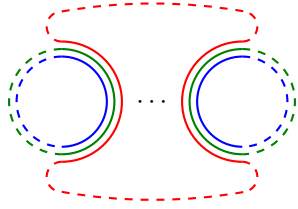
\begin{figure}
\centering
\begin{subfigure}[t]{\linewidth}
\centering
\def\c{0} \def\s{1.2}  \def\v1{1/2} \def\b{1.4}  
\begin{tikzpicture}[scale=1]

\draw[thick,color=blue] (-\s+\c,0.1) -- (\s+\c,0.1);
\draw[thick,color=blue] (-\s+\c,-\b-0.1) -- (\s+\c,-\b-0.1);
\draw[thick,color=blue,dashed] (\s+\c,0.1) arc (90:-90:\b/2+0.1);
\draw[thick,color=blue,dashed] (-\s+\c,0.1) arc (90:270:\b/2+0.1);

\draw[thick,color=red,dashed] (-\s+\c,0.0) arc (270:95:0.25);
\draw[thick,color=red,dashed] (\s+\c,0.0) arc (-90:90-5:0.25);
\draw[thick,color=red,dashed] (-\s+\c,0.5) arc (90+10:90-10:7);
\draw[thick,color=red,dashed] (\s+\c,-\b) arc (90:-90+5:0.25);
\draw[thick,color=red,dashed] (-\s+\c,-\b) arc (90:270-5:0.25);
\draw[thick,color=red,dashed] (-\s+\c,-\b-0.1-0.4) arc (-90-10:-90+10:7);

\draw[thick,color=red] (-\s+\c,0) arc  (90:-90:\b/2);
\draw[thick,color=red] (\s+\c,0) arc (90:270:\b/2);

\draw[thick,color=black!50!green,dashed] (-\s+\c,-0.1) arc (90:270:\b/2-0.1);
\draw[thick,color=black!50!green] (-\s+\c,-0.1) arc (90:-90:\b/2-0.1);
\draw[thick,color=black!50!green,dashed] (\s+\c,-0.1) arc (90:-90:\b/2-0.1);
\draw[thick,color=black!50!green] (\s+\c,-0.1) arc (90:270:\b/2-0.1);

\draw[thick, color=black]
{
(\c-0.35,-\b/2) node [right] {\small{$\cdots$}} 
};
\end{tikzpicture} 
\caption{Unbroken ladder diagram of different color in Pillow channel $\sim N^2$}
\label{fig:unbroken ladder diagram of different color  in pillow channel}	
\end{subfigure}\\
\vspace{0.5cm}

\begin{subfigure}[t]{\linewidth}
\centering
\def\c{0} \def\s{1.2}  \def\v1{1/2} \def\b{1.4}  
\begin{tikzpicture}[scale=1]

\draw[thick,color=red] (-\s+\c,0.1) arc (90:-90:\b/2+0.1);
\draw[thick,color=red] (\s+\c,0.1) arc (90:270:\b/2+0.1);
\draw[thick,color=red,dashed] (-\s+\c,0.1) arc (270:95:0.2);
\draw[thick,color=red,dashed] (\s+\c,0.1) arc (-90:90-5:0.2);
\draw[thick,color=red,dashed] (-\s+\c,0.5) arc (90+10:90-10:7);
\draw[thick,color=red,dashed] (\s+\c,0.1-\b-0.2) arc (90:-90+5:0.2);
\draw[thick,color=red,dashed] (-\s+\c,0.1-\b-0.2) arc (90:270-5:0.2);
\draw[thick,color=red,dashed] (-\s+\c,-\b-0.1-0.4) arc (-90-10:-90+10:7);

\draw[thick,color=black!50!green,dashed] (-\s+\c,0) arc (90:270:\b/2);
\draw[thick,color=black!50!green] (-\s+\c,0) arc  (90:-90:\b/2);
\draw[thick,color=black!50!green,dashed] (\s+\c,0) arc (90:-90:\b/2);
\draw[thick,color=black!50!green] (\s+\c,0) arc (90:270:\b/2);

\draw[thick,color=blue,dashed] (-\s+\c,-0.1) arc (90:270:\b/2-0.1);
\draw[thick,color=blue] (-\s+\c,-0.1) arc (90:-90:\b/2-0.1);
\draw[thick,color=blue,dashed] (\s+\c,-0.1) arc (90:-90:\b/2-0.1);
\draw[thick,color=blue] (\s+\c,-0.1) arc (90:270:\b/2-0.1);

\draw[thick, color=black]
{
(\c-0.35,-\b/2) node [right] {\small{$\cdots$}} 
};
\end{tikzpicture} 	
\caption{Broken ladder diagram in Pillow channel $\sim N^2$}
\label{fig:broken ladder diagram in pillow channel}
\end{subfigure}
\caption{Unbroken and broken ladder diagram contribution of order $O(N^2)$. The dashed line represents contraction of external gauge indices.}
\label{fig:unbroken and broken ladder diagram contribution of order n2}
\end{figure}

\subsection{Pillow Channel}
\label{sec:pillow channel}

Next, we consider Pillow contraction of external gauge indices in ladder diagrams. For any color $\mathfrak{c}$, we have
\begin{align}
F^{P,\mathfrak{c}}_{ab} (\tau_1,\tau_2,\tau_3,\tau_4)=& N^5 G(\tau_{12})G(\tau_{34}) + N^4 \mathcal{F}^{P,\mathfrak{c}}_{ab}(\tau_1,\tau_2,\tau_3,\tau_4)
\end{align}
The first term comes from the disconnected diagram as in the Cooper channel, but now it is of order~$\mathcal{O}(N^5)$ (See Fig.~\ref{fig:disconnected diagram in pillow channel} for $\mathfrak{c} = r$). Moreover, unlike the Cooper channel, unbroken and broken ladder diagrams have different order in $N$ under the Pillow contraction. For example, Pillow contraction of red color gives additional factor $N^6, N^4$ and $N^5$ to unbroken one of red color, unbroken one of green/blue colors and broken one, respectively (See Fig.~\ref{fig:unbroken ladder diagram of the same color in pillow channel} and \ref{fig:unbroken and broken ladder diagram contribution of order n2} ). Therefore, in the Pillow channel of a color $\col$, unbroken ladder diagrams of the color $\mathfrak{c}$ are of order~$\mathcal{O}(N^4)$ while others (either unbroken ones of different color or broken ones) are of order~$\mathcal{O}(N^2)$. This implies that (leading) ladder diagram $ \mathcal{F}^{P,\mathfrak{c}}$ can be represented by the following geometric series.
\begin{equation}
N^4 \mathcal{F}^{P,\mathfrak{c}}=N^4\sum_{n=0}^\infty  (\dipole^\mathfrak{c})^n\mathcal{F}_0\label{eq:geometric series for pillow 4pt ftn}
\end{equation}
where the product of dipoles is a matrix product defined in \eqref{def:matrix product}. For example, the common ratio of the geometric series for a specific color $\mathfrak{c}=\R$ is 
\begin{align}
\dipole^{\R}_{a_1 a_2} (\tau_1,\tau_2,\tau_3,\tau_4)=&-[(\local^2+\hop_\R^2)\delta_{a_1, a_2}+{1 \over 2 }(\hop_\G^2+\hop_\B^2)\delta_{a_1, a_2\pm1}]G(\tau_{13})G(\tau_{24})[G(\tau_{34})]^2\cr
\equiv&{1\over 3} S^{P,\R}_{a_1 a_2} \; \mathcal{K}(\tau_1,\tau_2,\tau_3,\tau_4)
\end{align}
where $S^{P,\R}_{a_1 a_2}$ is a hopping matrix defined by
\begin{equation}
S^{P,\R}_{a_1 a_2} = \delta_{a_1,a_2}+{\hop_{\G}^2+\hop_{\B}^2 \over 2\effcoupling^2}(\delta_{a_1, a_2\pm 1} -2 \delta_{a_1 a_2} )
\end{equation}
As in the Cooper channel, one can move to lattice momentum space to get the following expression for ${\cal F}^{P,\R}$:
\begin{equation}
{\cal F}^{P,\R}_p(\tau_1,\tau_2,\tau_3,\tau_4) = {1 \over 1 - {1 \over 3}s^{P,\R}(p) {\cal K}} {\cal F}_0  
\end{equation}
where the structure constant is given by $s^{P,\R}(p) = 1-{\hop_{\G}^2+\hop_{\B}^2   \over \effcoupling^2}(1- \cos p  )$. At this stage, note that the above equation is the same as \eqref{eq:FpinKTLattice} we obtained for the Cooper channel except for some trivial relabelling of couplings and the factor of ${1 \over 3}$ in the denominator. Hence, following the logic similar to \cite{Gu:2016oyy}, one can diagonalize ${\cal K}$ to obtain
\begin{align}
\mathcal{F}^{P,\R}_p(\chi)\equiv {\mathcal{F}^{P,\R}_p(\tau_1,\tau_2,\tau_3,\tau_4)\over G(\tau_{12})G(\tau_{34})}=&{4\pi \over 3}\left[\int_{-\infty}^\infty {ds\over 2\pi} \left.{h-{1\over 2}\over \pi \tan{\pi h\over 2}}{k_c(h)  \over 1-{1\over 3}s^{P,\R}(p) k_c(h)}\Psi_h(\chi)\right|_{h={1\over 2}+is }\right.\cr
&\hspace{0.5 cm}\left.+\left.\sum_{n=1}^\infty \left({2h-1\over \pi^2 }{k_c(h)\over 1- {1\over 3} s^{P,\R}(p) k_c(h)}\Psi_h(\chi) \right)\right|_{h=2n}\right]\label{eq:Pillow channel ladder diagram result}
\end{align}
%
%
%
One can also obtain the Pillow channels of other colors $\mathcal{F}^{P,\G}_p$ and $\mathcal{F}^{P,\B}_p$ by trivial permutation of couplings. The computation of Lyapunov exponent is, however, sensitive to this factor of ${1 \over 3}$ as we show in Appendix~\ref{app:large time behavior}. Because of additional factor ${1\over 3}$, we have
\begin{equation}
- {1\over 3} \;\leqq \; {1\over 3} s^{P,\mathfrak{c}}(p)\; \leqq \; {1\over 3} \, ,
\end{equation}
and therefore, $\mathcal{F}^{P,\mathfrak{c}}_p$ exhibits non-chaotic behavior according to analysis in Appendix~\ref{app:large time behavior}. This shows that the four point function does not grow exponentially up to term of order $\mathcal{O}(N^4)$. Note that this implies that in the Pillow channel the scrambling time $t_\ast^P>\log N$ because the leading disconnected diagram is of order $\mathcal{O}(N^5)$. 

This shows that to detect the exponential growth and hence extract the Lyapunov exponent and scrambling time, one has to look at subleading terms. There could be two sources for the subleading effects: $\frac{1}{\beta \effcoupling}$ or $\frac{1}{N}$ corrections, although the latter is much smaller than the former. To find ${1 \over \beta \effcoupling}$ effects, one needs to keep track of ${1 \over \beta \effcoupling}$ corrections to the propagator and \cite{Maldacena:2016hyu} has outlined a systematic procedure to do this although we will not pursue this here. The ${1 \over N}$ corrections are much harder, since now a host of other non-melonic diagrams as well as other class of diagrams \footnote{For example, in the Pillow channel, a vertical broken ladder diagram is also of order~$\mathcal{O}(N^3)$, and a tetrahedron diagram(or so-called `exceptional four point function' in $\rank=3$ case~\cite{Gurau:2016lzk}) is of order~$\mathcal{O}(N^{7\over 2})$} starts contributing. In fact, it was shown~\cite{Bonzom:2017pqs} that the leading non-melonic contribution scales like $\mathcal{O}(N^3)$. Hence, one cannot immediately conclude whether Pillow channel will also saturate the chaos bound or not.

In Appendix~\ref{app:spectrum}, we evaluated spectrum of the operators in the OPE limit. Let us mention the results for KT model explicitly here. Note that the zero momentum mode i.e ${\cal F}^{P,r}_{p=0}$ for KT chain model is actually just the corresponding connected four point function of KT model. Hence from the Pillow channel for KT model, one can read off the conformal dimensions of the operators are found to be
\begin{equation}
h_0\approx 1.7434\;,\quad h_1\approx 3.6018 \;,\quad h_2\approx 5.5627\;,\quad \cdots
\end{equation}
and, for large conformal dimension $h_n$, it asymptotes to 
\begin{equation}
h_n\approx 2n+{3\over 2} +{1\over 2\pi n}\hspace{1cm} (n\gg 1)
\end{equation}


\begin{figure}\begin{center}
\begin{tikzpicture}[scale=1]
\draw[ultra thick,color=black] (-2,0) -- (2,0);
\draw[ultra thick,color=black] (0,-2) -- (0,2);
\draw[thick, color=black]
{
(2,0) node [right] {\small{$c$}}
(-2,0) node [left] {\small{$a$}}
(0,-2) node [below] {\small{$d$}}
(0,2) node [above] {\small{$b$}}
};

\end{tikzpicture} 
\caption{Tetrahedron channel $\sim N^{9\over 2}$}
\label{fig:tetrahedron channel}
\end{center} \end{figure}
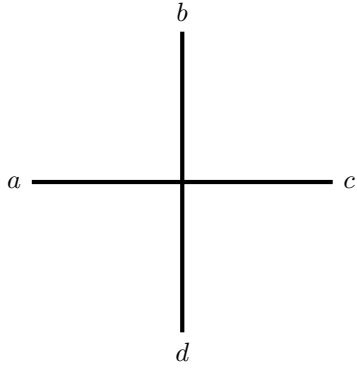
%

\subsection{Tetrahedron Channel}
\label{sec:tetrahedron channel}

Here, the leading contribution is qualitatively different from the previous other channels. The leading Tetrahedron channel does not come from disconnected diagram, but connected one in Fig.~\ref{fig:tetrahedron channel} which is of order $\mathcal{O}(N^{{9\over 2}})$
\begin{equation}
 \mathcal{F}^T_{a_1 a_2 a_3 a_4}(\tau_1,\tau_2,\tau_3,\tau_4)\equiv N^{-{9\over 2}} \langle \psi^{a_1}_{i_1 j_1 k_1}(\tau_1) \psi^{a_2}_{i_1 j_2 k_2}(\tau_2) \psi^{a_3}_{i_2 j_1 k_2}(\tau_3) \psi^{a_4}_{i_2 j_2 k_1}(\tau_4) \rangle 
\end{equation}
and is analogous to `exceptional four point function' in tensor models~\cite{Gurau:2016lzk}. From the diagram in Fig.~\ref{fig:tetrahedron channel}, we get
\begin{equation}
\mathcal{F}^T_{a_1a_2a_3a_4}(\tau_1,\tau_2,\tau_3,\tau_4) =\effcoupling^T_{a_1a_2a_3a_4}\int d\tau \; G(\tau_1,\tau)G(\tau_2,\tau)G(\tau_3,\tau)G(\tau_4,\tau)
\end{equation}
where
\begin{equation}
\effcoupling^T_{a_1a_2a_3a_4}= \begin{cases}
\quad \local & \quad a_1=a_2=a_3=a_4\\
\quad \hop_\R & \quad a_1=a_2\;,\; a_3=a_4\;\text{and}\; |a_1-a_3|=1\\
\quad \hop_\G & \quad a_1=a_3\;,\; a_2=a_4\;\text{and}\; |a_1-a_4|=1 \\
\quad \hop_\B & \quad  a_1=a_4\;,\; a_2=a_3\;\text{and}\; |a_1-a_2|=1 \\
\quad 0 & \quad \mbox{otherwise}\\
\end{cases}
\end{equation}
Using the two point function in~\eqref{eq: Two Point Function with effective coupling J KT model on lattice}, one can explicitly evaluate the leading Tetrahedron channel. For $\tau_1>\tau_2>\tau_3>\tau_4$, we obtain
\begin{align}
&\mathcal{F}^T_{a_1a_2a_3a_4}(\chi)\equiv{\mathcal{F}^T_{a_1a_2a_3a_4}(\tau_1,\tau_2,\tau_3,\tau_4) \over G(\tau_{12})G(\tau_{34})}\cr
=&{\effcoupling^T_{a_1a_2a_3a_4}\over \effcoupling }  \left({ 6\log 2 -\pi\over \sqrt{\pi} }  + {3\over 2\sqrt{\pi}} \log( 1-\chi )- {3\over 2\sqrt{\pi}} \log \chi \right) \chi^{1\over 2 } \; {}_2 F_1({1\over 2}, {1\over 2}; 1; \chi) \cr
&+ \sum_{n=1}^\infty \left[ \sqrt{{\effcoupling^T_{a_1a_2a_3a_4}\over \effcoupling }}   {\sqrt{3} (2n-1)!!\over \pi^{1\over 4} \sqrt{n} 2^{n} n! }\right]^2 \chi^{n+{1\over 2} }\;  {}_3F_2(n+{1\over 2 },n+ {1\over 2}, 1; n+1, n+1; \chi)\label{eq:tetrahedron channel result KT model}
\end{align}
where the cross ratio $\chi={\tau_{12}\tau_{34}\over \tau_{13}\tau_{24}}<1$. On the other hand, for $\tau_1>\tau_3>\tau_2>\tau_4$ where $\chi>1$, we have 
\begin{align}
&\mathcal{F}^T_{a_1a_2a_3a_4}(\chi)=-{\effcoupling^T_{a_1a_2a_3a_4}\over \effcoupling }  \left({ 8\log 2 - 3\pi\over 2\sqrt{\pi} }  + {1\over \sqrt{\pi}} \log{1-\chi\over \chi }\right)   {}_2 F_1({1\over 2}, {1\over 2}; 1; \chi) \cr
&\hspace{1cm}- \sum_{n=1}^\infty \left[ \sqrt{{\effcoupling^T_{a_1a_2a_3a_4}\over \effcoupling }}   {\sqrt{2} (2n-1)!!\over \pi^{1\over 4} \sqrt{n} 2^{n} n! }\right]^2 \chi^{n }  {}_3F_2(n+{1\over 2 },n+ {1\over 2}, 1; n+1, n+1; \chi)\label{eq:tetrahedron channel KT model bigger chi}
\end{align}
In order to evaluate the out-of-time-ordered correlator, we perform analytic continuation of $\chi>1$ expression in~\eqref{eq:tetrahedron channel KT model bigger chi} to $\chi<1$ as in~\cite{Maldacena:2016hyu}. Then, by conformal transformation in~\eqref{def:conformal transformation}, we take large $t$ limit (equivalently, $\chi\sim e^{-{2\pi  \over \beta}t}\longrightarrow  0$ ) to have
\begin{equation}\label{Tetrahedron4PointFn}
\mathcal{F}^T_{a_1a_2a_3a_4}(\chi) \sim  \log \chi  \sim t
\end{equation}
%

%
%
%
%
%
%
%
%

%
%
%
%
%
%
%

\section{Generalized Tensor Models on a Lattice}
\label{sec:models}

%
%
%
%
%
%

In this section, we will list some possible lattice generalizations of KT models. The most general translationally invariant Hamiltonian with a sublattice symmetry can be written as
\begin{align}\label{MostGenHamilt}
H^{n_2,n_3,n_4}  =   N^{-{3\over 2}} \sum_{a=1}^L \sum_{\alpha,\beta,\gamma,\delta=1}^M  \hop^{n_1,n_2,n_3}_{\alpha\beta\gamma\delta} \ \  \psi^{a,\alpha}_{i_1 j_1 k_1}\psi^{a+n_1,\beta}_{i_1 j_2 k_2}\psi^{a+n_2,\gamma}_{i_2 j_1 k_2}\psi^{a+n_3,\delta}_{i_2 j_2 k_1}
\end{align}
where the lattice index is $a=1,\cdots, L$, and the sublattice site index runs over $\alpha,\beta,\gamma,\delta = 1,\cdots, M$. For example, KT chain model given in \eqref{eq:hamiltonian of KT chain model} has no sublattice structure \ie $M=1$. Furthermore, we will henceforth only consider models with the following \emph{Unique Last Fermion (ULF) Property}
\begin{boxeD}{Unique Last Fermion (ULF) Property}
All the quartic interactions have the following property : Given three of the fermions participating in the interaction, the fourth is completely fixed
\end{boxeD}
Together with melonic dominance, \condition leads to a nice structure of two and four point functions. (\eg two point function is diagonal in the lattice space.) Some simple models with the above structure of tetrahedron quartic interactions are given below:

\begin{enumerate}
\item {\bf Klebanov-Tarnopolsky Chain Model (KT Chain model)} (Section~\ref{sec:large N KT model on lattice}):

\begin{align}\label{GQSModel}
H=&{\local N^{-{3\over 2}} \over 4} \sum_{a=1}^L  \psi^a_{i_1 j_1 k_1}\psi^a_{i_1 j_2 k_2}\psi^a_{i_2 j_1 k_2}\psi^a_{i_2 j_2 k_1}+{\hop_\R N^{-{3\over 2}} \over 2\sqrt 2} \sum_{a=1}^{L} \psi^{a}_{i_1 j_1 k_1}\psi^{a}_{i_1 j_2 k_2}\psi^{a+1}_{i_2 j_1 k_2}\psi^{a+1}_{i_2 j_2 k_1}\cr
&+{\hop_\G N^{-{3\over 2}} \over 2\sqrt 2}  \sum_{a=1}^{L} \psi^{a}_{i_1 j_1 k_1}\psi^{a+1}_{i_1 j_2 k_2}\psi^{a}_{i_2 j_1 k_2}\psi^{a+1}_{i_2 j_2 k_1}+{\hop_\B N^{-{3\over 2}} \over 2\sqrt 2}  \sum_{a=1}^{L} \psi^{a}_{i_1 j_1 k_1}\psi^{a+1}_{i_1 j_2 k_2}\psi^{a+1}_{i_2 j_1 k_2}\psi^{a}_{i_2 j_2 k_1}
\end{align}
KT chain model does not have sublattice symmetry (\ie $M=1$), and therefore we can drop the sublattice indices (\ie $\alpha, \beta,\cdots$). In the concise notation introduced in \eqref{MostGenHamilt}, this is given by the set of $\hop^{n_1,n_2,n_3} $:
\begin{align}
\hop^{011} = {\hop_\R \over 2\sqrt{2}} \hspace{10mm} \hop^{101} = {\hop_\G \over 2\sqrt{2}} \hspace{10mm} \hop^{110} = {\hop_\B \over 2\sqrt{2}}
\end{align}
%

\item {\bf Gurau-Witten Model (GW model)}~\cite{Witten:2016iux,Gurau:2016lzk} (Section~\ref{sec:GW model}): 
\begin{align}
H= \gw N^{-{3\over 2}} \psi^{1}_{i_1 j_1 k_1}\psi^{2}_{i_1 j_2 k_2}\psi^{3}_{i_2 j_1 k_2}\psi^{4}_{i_2 j_2 k_1}\label{def:hamiltonian of GW model}
\end{align}
As we will discuss in Section~\ref{sec:GW model}, for our purpose of computing correlators, GW model can be treated as a particular case of \eqref{MostGenHamilt}. Note that as written (\ref{def:hamiltonian of GW model}) does not have $\mathbb{Z}_4$  spatial translational symmetry and hence can be thought of as a sublattice $L=1$, $M=4$ case. But one of the the $\mathbb{Z}_2$ global symmetry, which we will see in Fig \ref{fig:symmetry of colored tetrahedron}, can play a role of translational symmetry so that GW model can be considered as $L=2$ and $M=2$ lattice.

%
%
%

\item {\bf Generalized Gurau-Witten Model} (Section~\ref{sec:generalized GW moel}):
\begin{align}
H\equiv&\local N^{-{3\over 2}}\sum_{a=1}^4  \psi^a_{i_1 j_1 k_1}\psi^a_{i_1 j_2 k_2}\psi^a_{i_2 j_1 k_2}\psi^a_{i_2 j_2 k_1}\cr
&+ \sum_{\sigma\in S_3 } \gw_{(1 \sigma(2) \sigma(3) \sigma(4) )} N^{-{3\over 2}} \psi^{1}_{i_1 j_1 k_1}\psi^{\sigma(2)}_{i_1 j_2 k_2}\psi^{\sigma(3)}_{i_2 j_1 k_2}\psi^{\sigma(4)}_{i_2 j_2 k_1}\label{def:hamiltonian of generalized GW model}
\end{align}
In generalized GW model, various hopping interactions as well as onsite interactions are added to GW model. Again, there is no $\mathbb{Z}_4$ spatial translational symmetry, but one can one can think of it as a sublattice like GW model (\ie $L=1$, $M=4$ or $L=2$, $M=2$). Note that the Hamiltonian for this model smoothly interpolates between that of KT model and GW model as we vary the couplings.
%
%

\end{enumerate}

Since we have already discussed KT model on a chain in section  \ref{sec:large N KT model on lattice}, we will next turn to analysing GW and generalized GW models in section \ref{sec:GW model} and section \ref{sec:generalized GW moel}.

\subsection{Gurau-Witten Colored Tensor Model}
\label{sec:GW model}

In this section, we will consider the Gurau-Witten colored tensor model~\cite{Witten:2016iux,Gurau:2016lzk}. The Hamiltonian for this model is given in \eqref{def:hamiltonian of GW model}, \ie looks the same as the KT model on a four-site lattice except for the lattice index. In spite of being described by the same Hamiltonian, the gauge symmetries of the two models are different: GW model has $O(N)^6$ gauge symmetry while the KT model on a four-site lattice has only  $O(N)^3$ gauge symmetry. However, since in quantum mechanics gauge fields are non-dynamical, the only role of gauging is to restrict the theory to the gauge invariant sector. In particular, computing observables outlined in~\eqref{def:KT Lattice 2 point function} and \eqref{All4PointFunc} which are gauge invariant in both models, will give identical results. Henceforth, we will not make a distinction between the two models and refer to both of them as GW model. We will also see that thinking of GW model as a KT model on lattice will serve as a useful bookkeeping device. For example, one may consider the diagonal $O(N)_\R$ subgroup of $O(N)_{12}\times O(N)_{34}$ gauge group\footnote{In GW model, for any pair $a,b$, there is a gauge group $O(N)_{ab}$ under which the $\psi^a_{ijk}$ and $\psi^b_{ijk}$ fermions transform in the fundamental representation, while all the other fermions are singlets of $O(N)_{ab}$} in GW model as one of the gauge group of four-site KT model, and similar for other colors. Then, the remaining becomes a global symmetry of the four-site KT model.

\begin{figure}[t!]
\centering
\begin{tikzpicture}[scale=0.8]
\def\xa{0}  \def\ya{0}   \def\xb{8}  \def\yb{0}    \def\xc{0}  \def\yc{6}    \def\xd{8}  \def\yd{6}    
 
    \coordinate (a1) at (\xa+4.4,\ya+2.5);
    \coordinate (b1) at (\xa+3,\ya+0.8);
    \coordinate (c1) at (\xa+4.8,\ya-0.2);
    \coordinate (d1) at (\xa+5.8,\ya+1.2);
    
      \draw[very thick,color=red] (b1) -- (a1) -- cycle;
    \draw[very thick,color=black!50!green]  (a1) -- (d1) -- cycle;       
    \draw[very thick,color=red]  (d1) -- (c1) -- cycle;
    \draw[very thick,color=blue] (a1) -- (c1);
    \draw[very thick,color=black!50!green] (c1) -- (b1)  -- cycle;
    \draw[very thick,dash dot dot,color=blue] (b1) -- (d1);
    
	\fill[black!20, draw=black, thick] (a1) circle (2pt) node[black, above right] {$1$};
    \fill[black!20, draw=black, thick] (b1) circle (2pt) node[black, above left] {$2$};
    \fill[black!20, draw=black, thick] (c1) circle (2pt) node[black, below right] {$4$};
    \fill[black!20, draw=black, thick] (d1) circle (2pt) node[black, above right] {$3$};
    
    \coordinate (a2) at (\xb+4.4,\yb+2.5);
    \coordinate (b2) at (\xb+3,\yb+0.8);
    \coordinate (c2) at (\xb+4.8,\yb-0.2);
    \coordinate (d2) at (\xb+5.8,\yb+1.2);

      \draw[very thick,color=red] (b2) -- (a2) -- cycle;
    \draw[very thick,color=black!50!green]  (a2) -- (d2) -- cycle;       
    \draw[very thick,color=red]  (d2) -- (c2) -- cycle;
    \draw[very thick,color=blue] (a2) -- (c2);
    \draw[very thick,color=black!50!green] (c2) -- (b2)  -- cycle;
    \draw[very thick,dash dot dot,color=blue] (b2) -- (d2);
    
	\fill[black!20, draw=black, thick] (a2) circle (2pt) node[black, above right] {$3$};
    \fill[black!20, draw=black, thick] (b2) circle (2pt) node[black, above left] {$4$};
    \fill[black!20, draw=black, thick] (c2) circle (2pt) node[black, below right] {$2$};
    \fill[black!20, draw=black, thick] (d2) circle (2pt) node[black, above right] {$1$};
    
    \coordinate (a3) at (\xc+4.4,\yc+2.5);
    \coordinate (b3) at (\xc+3,\yc+0.8);
    \coordinate (c3) at (\xc+4.8,\yc-0.2);
    \coordinate (d3) at (\xc+5.8,\yc+1.2);

  \draw[very thick,color=red] (b3) -- (a3) -- cycle;
    \draw[very thick,color=black!50!green]  (a3) -- (d3) -- cycle;       
    \draw[very thick,color=red]  (d3) -- (c3) -- cycle;
    \draw[very thick,color=blue] (a3) -- (c3);
    \draw[very thick,color=black!50!green] (c3) -- (b3)  -- cycle;
    \draw[very thick,dash dot dot,color=blue] (b3) -- (d3);
    
	\fill[black!20, draw=black, thick] (a3) circle (2pt) node[black, above right] {$2$};
    \fill[black!20, draw=black, thick] (b3) circle (2pt) node[black, above left] {$1$};
    \fill[black!20, draw=black, thick] (c3) circle (2pt) node[black, below right] {$3$};
    \fill[black!20, draw=black, thick] (d3) circle (2pt) node[black, above right] {$4$};

    \coordinate (a4) at (\xd+4.4,\yd+2.5);
    \coordinate (b4) at (\xd+3,\yd+0.8);
    \coordinate (c4) at (\xd+4.8,\yd-0.2);
    \coordinate (d4) at (\xd+5.8,\yd+1.2);
    
      \draw[very thick,color=red] (b4) -- (a4) -- cycle;
    \draw[very thick,color=black!50!green]  (a4) -- (d4) -- cycle;       
    \draw[very thick,color=red]  (d4) -- (c4) -- cycle;
    \draw[very thick,color=blue] (a4) -- (c4);
    \draw[very thick,color=black!50!green] (c4) -- (b4)  -- cycle;
    \draw[very thick,dash dot dot,color=blue] (b4) -- (d4);
    
	\fill[black!20, draw=black, thick] (a4) circle (2pt) node[black, above right] {$4$};
    \fill[black!20, draw=black, thick] (b4) circle (2pt) node[black, above left] {$3$};
    \fill[black!20, draw=black, thick] (c4) circle (2pt) node[black, below right] {$1$};
    \fill[black!20, draw=black, thick] (d4) circle (2pt) node[black, above right] {$ 2$};
    
    \draw[stealth-stealth,thick] (7,1) -- node[above] {$(13)(24)$} (9.6,1)  ;
    \draw[stealth-stealth,thick] (7,1+6) -- node[above] {$(13)(24)$} (9.6,1+6) ;    
    \draw[stealth-stealth,thick] (4.4,3.3) -- node[left] {$(12)(34)$} (4.4,5) ;       
    \draw[stealth-stealth,thick] (4.4+8,3.3) -- node[right] {$(12)(34)$} (4.4+8,5) ;        
   
\end{tikzpicture}
\caption{Symmetry of colored tetrahedron, $\mathbb{Z}_2\times \mathbb{Z}_2$. Each vertex represent the lattice index of fermion. (or, ``color'' in~\cite{Gurau:2016lzk}) The colored edge corresponds to contraction of gauge indices. Each vertex should be connected to others via specific color, which is actually equivalent to $O(N)^6$ gauge invariance.}
\label{fig:symmetry of colored tetrahedron}
\end{figure}
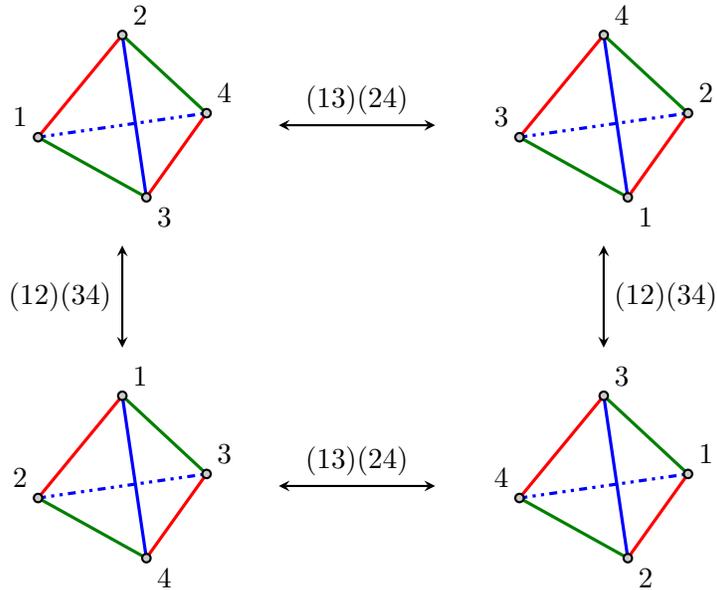
%

Moreover, the GW model has $\mathbb{Z}_2\times \mathbb{Z}_2$ symmetry which is related to symmetry of colored tetrahedron. The Hamiltonian~\eqref{def:hamiltonian of GW model} of the GW model is represented by colored tetrahedron, which is invariant under $\mathbb{Z}_2\times \mathbb{Z}_2$ transformation (See Fig.~\ref{fig:symmetry of colored tetrahedron}).

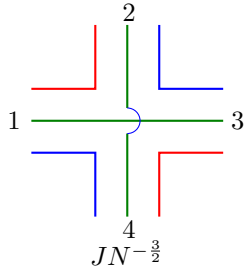
\begin{figure}[t!]
\centering
\begin{tikzpicture}[scale=0.85]
\draw[thick,color=red] (-1.5,0.5) -- (-0.5,0.5) -- (-0.5,1.5);
\draw[thick,color=blue] (-1.5,-0.5) -- (-0.5,-0.5) -- (-0.5,-1.5);
\draw[thick,color=red] (1.5,-0.5) -- (0.5,-0.5) -- (0.5,-1.5);
\draw[thick,color=blue] (1.5,0.5) -- (0.5,0.5) -- (0.5,1.5);

\draw[thick,color=black!50!green] (-1.5,0.0) -- (1.5,0.0) ;
\draw[thick,color=black!50!green] (0.0,1.5) -- (0.0,0.2) ;
\draw[thick,color=black!50!green] (0.0,-1.5) -- (0.0,-0.2) ;
\draw[color=blue] (0.0,0.2) arc (90:-90:0.2);
\draw[thick, color=black]
{
(0.0,-1.7) node [below] {\small{$ \gw N^{-{3 \over 2}}$}}
(-1.5,0) node [left] {\small{$ 1  $}}
(2.0,0) node [left] {\small{$ 3  $}}
(0.3,1.7) node [left] {\small{$ 2  $}}
(0.3,-1.7) node [left] {\small{$ 4  $}}
};
\end{tikzpicture}
\caption{The basic vertices of GW model}
\label{fig:vertex for GW model}
\end{figure}

The large $N$ diagrammatics of GW model is the same as the KT chain model except that GW model has only one vertex illustrated in Fig.~\ref{fig:vertex for GW model}. Hence, one may repeat the same large $N$ diagrammatics as before, especially for two point functions. Note that although GW model does not have $\mathbb{Z}_2^4$ symmetry as in the 4-site KT chain model, $O(N)^6$ symmetry implies that two point function is diagonal in the lattice space. Since there is only one vertex, one can easily find recursion like Fig.~\ref{fig:SD eq for two point function strand/single line}, and the corresponding Schwinger-Dyson equation for two point function reads
\begin{equation}
G(\tau_1,\tau_2)=G_0(\tau_1,\tau_2)+\gw^2\int d\tau_3 d\tau_4 G_0(\tau_1,\tau_3)[G(\tau_3,\tau_4)]^3G(\tau_4,\tau_2)\label{eq:SD eq for two point function GW model}
\end{equation}
And, as before, a solution in strong coupling limit $|\gw \tau_{12}|\gg 1$ is
\begin{equation}\label{Two Point Function with effective coupling J}
G(\tau_1,\tau_2)=b{\sgn(\tau_{12})\over |\gw \tau_{12}|^{1\over 2}}
\end{equation} 
where $b=-(4\pi)^{-{1\over 4}}$.

\begin{figure}[t!]
\centering
\begin{tikzpicture}[scale=1]
\def\xa{0}

\draw[thick,color=red] (\xa+-1.2,1) -- (\xa+1.2,1);

\draw[thick,color=blue]  (\xa+-1.2,0.7) -- (\xa+-0.4,0.7);
\draw[thick,color=blue]  (\xa+1.2,0.7) --  (\xa+0.4,0.7);
\draw[thick,color=blue]  (\xa+0.4,0.7) arc (45:-45:1);
\draw[thick,color=blue]  (\xa+-0.4,0.7) arc (135:225:1);

\draw[thick,color=black!50!green]  (\xa+-1.2,0.85)  -- (\xa+-0.1,0.85);
\draw[thick,color=black!50!green] (\xa+1.2,0.85)  -- (\xa+0.1,0.85);
\draw[thick,color=black!50!green] (\xa+-1.2,-0.85)  -- (\xa+-0.1,-0.85);
\draw[thick,color=black!50!green] (\xa+1.2,-0.85)  -- (\xa+0.1,-0.85);

\draw[thick,color=black!50!green] (\xa+-0.1,0.85) arc (65:0:1.0) ;
\draw[thick,color=black!50!green] (\xa+0.1,0.85) arc (-45:-45-45:0.3) ;
\draw[thick,color=black!50!green] (\xa+-0.1,0.78) arc (135:180+45:1.1) ;

\draw[thick,color=black!50!green] (\xa+-0.1,-0.85) arc (-65:0:1.0) ;
\draw[thick,color=black!50!green] (\xa+0.1,-0.85) arc (45:45+45:0.3) ;

\draw[thick,color=blue] (\xa+-1.2,-0.7) -- (\xa+-0.4,-0.7);
\draw[thick,color=blue] (\xa+1.2,-0.7) -- (\xa+0.4,-0.7);

\draw[thick,color=red] (\xa+0,0.6) arc (30:-30:1.3) ;
\draw[thick,color=red] (\xa+0,0.6) arc (180-30:180+30:1.3) ;
\draw[thick,color=red] (\xa+-1.2,-1) -- (\xa+1.2,-1);
\draw[thick, color=black]{
(\xa,-1) node [below] {\small{$ \gw^2 N^{-2}$}}
};
\draw[thick, color=black]
{
(-1.2,0.85) node [left] {\small{$ 1 $}}
(1.2,0.85) node [right] {\small{$ 2 $}}
(-1.4, -0.65) node [below] {\small{$ 1 $}}
(1.4, -1.05) node [above] {\small{$ 2 $}}
(-1.0,-0.3) node [above] {\small{$ 4 $}}
(0.95,-0.3) node [above] {\small{$ 3 $}}
};
\end{tikzpicture}\hspace{2cm}
\begin{tikzpicture}[scale=1]
\def\xa{0}

\draw[thick,color=red] (\xa+-1.2,1) -- (\xa+1.2,1);

\draw[thick,color=blue]  (\xa+-1.2,0.7) -- (\xa+-0.4,0.7);
\draw[thick,color=blue]  (\xa+1.2,0.7) --  (\xa+0.4,0.7);
\draw[thick,color=blue]  (\xa+0.4,0.7) arc (45:-45:1);
\draw[thick,color=blue]  (\xa+-0.4,0.7) arc (135:225:1);

\draw[thick,color=black!50!green]  (\xa+-1.2,0.85)  -- (\xa+-0.1,0.85);
\draw[thick,color=black!50!green] (\xa+1.2,0.85)  -- (\xa+0.1,0.85);
\draw[thick,color=black!50!green] (\xa+-1.2,-0.85)  -- (\xa+-0.1,-0.85);
\draw[thick,color=black!50!green] (\xa+1.2,-0.85)  -- (\xa+0.1,-0.85);

\draw[thick,color=black!50!green] (\xa+-0.1,0.85) arc (65:0:1.0) ;
\draw[thick,color=black!50!green] (\xa+0.1,0.85) arc (-45:-45-45:0.3) ;
\draw[thick,color=black!50!green] (\xa+-0.1,0.78) arc (135:180+45:1.1) ;

\draw[thick,color=black!50!green] (\xa+-0.1,-0.85) arc (-65:0:1.0) ;
\draw[thick,color=black!50!green] (\xa+0.1,-0.85) arc (45:45+45:0.3) ;

\draw[thick,color=blue] (\xa+-1.2,-0.7) -- (\xa+-0.4,-0.7);
\draw[thick,color=blue] (\xa+1.2,-0.7) -- (\xa+0.4,-0.7);

\draw[thick,color=red] (\xa+0,0.6) arc (30:-30:1.3) ;
\draw[thick,color=red] (\xa+0,0.6) arc (180-30:180+30:1.3) ;
\draw[thick,color=red] (\xa+-1.2,-1) -- (\xa+1.2,-1);
\draw[thick, color=black]{
(\xa,-1) node [below] {\small{$ \gw^2 N^{-2} $}}
};
\draw[thick, color=black]
{
(-1.2,0.85) node [left] {\small{$ 2 $}}
(1.2,0.85) node [right] {\small{$ 1 $}}
(-1.4, -0.65) node [below] {\small{$ 2 $}}
(1.4, -1.05) node [above] {\small{$ 1 $}}
(-1.0,-0.3) node [above] {\small{$ 3 $}}
(0.95,-0.3) node [above] {\small{$ 4 $}}
};
\end{tikzpicture}\\\vspace{0.5cm}
\begin{tikzpicture}[scale=1]
\def\xa{0}

\draw[thick,color=red] (\xa+-1.2,1) -- (\xa+1.2,1);

\draw[thick,color=blue]  (\xa+-1.2,0.7) -- (\xa+-0.4,0.7);
\draw[thick,color=blue]  (\xa+1.2,0.7) --  (\xa+0.4,0.7);
\draw[thick,color=blue]  (\xa+0.4,0.7) arc (45:-45:1);
\draw[thick,color=blue]  (\xa+-0.4,0.7) arc (135:225:1);

\draw[thick,color=black!50!green]  (\xa+-1.2,0.85)  -- (\xa+-0.1,0.85);
\draw[thick,color=black!50!green] (\xa+1.2,0.85)  -- (\xa+0.1,0.85);
\draw[thick,color=black!50!green] (\xa+-1.2,-0.85)  -- (\xa+-0.1,-0.85);
\draw[thick,color=black!50!green] (\xa+1.2,-0.85)  -- (\xa+0.1,-0.85);

\draw[thick,color=black!50!green] (\xa+-0.1,0.85) arc (65:0:1.0) ;
\draw[thick,color=black!50!green] (\xa+0.1,0.85) arc (-45:-45-45:0.3) ;
\draw[thick,color=black!50!green] (\xa+-0.1,0.78) arc (135:180+45:1.1) ;

\draw[thick,color=black!50!green] (\xa+-0.1,-0.85) arc (-65:0:1.0) ;
\draw[thick,color=black!50!green] (\xa+0.1,-0.85) arc (45:45+45:0.3) ;

\draw[thick,color=blue] (\xa+-1.2,-0.7) -- (\xa+-0.4,-0.7);
\draw[thick,color=blue] (\xa+1.2,-0.7) -- (\xa+0.4,-0.7);

\draw[thick,color=red] (\xa+0,0.6) arc (30:-30:1.3) ;
\draw[thick,color=red] (\xa+0,0.6) arc (180-30:180+30:1.3) ;
\draw[thick,color=red] (\xa+-1.2,-1) -- (\xa+1.2,-1);
\draw[thick, color=black]{
(\xa,-1) node [below] {\small{$ \gw^2 N^{-2} $}}
};
\draw[thick, color=black]
{
(-1.2,0.85) node [left] {\small{$ 3 $}}
(1.2,0.85) node [right] {\small{$ 4 $}}
(-1.4, -0.65) node [below] {\small{$ 3 $}}
(1.4, -1.05) node [above] {\small{$ 4 $}}
(-1.0,-0.3) node [above] {\small{$ 2 $}}
(0.95,-0.3) node [above] {\small{$ 1 $}}
};
\end{tikzpicture}\hspace{2cm}
\begin{tikzpicture}[scale=1]
\def\xa{0}

\draw[thick,color=red] (\xa+-1.2,1) -- (\xa+1.2,1);

\draw[thick,color=blue]  (\xa+-1.2,0.7) -- (\xa+-0.4,0.7);
\draw[thick,color=blue]  (\xa+1.2,0.7) --  (\xa+0.4,0.7);
\draw[thick,color=blue]  (\xa+0.4,0.7) arc (45:-45:1);
\draw[thick,color=blue]  (\xa+-0.4,0.7) arc (135:225:1);

\draw[thick,color=black!50!green]  (\xa+-1.2,0.85)  -- (\xa+-0.1,0.85);
\draw[thick,color=black!50!green] (\xa+1.2,0.85)  -- (\xa+0.1,0.85);
\draw[thick,color=black!50!green] (\xa+-1.2,-0.85)  -- (\xa+-0.1,-0.85);
\draw[thick,color=black!50!green] (\xa+1.2,-0.85)  -- (\xa+0.1,-0.85);

\draw[thick,color=black!50!green] (\xa+-0.1,0.85) arc (65:0:1.0) ;
\draw[thick,color=black!50!green] (\xa+0.1,0.85) arc (-45:-45-45:0.3) ;
\draw[thick,color=black!50!green] (\xa+-0.1,0.78) arc (135:180+45:1.1) ;

\draw[thick,color=black!50!green] (\xa+-0.1,-0.85) arc (-65:0:1.0) ;
\draw[thick,color=black!50!green] (\xa+0.1,-0.85) arc (45:45+45:0.3) ;

\draw[thick,color=blue] (\xa+-1.2,-0.7) -- (\xa+-0.4,-0.7);
\draw[thick,color=blue] (\xa+1.2,-0.7) -- (\xa+0.4,-0.7);

\draw[thick,color=red] (\xa+0,0.6) arc (30:-30:1.3) ;
\draw[thick,color=red] (\xa+0,0.6) arc (180-30:180+30:1.3) ;
\draw[thick,color=red] (\xa+-1.2,-1) -- (\xa+1.2,-1);
\draw[thick, color=black]{
(\xa,-1) node [below] {\small{$ \gw^2 N^{-2} $}}
};
\draw[thick, color=black]
{
(-1.2,0.85) node [left] {\small{$ 4 $}}
(1.2,0.85) node [right] {\small{$ 3 $}}
(-1.4, -0.65) node [below] {\small{$ 4 $}}
(1.4, -1.05) node [above] {\small{$ 3 $}}
(-1.0,-0.3) node [above] {\small{$ 1 $}}
(0.95,-0.3) node [above] {\small{$ 2 $}}
};
\end{tikzpicture}
\caption{\bf Dipoles in the GW model}
\label{fig:dipole contribution from interaction GW model}
\end{figure}
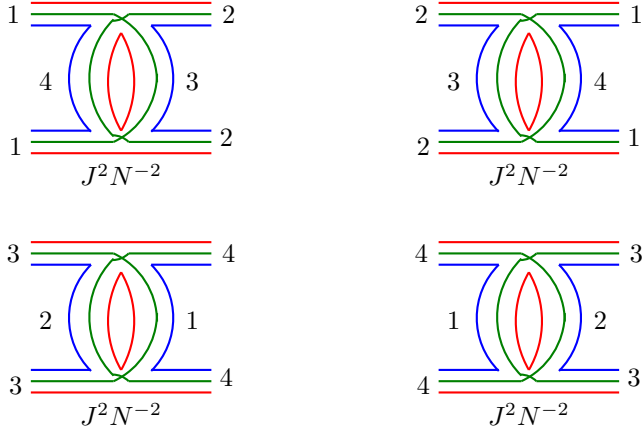

\paragraph{Four Point function:}We begin with evaluating dipoles which are now a $4\times4$ matrices due to the four sites. For example, considering the vertex in the Fig.~\ref{fig:vertex for GW model}, one can find four diagrams contributing to a dipole of red color (See Fig.~\ref{fig:dipole contribution from interaction GW model}), and similarly for other dipoles. Hence, we have 
\begin{align}
N^{-2}\dipole^\R(\tau_1,\tau_2,\tau_3,\tau_4)=&- N^{-2} \gw^2\begin{pmatrix}
0 & 1 & 0 & 0 \\
1 & 0 & 0 & 0\\
0 & 0&  0 &  1\\
0 & 0 & 1  & 0 \\
\end{pmatrix} G(\tau_{13}) G(\tau_{24}) [G(\tau_{34})]^2 ={1\over 3}\mathbb{1} \otimes \sigma_1\mathcal{K}\\
N^{-2} \dipole^\G(\tau_1,\tau_2,\tau_3,\tau_4)=&- N^{-2} \gw^2
 \begin{pmatrix}
0 & 0 & 1 & 0 \\
0 & 0 & 0 & 1\\
1 & 0 & 0 & 0\\
0 & 1 & 0 & 0 \\
\end{pmatrix}
G(\tau_{13}) G(\tau_{24}) [G(\tau_{34})]^2={1\over 3}\sigma_1 \otimes \mathbb{1}\mathcal{K}\\
N^{-2} \dipole^\B(\tau_1,\tau_2,\tau_3,\tau_4)=&- N^{-2} \gw^2  \begin{pmatrix}
0 & 0 & 0 & 1 \\
0 & 0 & 1 & 0\\
0 & 1 & 0 & 0\\
1 & 0 & 0  & 0 \\
\end{pmatrix} 
 G(\tau_{13}) G(\tau_{24}) [G(\tau_{34})]^2={1\over 3}  \sigma_1 \otimes \sigma_1  \mathcal{K}
\end{align}
where the kernel $\mathcal{K}(\tau_1,\tau_2,\tau_3,\tau_4)$ is given by
\begin{equation}
\mathcal{K}(\tau_1,\tau_2,\tau_3,\tau_4)\equiv -3 \gw^2G(\tau_{13})G(\tau_{24})[G(\tau_{34})]^2\label{eq:SYK kernel 3}
\end{equation}
Note that the above $4\times 4$ hopping matrices are symmetric because the dipole diagram is also symmetric under flip (See in Fig.~\ref{fig:dipole contribution from interaction GW model}). Also, note the $4\times 4$ matrix can be expressed as a Kronecker product of two $2\times 2 $ matrices. This naturally comes from the fact that the GW model can be interpreted as a tensor model on the $L=2$ $M=2$ lattice. From this point of view, one of the $2\times 2$ matrix corresponds to the hopping matrix of the $L=2$ lattice while the other is the hopping matrix of the $M=2$ sublattice.
%
%

\paragraph{Cooper Channel:} In large $N$ limit, like KT model, the leading four point function $F^C_{ab} (\tau_1,\tau_2,\tau_3,\tau_4)$ in Cooper channel comes from disconnected diagram and is of order $\mathcal{O}(N^6)$
\begin{equation}
F^C_{ab} (\tau_1,\tau_2,\tau_3,\tau_4)= N^6 G(\tau_{12})G(\tau_{34}) + N^3 \mathcal{F}^C_{ab} (\tau_1,\tau_2,\tau_3,\tau_4)\label{eq:four point function GW model}
\end{equation}
In the sub-leading order, both unbroken and broken ladder diagram in the Cooper channel are of the same order $\mathcal{O}(N^3)$. Hence, $\mathcal{F}^C_{ab}$ is composed of all combination of dipoles, which leads to a geometric series:
\begin{equation}
N^3 \mathcal{F}^C=N^3 \sum_{n=0}^\infty (\dipole^\R+\dipole^\G+\dipole^B)^n \mathcal{F}_0\label{eq:geometric series for Cooper channel GW model}
\end{equation}
The common ratio of the geometric series is
\begin{equation}
\dipole^\R+\dipole^\G+\dipole^\B=S^C \mathcal{K}
\end{equation}
where a hopping matrix is defined by
\begin{equation}
S^C\equiv {1\over 3}\begin{pmatrix}
0 & 1 & 1 &1 \\
1 & 0 & 1 &1 \\
1 & 1 & 0 &1 \\
1 & 1 & 1 &0 \\
\end{pmatrix}
\end{equation}
To evaluate the geometric series, one needs to diagonalize the common ratio. Since the kernel $\mathcal{K}(\tau_1,\tau_2,\tau_3,\tau_4)$ can be diagonalized in the same ways as before, it is sufficient to consider the eigenvalues of the hopping matrix:
\begin{equation}
S^C=U  \mbox{diag}(1,-{1\over 3},-{1\over 3},-{1\over 3})U^{-1} \hspace{0.5cm}\mbox{where}\hspace{0.5cm} U={1\over 2} \begin{pmatrix}
1 & 1  & 1 & 1  \\
1 & 1  & -1  & -1  \\
1 & -1  & 1  & -1  \\
1 & -1 &  -1  & 1  \\
\end{pmatrix}\label{eq:matrix U GW model}
\end{equation}
These modes, in fact, have distinct $\mathbb{Z}_2\times \mathbb{Z}_2$ charges explained in Fig.~\ref{fig:symmetry of colored tetrahedron}.
For each mode corresponding to eigenvalue $\xi=1$ or $-{1\over 3}$, one can evaluate its contribution to the Cooper channel:
\begin{align}
\mathcal{F}^C_\xi(\chi)\equiv{\mathcal{F}^C_\xi(\tau_1,\tau_2,\tau_3,\tau_4)\over G(\tau_{12})G(\tau_{34})}=&{4\pi \over 3}\left[\int_{-\infty}^\infty {ds\over 2\pi} \left.{h-{1\over 2}\over \pi \tan{\pi h\over 2}}{k_c(h)\over 1- \xi k_c(h)}\Psi_h(\chi)\right|_{h={1\over 2}+is }\right.\cr
&\hspace{1cm}\left.+\left.\sum_{n=1}^\infty \left({2h - 1\over \pi^2 }{k_c(h)\over 1- \xi k_c(h)}\Psi_h(\chi) \right)\right|_{h=2n}\right]\label{eq:Cooper channel ladder diagram result GW model}
\end{align}
In this basis, the leading term in \eqref{eq:four point function GW model} is also diagonalized:
\begin{equation}
G(\tau_{12})G(\tau_{34})\; \text{diag}(1,0,0,0)
\end{equation}
As before, $\xi=1$ mode is maximally chaotic. On the other hand, for $\xi=-{1\over 3}$ modes, the leading term vanishes as well as $\mathcal{F}^C_{-{1\over 3}}$ does not grow exponentially according to Appendix~\ref{app:large time behavior}.

\paragraph{Pillow Channel:} As in KT chain model, the leading Pillow channel comes from disconnected diagram of order~$\mathcal{O}(N^5)$
\begin{align}\label{eq:four point function Pillow channel GW model}
F^{P,\col}_{ab} (\tau_1,\tau_2,\tau_3,\tau_4)=& N^5 G(\tau_{12})G(\tau_{34}) + N^4 \mathcal{F}^{P,\col}_{ab}(\tau_1,\tau_2,\tau_3,\tau_4)\ ,
\end{align}
and the sub-leading contribution is given by unbroken ladder diagram with the same color as the Pillow channel. For example, the sub-leading contribution of Pillow channel of red color is given by a geometric series
\begin{equation}
N^4 \mathcal{F}^{P,\R}=N^4\sum_{n=0}^\infty  (\dipole^\R)^n\mathcal{F}_0\label{eq:geometric series for pillow 4pt ftn GW model}
\end{equation}
Its common ratio is 
\begin{equation}
\dipole^\R=S^{P,\R} \mathcal{K}
\end{equation}
where a hopping matrix is defined by
\begin{equation}
S^{P,\R}\equiv {1\over 3}\begin{pmatrix}
0 & 1 & 0 & 0 \\
1 & 0 & 0 & 0\\
0 & 0&  0 &  1\\
0 & 0 & 1  & 0 \\
\end{pmatrix}
\end{equation}
The eigenvalues of $S^{P,\R}$ are $\pm{1\over 3}$, and according to Appendix~\ref{app:large time behavior}, no eigenvalues leads to exponential behavior up to order $\mathcal{O}(N^4)$. Moreover, one can also simultaneously diagonalize the leading term in \eqref{eq:four point function Pillow channel GW model}. Namely, for one of $\xi=-{1\over 3}$ modes which has $(+,+)$ charge of $\mathbb{Z}_2\times \mathbb{Z}_2$, the leading term is $N^5 G(\tau_{12})G(\tau_{34})$ while it vanishes in the other three modes with $(-,\pm)$ or $(+,-)$ charges. Note that the mode with $(+,+)$ charge is analogous to $\xi=1$ mode which saturates chaos bound in the Cooper channel.

\paragraph{Tetrahedron Channel: } In GW model, there is only one non-vanishing tetrahedron channel which corresponds to `exceptional four point function' in the tensor model~\cite{Gurau:2016lzk}:
\begin{equation}
\langle \psi^{1}_{i_1 j_1 k_1}(\tau_1) \psi^{2}_{i_1 j_2 k_2}(\tau_2) \psi^{3}_{i_2 j_1 k_2}(\tau_3) \psi^{4}_{i_2 j_2 k_1}(\tau_4) \rangle = N^{9\over 2}\gw \int d\tau \; G(\tau_1,\tau)G(\tau_2,\tau)G(\tau_3,\tau)G(\tau_4,\tau)
\end{equation}
This gives the same result as the KT chain model in~\eqref{eq:tetrahedron channel result KT model}.

\subsection{Generalized Gurau-Witten Model : 4 Sites}
\label{sec:generalized GW moel}

%
\begin{figure}[t!]
\centering
\begin{subfigure}[t]{\linewidth}
\centering
\begin{tikzpicture}[scale=0.85]
\draw[thick,color=red] (-1.5,0.5) -- (-0.5,0.5) -- (-0.5,1.5);
\draw[thick,color=blue] (-1.5,-0.5) -- (-0.5,-0.5) -- (-0.5,-1.5);
\draw[thick,color=red] (1.5,-0.5) -- (0.5,-0.5) -- (0.5,-1.5);
\draw[thick,color=blue] (1.5,0.5) -- (0.5,0.5) -- (0.5,1.5);

\draw[thick,color=black!50!green] (-1.5,0.0) -- (1.5,0.0) ;
\draw[thick,color=black!50!green] (0.0,1.5) -- (0.0,0.2) ;
\draw[thick,color=black!50!green] (0.0,-1.5) -- (0.0,-0.2) ;
\draw[color=blue] (0.0,0.2) arc (90:-90:0.2);
\draw[thick, color=black]
{
(0.0,-1.7) node [below] {\small{$ \local N^{-{3 \over 2}}$}}
(-1.5,0) node [left] {\small{$ a  $}}
(2.0,0) node [left] {\small{$ a  $}}
(0.3,1.7) node [left] {\small{$ a  $}}
(0.3,-1.7) node [left] {\small{$ a  $}}
};
\end{tikzpicture}
\caption{On-site vertex of generalized GW model ($a=1,2,3,4$)}
\label{fig:on-site vertex for generalized GW model}
\end{subfigure}\\
\vspace{0.5cm}

\begin{subfigure}[t]{\linewidth}
\centering
\begin{tikzpicture}[scale=0.85]
\draw[thick,color=red] (-1.5,0.5) -- (-0.5,0.5) -- (-0.5,1.5);
\draw[thick,color=blue] (-1.5,-0.5) -- (-0.5,-0.5) -- (-0.5,-1.5);
\draw[thick,color=red] (1.5,-0.5) -- (0.5,-0.5) -- (0.5,-1.5);
\draw[thick,color=blue] (1.5,0.5) -- (0.5,0.5) -- (0.5,1.5);

\draw[thick,color=black!50!green] (-1.5,0.0) -- (1.5,0.0) ;
\draw[thick,color=black!50!green] (0.0,1.5) -- (0.0,0.2) ;
\draw[thick,color=black!50!green] (0.0,-1.5) -- (0.0,-0.2) ;
\draw[color=blue] (0.0,0.2) arc (90:-90:0.2);
\draw[thick, color=black]
{
(0.0,-1.7) node [below] {\small{$ \hop_{(1234)} N^{-{3 \over 2}}$}}
(-1.5,0) node [left] {\small{$ 1  $}}
(2.0,0) node [left] {\small{$ 3  $}}
(0.3,1.7) node [left] {\small{$ 2  $}}
(0.3,-1.7) node [left] {\small{$ 4  $}}
};
\end{tikzpicture}
\quad  \quad
\begin{tikzpicture}[scale=0.85]
\draw[thick,color=red] (-1.5,0.5) -- (-0.5,0.5) -- (-0.5,1.5);
\draw[thick,color=blue] (-1.5,-0.5) -- (-0.5,-0.5) -- (-0.5,-1.5);
\draw[thick,color=red] (1.5,-0.5) -- (0.5,-0.5) -- (0.5,-1.5);
\draw[thick,color=blue] (1.5,0.5) -- (0.5,0.5) -- (0.5,1.5);

\draw[thick,color=black!50!green] (-1.5,0.0) -- (1.5,0.0) ;
\draw[thick,color=black!50!green] (0.0,1.5) -- (0.0,0.2) ;
\draw[thick,color=black!50!green] (0.0,-1.5) -- (0.0,-0.2) ;
\draw[color=blue] (0.0,0.2) arc (90:-90:0.2);
\draw[thick, color=black]
{
(0.0,-1.7) node [below] {\small{$ \hop_{(1243)} N^{-{3 \over 2}}$}}
(-1.5,0) node [left] {\small{$ 1  $}}
(2.0,0) node [left] {\small{$ 4  $}}
(0.3,1.7) node [left] {\small{$ 2  $}}
(0.3,-1.7) node [left] {\small{$ 3  $}}
};
\end{tikzpicture}
\quad\quad
\begin{tikzpicture}[scale=0.85]
\draw[thick,color=red] (-1.5,0.5) -- (-0.5,0.5) -- (-0.5,1.5);
\draw[thick,color=blue] (-1.5,-0.5) -- (-0.5,-0.5) -- (-0.5,-1.5);
\draw[thick,color=red] (1.5,-0.5) -- (0.5,-0.5) -- (0.5,-1.5);
\draw[thick,color=blue] (1.5,0.5) -- (0.5,0.5) -- (0.5,1.5);

\draw[thick,color=black!50!green] (-1.5,0.0) -- (1.5,0.0) ;
\draw[thick,color=black!50!green] (0.0,1.5) -- (0.0,0.2) ;
\draw[thick,color=black!50!green] (0.0,-1.5) -- (0.0,-0.2) ;
\draw[color=blue] (0.0,0.2) arc (90:-90:0.2);
\draw[thick, color=black]
{
(0.0,-1.7) node [below] {\small{$ \hop_{(1324)} N^{-{3 \over 2}}$}}
(-1.5,0) node [left] {\small{$ 1  $}}
(2.0,0) node [left] {\small{$ 2  $}}
(0.3,1.7) node [left] {\small{$ 3  $}}
(0.3,-1.7) node [left] {\small{$ 4  $}}
};
\end{tikzpicture}
 \\
\vspace{0.5cm}
\begin{tikzpicture}[scale=0.85]
\draw[thick,color=red] (-1.5,0.5) -- (-0.5,0.5) -- (-0.5,1.5);
\draw[thick,color=blue] (-1.5,-0.5) -- (-0.5,-0.5) -- (-0.5,-1.5);
\draw[thick,color=red] (1.5,-0.5) -- (0.5,-0.5) -- (0.5,-1.5);
\draw[thick,color=blue] (1.5,0.5) -- (0.5,0.5) -- (0.5,1.5);

\draw[thick,color=black!50!green] (-1.5,0.0) -- (1.5,0.0) ;
\draw[thick,color=black!50!green] (0.0,1.5) -- (0.0,0.2) ;
\draw[thick,color=black!50!green] (0.0,-1.5) -- (0.0,-0.2) ;
\draw[color=blue] (0.0,0.2) arc (90:-90:0.2);
\draw[thick, color=black]
{
(0.0,-1.7) node [below] {\small{$ \hop_{(1342)} N^{-{3 \over 2}}$}}
(-1.5,0) node [left] {\small{$ 1  $}}
(2.0,0) node [left] {\small{$ 4  $}}
(0.3,1.7) node [left] {\small{$ 3  $}}
(0.3,-1.7) node [left] {\small{$ 2  $}}
};
\end{tikzpicture}
\quad  \quad
\begin{tikzpicture}[scale=0.85]
\draw[thick,color=red] (-1.5,0.5) -- (-0.5,0.5) -- (-0.5,1.5);
\draw[thick,color=blue] (-1.5,-0.5) -- (-0.5,-0.5) -- (-0.5,-1.5);
\draw[thick,color=red] (1.5,-0.5) -- (0.5,-0.5) -- (0.5,-1.5);
\draw[thick,color=blue] (1.5,0.5) -- (0.5,0.5) -- (0.5,1.5);

\draw[thick,color=black!50!green] (-1.5,0.0) -- (1.5,0.0) ;
\draw[thick,color=black!50!green] (0.0,1.5) -- (0.0,0.2) ;
\draw[thick,color=black!50!green] (0.0,-1.5) -- (0.0,-0.2) ;
\draw[color=blue] (0.0,0.2) arc (90:-90:0.2);
\draw[thick, color=black]
{
(0.0,-1.7) node [below] {\small{$ \hop_{(1423)} N^{-{3 \over 2}}$}}
(-1.5,0) node [left] {\small{$ 1  $}}
(2.0,0) node [left] {\small{$ 2  $}}
(0.3,1.7) node [left] {\small{$ 4  $}}
(0.3,-1.7) node [left] {\small{$ 3  $}}
};
\end{tikzpicture}
\quad\quad
\begin{tikzpicture}[scale=0.85]
\draw[thick,color=red] (-1.5,0.5) -- (-0.5,0.5) -- (-0.5,1.5);
\draw[thick,color=blue] (-1.5,-0.5) -- (-0.5,-0.5) -- (-0.5,-1.5);
\draw[thick,color=red] (1.5,-0.5) -- (0.5,-0.5) -- (0.5,-1.5);
\draw[thick,color=blue] (1.5,0.5) -- (0.5,0.5) -- (0.5,1.5);

\draw[thick,color=black!50!green] (-1.5,0.0) -- (1.5,0.0) ;
\draw[thick,color=black!50!green] (0.0,1.5) -- (0.0,0.2) ;
\draw[thick,color=black!50!green] (0.0,-1.5) -- (0.0,-0.2) ;
\draw[color=blue] (0.0,0.2) arc (90:-90:0.2);
\draw[thick, color=black]
{
(0.0,-1.7) node [below] {\small{$ \hop_{(1432)} N^{-{3 \over 2}}$}}
(-1.5,0) node [left] {\small{$ 1  $}}
(2.0,0) node [left] {\small{$ 3  $}}
(0.3,1.7) node [left] {\small{$ 4  $}}
(0.3,-1.7) node [left] {\small{$ 2  $}}
};
\end{tikzpicture}

\caption{6 hopping vertices of generalized GW model}
\label{fig:hopping vertices for generalized gw model}
\end{subfigure}
\caption{The Basic Vertices of generalized GW model}
\label{fig:vertices for generalized gw model}
\end{figure}
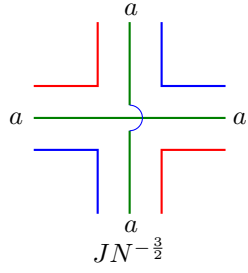
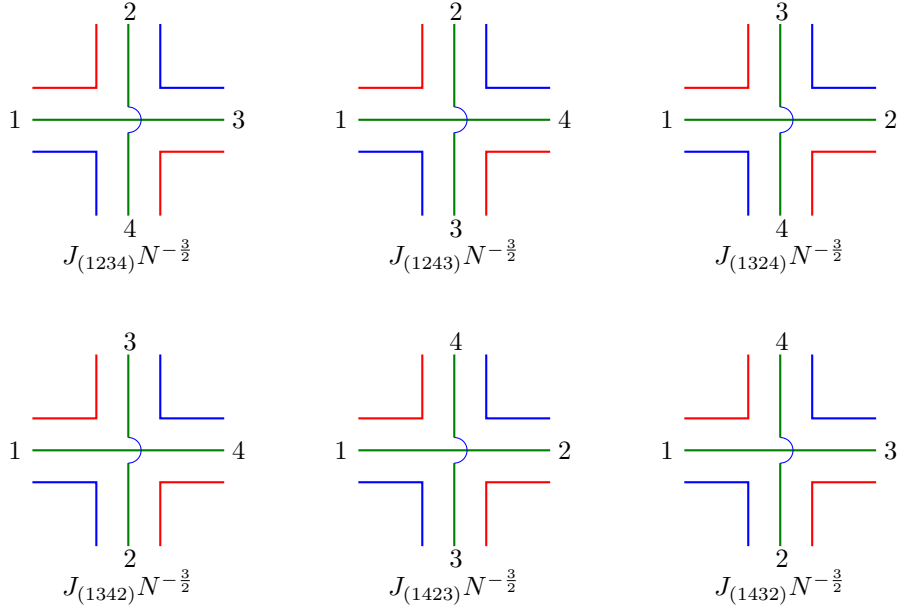
%

In the generalized GW model given in \eqref{def:hamiltonian of generalized GW model}, there are six interactions among four sites in addition to on-site interaction, and the corresponding vertices are shown in Figure~\ref{fig:vertices for generalized gw model}. Note that due to the on-site interaction, the generalized GW model  has $O(N)^3$ gauge symmetry instead of $O(N)^6$ gauge symmetry in the GW model. Like GW model, the generalized GW model also has the same $\mathbb{Z}_2\times \mathbb{Z}_2$ symmetry in Fig.~\ref{fig:symmetry of colored tetrahedron}. 

For the two point function in large $N$, one has seven melonic diagrams corresponding to the seven vertices. Therefore, the effective coupling constant $\effcoupling$ which appears in the two point function is given by
\begin{equation}
\effcoupling^2\equiv \local^2+\effhop^2\quad\mbox{and}\quad \effhop^2\equiv \hop_{(1234)}^2+\hop_{(1243)}^2+ \hop_{(1324)}^2+\hop_{(1342)}^2+\hop_{(1432)}^2+\hop_{(1423)}^2
\end{equation}
In a similar way to GW model, one can evaluate dipoles
\begin{align}
\dipole^\R= &\begin{pmatrix}
\local^2 & \hop_{(1234)}^2+\hop_{(1243)}^2 & \hop_{(1324)}^2+\hop_{(1342)}^2 & \hop_{(1432)}^2+\hop_{(1423)}^2\\
\hop_{(1234)}^2+\hop_{(1243)}^2 & \local^2 & \hop_{(1423)}^2+\hop_{(1432)}^2 & \hop_{(1324)}^2+\hop_{(1342)}^2\\
\hop_{(1324)}^2+\hop_{(1342)}^2 & \hop_{(1423)}^2+\hop_{(1432)}^2 &  \local^2 &   \hop_{(1234)}^2+\hop_{(1243)}^2\\
\hop_{(1432)}^2+\hop_{(1423)}^2 & \hop_{(1324)}^2+\hop_{(1342)}^2 & \hop_{(1234)}^2+\hop_{(1243)}^2  & \local^2  \\
\end{pmatrix}{1\over 3 \effcoupling^2}\mathcal{K}\\
\dipole^\G=& \begin{pmatrix}
\local^2 & \hop_{(1342)}^2+\hop_{(1432)}^2 & \hop_{(1423)}^2+\hop_{(1243)}^2 & \hop_{(1234)}^2+\hop_{(1324)}^2\\
\hop_{(1342)}^2+\hop_{(1432)}^2 & \local^2 & \hop_{(1324)}^2+\hop_{(1234)}^2 & \hop_{(1423)}^2+\hop_{(1243)}^2\\
\hop_{(1423)}^2+\hop_{(1243)}^2 & \hop_{(1324)}^2+\hop_{(1234)}^2 &  \local^2 &   \hop_{(1432)}^2+\hop_{(1342)}^2\\
\hop_{(1234)}^2+\hop_{(1324)}^2 & \hop_{(1423)}^2+\hop_{(1243)}^2 & \hop_{(1432)}^2+\hop_{(1342)}^2  & \local^2  \\
\end{pmatrix}{1\over 3 \effcoupling^2}\mathcal{K}\\
\dipole^\B=& \begin{pmatrix}
\local^2 & \hop_{(1423)}^2+\hop_{(1324)}^2 & \hop_{(1234)}^2+\hop_{(1432)}^2 & \\
\hop_{(1423)}^2+\hop_{(1324)}^2 & \local^2 & \hop_{(1243)}^2+\hop_{(1342)}^2 & \hop_{(1234)}^2+\hop_{(1432)}^2\\
\hop_{(1234)}^2+\hop_{(1432)}^2 & \hop_{(1243)}^2+\hop_{(1342)}^2 &  \local^2 &   \hop_{(1324)}^2+\hop_{(1423)}^2\\
\hop_{(1243)}^2+\hop_{(1342)}^2 & \hop_{(1234)}^2+\hop_{(1432)}^2 & \hop_{(1324)}^2+\hop_{(1423)}^2  & \local^2  \\
\end{pmatrix}{1\over 3 \effcoupling^2 }\mathcal{K}
\end{align}
As in GW model, we recognize that these $4\times 4$ matrices can be expressed as Kronecker product of two $2\times 2$ matrices. For example,
\begin{equation}
\dipole^\R=\local^2\mathbb{1}\otimes \mathbb{1}+(\hop_{(1234)}^2+\hop_{(1243)}^2)\mathbb{1}\otimes \sigma_1+(\hop_{(1324)}^2+\hop_{(1342)}^2)\sigma_1\otimes \mathbb{1}+( \hop_{(1432)}^2+\hop_{(1423)}^2)\sigma_1 \otimes \sigma_1
\end{equation}
which is also natural from the point of view of $L=2$, $M=2$ lattice. 
\paragraph{Cooper Channel:}For Cooper channel
\begin{equation}
F^C_{ab} (\tau_1,\tau_2,\tau_3,\tau_4)= N^6 G(\tau_{12})G(\tau_{34}) + N^3 \mathcal{F}^C_{ab} (\tau_1,\tau_2,\tau_3,\tau_4) \ ,\label{eq:four point function generalized GW model}
\end{equation}
we evaluate eigenvalues of a hopping matrix in $\dipole^\R+\dipole^\G+\dipole^\B$:
\begin{align}
&\dipole^\R+\dipole^\G+\dipole^\B=\begin{pmatrix}
3\local^2 & \effhop^2 & \effhop^2 & \effhop^2\\
\effhop^2 & 3\local^2 & \effhop^2 & \effhop^2\\
\effhop^2 & \effhop^2 & 3\local^2 &  \effhop^2\\
\effhop^2 & \effhop^2 & \effhop^2 & 3\local^2 \\
\end{pmatrix}{1\over 3 \effcoupling^2 }\mathcal{K}= U\mbox{diag}(1, \zeta , \zeta, \zeta )U^{-1}\mathcal{K}
\end{align}
where the matrix $U$ is the same as \eqref{eq:matrix U GW model}, and three eigenvalues are found to be
\begin{equation}
\zeta\equiv{ 3\local^2-\effhop^2\over 3\local^2+3\effhop^2}=-{1\over 3} + {{4\over 3} \over (\effhop^2/\local^2)+1}
\end{equation}
Note that as ${ \effhop^2 \over \local^2}$ goes from $0$ to $\infty$, the result goes smoothly from $\zeta = 1$ of KT model to $\zeta = -{1 \over 3}$ of GW model. 

According to Appendix~\ref{app:large time behavior}, $\xi=1$ mode, which has $(+,+)$~charge of $\mathbb{Z}_2\times \mathbb{Z}_2$, saturates chaos bound. On the other hand, other modes, which have $(-,\pm)$ or $(+,-)$ charge, are either non-maximally chaotic or non-chaotic depending on the ratio of on-site and hopping coupling constants:
\begin{alignat}{4}
&0<\effhop^2/ \local^2<1&&\qquad\Longrightarrow&&\qquad {1\over 3}<\zeta <1 &&\quad: \quad \mbox{non-maximal chaotic}\\
&1<\effhop^2/ \local^2<3 &&\qquad\Longrightarrow&&\qquad 0< \zeta<{1\over 3} &&\quad: \quad \mbox{non-chaotic}\\
&3<\effhop^2/ \local^2 &&\qquad\Longrightarrow&&\qquad  -{1\over 3} <\zeta<0&&\quad: \quad \mbox{non-chaotic}
\end{alignat}
Note that for $\zeta <{1 \over 3}$ or  ${ \effhop^2 \over \local^2} >1 $, the result is qualitatively similar to that of the GW model in that three of the four modes do not have exponential growth (up to this order). But for ${ \effhop^2 \over \local^2} < 1$, we have exponential growth but it is not maximal.

\paragraph{Pillow Channels:}
Like Pillow channels in previous models, Pillow channel in generalized GW model is again non-chaotic up to the of order $\mathcal{O}(N^4)$, which can be seen from the eigenvalues of dipoles:
\begin{equation}
\dipole^\R=\begin{pmatrix}
\local^2 & \hop_1^2 & \hop_2^2 & \hop_3^2\\
\hop_1^2 & \local^2 & \hop_3^2 & \hop_2^2\\
\hop_2^2 & \hop_3^2 & \local^2 &  \hop_1^2\\
\hop_3^2 & \hop_2^2 & \hop_1^2 & \local^2 \\
\end{pmatrix}{1\over 3\effhop^2}\mathcal{K}= U\mbox{diag}({1\over 3}, \zeta_1 , \zeta_2, \zeta_3 )U^{-1}K
\end{equation}
where
\begin{align}
\zeta_i={ \local^2-\effhop^2+2\hop_i^2\over 3\local^2+3\effhop^2}={1-\effhop^2/\local^2 +2\hop_i^2/\local^2\over 3+3\effhop^2/\local^2}
\end{align}
with $\hop_1^2\equiv  \hop_{(1234)}^2+\hop_{(1243)}^2$, $\hop_2^2\equiv \hop_{(1324)}^2+\hop_{(1342)}^2$ and $\hop_3^2\equiv  \hop_{(1432)}^2+\hop_{(1423)}^2 $.
In addition, the matrix $U$ is defined in \eqref{eq:matrix U GW model}. The $\zeta_i$ mode $(i=1,2,3)$ corresponds to the $(+,-)$, $(-,+)$ and $(-,-)$ charge under ${\mathbb Z}_2\times {\mathbb Z}_2$ global symmetry and the ${1\over 3}$ mode carries $(+,+)$ charge. One can show that $\mathcal{F}^{P,c}$ ($c=\R, \G, \B$) will never exhibit chaotic behavior for any value of coupling constants. To see this, one can rewrite $\zeta_1$ as follows.
%
%
\begin{equation}
\hop^2_2+\hop_3^2={1-3\zeta_1\over 1+3\zeta_1}(1+\hop_1^2)
\end{equation}
For chaotic behavior, one must have ${1\over 3}< \zeta_1\leqq 1$, which is inconsistent because the LHS is non-negative.

\section{Rank-$\rank$ Tensor Model}
\label{sec:rank d tensor model}

In this section, we will briefly generalize the large $N$ diagrammatics that we have discussed to study rank-$\rank$ tensor model. The higher rank tensor model has much richer structure than rank-3 tensor model. For example, there will be many more Pillow-like four point function channels possible due to various choices for the external indices contractions.  However, we will concentrate on Cooper channel and the simplest possible Pillow channel. It turns out that for these observables, the dipole we introduced before (where only one colour gets transmitted across) suffices and therefore, we will not introduce a new notation.

We will study the KT model of rank-$\rank$~\cite{Bonzom:2012hw,Klebanov:2016xxf} because the generalization to other lattice models is straightforward. It is convenient to understand the rank-$\rank$ KT model via ``uncoloring'' the GW model of rank-$\rank$ which has been more explicitly studied in the literature~\cite{Witten:2016iux,Gurau:2016lzk,Bonzom:2017pqs}.

The GW model has $(\rank+1)$ Majorana fermions $\psi^a$ $(a=0,1,\cdots, \rank)$. It has ${\rank(\rank+1)\over 2}$ number of $O(N)$ gauge groups, \ie
\begin{equation}
\prod_{0 \leqq a< b\leqq \rank} O_{(ab)}(N)
\end{equation}
where $\psi^a$ transforms in the vector representation under $O_{(ab)}$ for all $b\ne a$, and is invariant under other gauge groups. Note that we use the parenthesis in the subscript of the gauge group to clarify that the order of $a$ and $b$ is irrelevant (\ie $O_{(ab)}=O_{(ba)}$). Accordingly, each fermion $\psi^a$ has $\rank$ number of gauge indices:
\begin{equation}
\psi^a_{i_{(a0)}i_{(a1)}\cdots i_{(a a-1)}i_{(a a+1)} \cdots i_{(a\rank)}}
\end{equation}
where $i_{(ab)}$ corresponds to gauge index of $O_{(ab)}$.

``Uncoloring'' process is to strip the fermion species index\footnote{As in the rank-3 case, this index can be thought as lattice index. In the literature, this index is called ``color''. But, we use fermion species index or lattice index in order to avoid confusion with RGB color. $a(=0,1,2,\cdots, \rank)$ from the fermions,} $a$ and at the same time, to reduce $O(N)^{\rank+1\over 2}$ gauge groups into its diagonal subgroup $O(N)$ as we have discussed for $\rank=3$ case in Section~\ref{sec:GW model}. Equivalently, this can also be thought of as coloring the ${\rank(\rank+1)\over 2}$ edges of regular $\rank$-simplex with $\rank$ colors in a way that no vertex is connected to two edges of the identical color.\footnote{Or, let us consider all possible pairs of two numbers among $\{0,1,2,\cdots, \rank\}$. There are the ${}_{\rank+1}C_{2}$ number of such pairs, and we will group them into $\rank$ sets in such a way that no pairs in each set share common number. Hence, each set has the ${\rank+1\over 2}$  number of pairs. Then, for each set, one can find a diagonal subgroup of $\rank$ groups corresponding to $\rank$ pairs in the set. That is, for the given set $\{(a_1,a_2)\;,\;(a_3,a_4)\;,\; \cdots \;,\; (a_{\rank},a_{\rank+1})\}$, 
\begin{equation*}
O_{(a_1a_2)}(N)\times \cdots O_{(a_\rank a_{\rank+1})}(N)\quad\longrightarrow \quad O(N)
\end{equation*}} This is also equivalent to a schedule for a so-called ``Round-robin Tournament''\footnote{We thank Aaditya Salgarkar for pointing out this.} with $\rank+1$ teams. At the end of this process, we have $O(N)^\rank$ gauge group, and therefore, the fermion has only $\rank$ indices \ie $\psi_{i_1\cdots i_\rank}$.

The ``uncoloring'' process enable us to apply the large $N$ diagrammatics of GW model to KT model in a straightforward way as we have seen in the previous sections for rank-3. Of course, one evaluate correlators of GW model of rank-$\rank$ in the same way as in this section, which will lead to a similar result except for $(\rank+1)\times (\rank+1) $ hopping matrix in the four point functions. Hence, we will present the rank-$\rank$ KT model which is the simplest one but captures the essence of the calculations.

Two point function in rank-$\rank$ was already shown to be the same as that of SYK model with $(D+1)$-fold random coupling in~\cite{Witten:2016iux}. Thus, we will summarize the Cooper/Pillow channel of four point functions in the KT model of rank-$\rank$.

In rank-$\rank$ tensor model, dipoles\footnote{\cite{Gurau:2016lzk} defined $k$-dipole more generally. However, since we consider only $(\rank-1)$-dipole, we, for simplicity, mean $(\rank-1)$-dipole by dipole in this paper.} consist of four external legs and $\rank-2$ number of legs in the rung. As in rank-$3$ case, the rank-$\rank$ tensor model is defined with the scaled coupling with $N$ so that each vertex gives a contribution $N^{-{\rank(\rank-1)\over 4}}\local$~\cite{Witten:2016iux,Gurau:2016lzk}. Hence, dipoles are of order $\mathcal{O}(N^{-D+1})$~\cite{Gurau:2016lzk} in large $N$. As in the rank-3 tensor, one can construct unbroken and broken ladder diagram by connecting these dipoles. 
%

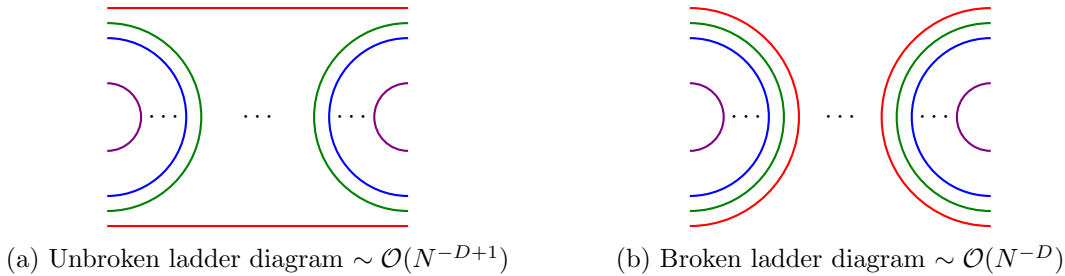
\begin{figure}
\centering

\begin{subfigure}[b]{0.49\linewidth}
\centering
 \def\c{1} \def\s{2}  \def\v1{1/2} \def\b{2.5}  

\begin{tikzpicture}[scale=1]
\draw[thick,color=red] (-\s+\c,0.2) -- (\s+\c,0.2);
\draw[thick,color=red] (-\s+\c,-\b-0.2) -- (\s+\c,-\b-0.2);

\draw[thick,color=black!50!green] (-\s+\c,0) arc  (90:-90:\b/2);
\draw[thick,color=black!50!green] (\s+\c,0) arc (90:270:\b/2);

\draw[thick,color=blue] (-\s+\c,-0.2) arc (90:-90:\b/2-0.2);
\draw[thick,color=blue] (\s+\c,-0.2) arc (90:270:\b/2-0.2);

\draw[thick,color=violet] (-\s+\c,-0.8) arc (90:-90:\b/2-0.8);
\draw[thick,color=violet] (\s+\c,-0.8) arc (90:270:\b/2-0.8);

\draw[thick, color=black]
{
(-\s+\b/2+\c+0.4,-\b/2) node [right] {\small{$\cdots$}} 
(\s+\b/2-\c - 0.35,-\b/2) node [right] {\small{$\bf \cdots$}}
(-\s+\b/2+\c - 0.85 ,-\b/2) node [right] {\small{$\bf \cdots$}} 
};
\end{tikzpicture}
\caption{Unbroken ladder diagram $\sim \mathcal{O}(N^{-D+1})$}
\label{fig:unbroken ladder diagram in rank D}
\end{subfigure}  
\begin{subfigure}[b]{0.49\linewidth}
\centering
\def\c{1} \def\s{2}  \def\v1{1/2} \def\b{2.5}  
\begin{tikzpicture}[scale=1]

\draw[thick,color=red] (-\s+\c,0.2) arc (90:-90:\b/2+0.2);
\draw[thick,color=red] (\s+\c,0.2) arc (90:270:\b/2+0.2);

\draw[thick,color=black!50!green] (-\s+\c,0) arc  (90:-90:\b/2);
\draw[thick,color=black!50!green] (\s+\c,0) arc (90:270:\b/2);

\draw[thick,color=blue] (-\s+\c,-0.2) arc (90:-90:\b/2-0.2);
\draw[thick,color=blue] (\s+\c,-0.2) arc (90:270:\b/2-0.2);

\draw[thick,color=violet] (-\s+\c,-0.8) arc (90:-90:\b/2-0.8);
\draw[thick,color=violet] (\s+\c,-0.8) arc (90:270:\b/2-0.8);

\draw[thick, color=black]
{
(-\s+\b/2+\c+0.4,-\b/2) node [right] {\small{$\cdots$}} 
(\s+\b/2-\c - 0.35,-\b/2) node [right] {\small{$\bf \cdots$}}
(-\s+\b/2+\c - 0.85 ,-\b/2) node [right] {\small{$\bf \cdots$}} 
};
\end{tikzpicture} 
\caption{Broken ladder diagram $\sim\mathcal{O}(N^{-D})$}
\label{fig:broken ladder diagram in rank D}
\end{subfigure}
\caption{Unbroken and broken ladder diagrams (without the contraction of gauge indices) in rank-$\rank$ tensor model}
\label{fig:unbroken and broken ladder diagram in rank D}
\end{figure}

\begin{itemize}
\item In unbroken ladder diagram, one of $\rank$ colors passes through the ladder while others go back (\ie Fig.~\ref{fig:unbroken ladder diagram in rank D}). By induction, one can easily show that the unbroken one is of order $\mathcal{O}(N^{-\rank+1})$. By attaching the same color of dipole to the unbroken ladder diagram, one gets additional $N^{\rank-1}$, which will be cancelled with $N$ scale of the attached dipole.

\item All $\rank$ colors return in the broken ladder diagram (\ie Fig.~\ref{fig:broken ladder diagram in rank D}). One can generate broken ladder diagrams by connecting dipoles in a way that at least one color of dipoles is different from others. One can also prove that broken ladder diagrams are of order $\mathcal{O}(N^{-\rank})$ in a similar way.

\end{itemize}
%

\subsection{Cooper Channel in Rank-$\rank$}
\label{sec:cooper channel rank d}

\begin{figure}
\centering

\begin{subfigure}[b]{\linewidth}
\centering
 \def\c{1} \def\s{2}  \def\v1{1/2} \def\b{2.5}  

\begin{tikzpicture}[scale=1]
\draw[thick,color=red] (-\s+\c,0.2) -- (\s+\c,0.2);
\draw[thick,color=red] (-\s+\c,-\b-0.2) -- (\s+\c,-\b-0.2);
\draw[thick,color=red,dashed] (-\s+\c,0.2) arc (90:270:\b/2+0.2);
\draw[thick,color=red,dashed] (\s+\c,0.2) arc (90:-90:\b/2+0.2);

\draw[thick,color=black!50!green,dashed] (-\s+\c,0) arc (90:270:\b/2);
\draw[thick,color=black!50!green] (-\s+\c,0) arc  (90:-90:\b/2);
\draw[thick,color=black!50!green,dashed] (\s+\c,0) arc (90:-90:\b/2);
\draw[thick,color=black!50!green] (\s+\c,0) arc (90:270:\b/2);

\draw[thick,color=blue,dashed] (-\s+\c,-0.2) arc (90:270:\b/2-0.2);
\draw[thick,color=blue] (-\s+\c,-0.2) arc (90:-90:\b/2-0.2);
\draw[thick,color=blue,dashed] (\s+\c,-0.2) arc (90:-90:\b/2-0.2);
\draw[thick,color=blue] (\s+\c,-0.2) arc (90:270:\b/2-0.2);

\draw[thick,color=violet,dashed] (-\s+\c,-0.8) arc (90:270:\b/2-0.8);
\draw[thick,color=violet] (-\s+\c,-0.8) arc (90:-90:\b/2-0.8);
\draw[thick,color=violet,dashed] (\s+\c,-0.8) arc (90:-90:\b/2-0.8);
\draw[thick,color=violet] (\s+\c,-0.8) arc (90:270:\b/2-0.8);

\draw[thick, color=black]
{
(-\s+\b/2+\c+0.4,-\b/2) node [right] {\small{$\cdots$}} 
(-\s-\b/2+\c + 0.15,-\b/2) node [right] {\small{$\bf \cdots$}} 
(\s+\b/2-\c - 0.35,-\b/2) node [right] {\small{$\bf \cdots$}}
(-\s+\b/2+\c - 0.85 ,-\b/2) node [right] {\small{$\bf \cdots$}} 
(\s+\b/2+\c- 0.85,-\b/2) node [right] {\small{$\bf \cdots$}} 
};
\end{tikzpicture}
\caption{Unbroken ladder diagram of order $\mathcal{O}(N^D)$ in Cooper channel in rank-$\rank$ tensor model}
\label{fig:unbroken ladder diagram in cooper channel in rank D}
\end{subfigure} \hspace{10mm} \\
\vspace{0.5cm}
\begin{subfigure}[b]{\linewidth}
\centering
\def\c{1} \def\s{2}  \def\v1{1/2} \def\b{2.5}  
\begin{tikzpicture}[scale=1]

\draw[thick,color=red] (-\s+\c,0.2) arc (90:-90:\b/2+0.2);
\draw[thick,color=red] (\s+\c,0.2) arc (90:270:\b/2+0.2);
\draw[thick,color=red,dashed] (-\s+\c,0.2) arc (90:270:\b/2+0.2);
\draw[thick,color=red,dashed] (\s+\c,0.2) arc (90:-90:\b/2+0.2);

\draw[thick,color=black!50!green,dashed] (-\s+\c,0) arc (90:270:\b/2);
\draw[thick,color=black!50!green] (-\s+\c,0) arc  (90:-90:\b/2);
\draw[thick,color=black!50!green,dashed] (\s+\c,0) arc (90:-90:\b/2);
\draw[thick,color=black!50!green] (\s+\c,0) arc (90:270:\b/2);

\draw[thick,color=blue,dashed] (-\s+\c,-0.2) arc (90:270:\b/2-0.2);
\draw[thick,color=blue] (-\s+\c,-0.2) arc (90:-90:\b/2-0.2);
\draw[thick,color=blue,dashed] (\s+\c,-0.2) arc (90:-90:\b/2-0.2);
\draw[thick,color=blue] (\s+\c,-0.2) arc (90:270:\b/2-0.2);

\draw[thick,color=violet,dashed] (-\s+\c,-0.8) arc (90:270:\b/2-0.8);
\draw[thick,color=violet] (-\s+\c,-0.8) arc (90:-90:\b/2-0.8);
\draw[thick,color=violet,dashed] (\s+\c,-0.8) arc (90:-90:\b/2-0.8);
\draw[thick,color=violet] (\s+\c,-0.8) arc (90:270:\b/2-0.8);

\draw[thick, color=black]
{
(-\s+\b/2+\c+0.4,-\b/2) node [right] {\small{$\cdots$}} 
(-\s-\b/2+\c + 0.15,-\b/2) node [right] {\small{$\bf \cdots$}} 
(\s+\b/2-\c - 0.35,-\b/2) node [right] {\small{$\bf \cdots$}}
(-\s+\b/2+\c - 0.85 ,-\b/2) node [right] {\small{$\bf \cdots$}} 
(\s+\b/2+\c- 0.85,-\b/2) node [right] {\small{$\bf \cdots$}} 
};
\end{tikzpicture} 
\caption{Broken ladder diagram of order $\mathcal{O}(N^D)$ in Cooper channel in rank-$\rank$ tensor model}
\label{fig:broken ladder diagram in cooper channel in rank D}
\end{subfigure}
\caption{Unbroken and broken ladder diagrams of order $\mathcal{O}(N^D)$ in Cooper channel in rank-$\rank$ tensor model}
\label{fig:unbroken and broken ladder diagram in cooper channel in rank D}
\end{figure}
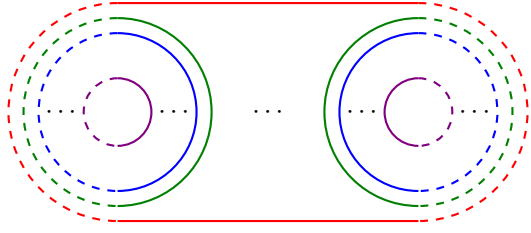
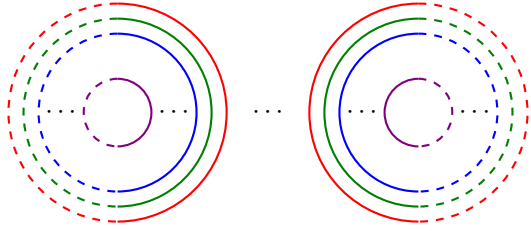

As we have seen in rank-3 tensor, depending on the contraction of external gauge indices, $N$ scalings of ladder diagrams become different. Though $N$ scalings of unbroken and broken ladder diagrams are different, Cooper contraction of gauge indices make them same order. As shown in Fig.~\ref{fig:unbroken and broken ladder diagram in cooper channel in rank D}, Cooper contraction of unbroken ladder diagram gives $(2D-1)$ loops while broken one gains $2D$ loops. Hence, all ladder diagrams in the Cooper channel are of order $\mathcal{O}(N^{D})$.

As in the rank-3 tensor model, Cooper channel is 
\begin{align}
F^C (\tau_1,\tau_2,\tau_3,\tau_4)\equiv& \langle \psi_{I_1}(\tau_1) \psi_{I_1}(\tau_2) \psi_{I_2}(\tau_3) \psi_{I_2}(\tau_4) \rangle\cr
=& N^{2D} G(\tau_{12})G(\tau_{34}) + N^D \mathcal{F}^C (\tau_1,\tau_2,\tau_3,\tau_4) \label{eq:Cooper channer app}
\end{align}
where $I_1$ and $I_2$ are a collection of indices. \ie $I_1=(i_1,i_2,\cdots, i_\rank)$. Note that the leading disconnected diagram is of order $\mathcal{O}(N^{2\rank})$. As we have discussed, unbroken/broken ladder diagrams are of the same order $\mathcal{O}(N^\rank)$. Hence, $\mathcal{F}^C$ can be written as a geometric series of dipoles:
\begin{equation}
\mathcal{F}^C= \sum_{n=0}^\infty \left[\sum_{j=1}^\rank\dipole^{\col_j}\right]^n \mathcal{F}_0\label{eq:geometric series leading Cooper rank d}
\end{equation}
The common ratio can be found to be
\begin{equation}
\left(\sum_{j=1}^\rank\dipole^{\col_j}\right)(\tau_1,\tau_2,\tau_3,\tau_4)=\mathcal{K}(\tau_1,\tau_2,\tau_3,\tau_4) 
\end{equation}
where the kernel
\begin{equation}
\mathcal{K}(\tau_1,\tau_2,\tau_3,\tau_4) =- \rank \local^2 G(\tau_{13})G(\tau_{24})[G(\tau_{34})]^{D-1}
\end{equation}
is exactly the same as that in SYK model with $(D+1)$-fold random coupling. Hence, one can diagonalize the kernel $\mathcal{K}$ as in \cite{Maldacena:2016hyu} to evaluate the geometric series:
\begin{align}
\mathcal{F}^C_{h\ne 2}(\chi)\equiv{\mathcal{F}_{h\ne 2}^C(\tau_1,\tau_2,\tau_3,\tau_4)\over G(\tau_{12})G(\tau_{34})}=&{4\pi \over 3}\left[\int_{-\infty}^\infty {ds\over 2\pi} \left.{h-{1\over 2}\over \pi \tan{\pi h\over 2}}{k_c(h)\over 1- k_c(h)}\Psi_h(\chi)\right|_{h={1\over 2}+is }\right.\cr
&\hspace{1cm}\left.+\left.\sum_{n=1}^\infty \left({2h-1\over \pi^2 }{k_c(h)\over 1-  k_c(h)}\Psi_h(\chi) \right)\right|_{h=2n}\right]
\end{align}
where $k_c(h)$ is an eigenvalue of $\mathcal{K}$ given by~\cite{Maldacena:2016hyu}
\begin{equation}
k_c(h)\equiv -\rank {\Gamma\left({3\over 2} -{1\over \rank+1}\right)\Gamma\left(1 -{1\over \rank+1}\right)\Gamma\left({1\over \rank+1}+{h\over 2} \right)\Gamma\left({1\over 2} + {1\over \rank+1}-{h\over 2} \right)\over \Gamma\left({1\over 2} + {1\over \rank+1}\right)\Gamma\left({1\over \rank+1}\right)\Gamma\left({3\over 2} -{1\over \rank+1}-{h\over 2} \right)\Gamma\left(1-{1\over \rank+1}+{h\over 2} \right)}
\end{equation}
 This result is exactly the same as four point function of SYK model with $(D+1)$-fold random coupling and as in \cite{Maldacena:2016hyu} the Cooper channel also saturate chaos bound. In addition, comparing the leading and sub-leading term in \eqref{eq:Cooper channer app}, one can see that the scrambling time reads
\begin{equation}
t_\ast^C \sim \log N^\rank\label{eq:scrambling time cooper rank d}
\end{equation}

\subsection{Pillow Channel in Rank-$\rank$}
\label{sec:pillow channel rank d}

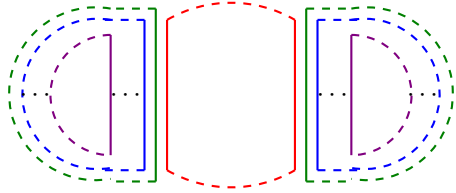
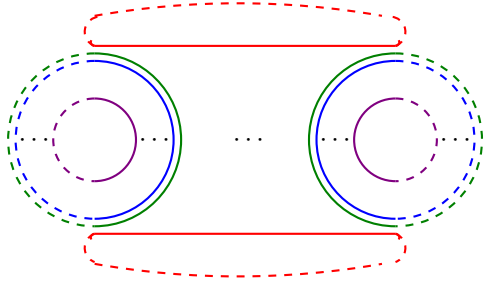
\begin{figure}
\centering
\begin{subfigure}[t]{\linewidth}
\centering
\def\c{-2} \def\s{1}  \def\gap{0.15}  \def\v{1} \def\ra{2}
\begin{tikzpicture}[scale=1]

\draw[thick,color=black!50!green] (-\s,-\v-0.15) -- (-\s,\v+0.15);
\draw[thick,color=blue] (-\s-\gap,-\v) -- (-\s-\gap,\v);
\draw[thick,color=red] (-\s+\gap,-\v) -- (-\s+\gap,\v);
\draw[thick,color=violet] (-\s-4*\gap,-\v+0.2) -- (-\s-4*\gap,\v-0.2);
\draw[thick,color=black!50!green] (\s,-\v-0.15) -- (\s,\v+0.15);
\draw[thick,color=red] (\s-\gap,-\v) -- (\s-\gap,\v);
\draw[thick,color=blue] (\s+\gap,-\v) -- (\s+\gap,\v);
\draw[thick,color=violet] (\s+4*\gap,-\v+0.2) -- (\s+4*\gap,\v-0.2);

\draw[thick,color=black!50!green,dashed]  (\s,\v+0.15) -- (\s+4*\gap,\v+0.15);
\draw[thick,color=black!50!green,dashed]  (\s,-\v-0.15) -- (\s+4*\gap,-\v-0.15);
\draw[thick,color=black!50!green,dashed]  (-\s,-\v-0.15) --  (-\s-4*\gap,-\v-0.15);
\draw[thick,color=black!50!green,dashed]  (-\s,\v+0.15) -- (-\s-4*\gap,\v+0.15);
\draw[thick,color=blue,dashed] (\s+\gap,-\v) -- (\s+\gap+4*\gap,-\v);
\draw[thick,color=blue,dashed]  (\s+\gap,\v)  --  (\s+\gap+4*\gap,\v);
\draw[thick,color=blue,dashed]  (-\s-\gap,-\v) -- (-\s-\gap-4*\gap,-\v);
\draw[thick,color=blue,dashed] (-\s-\gap,\v) -- (-\s-\gap-4*\gap,\v);

\draw[thick,color=red,dashed] (-\s+\gap,\v) arc (90+30:90-30:\ra-2*\gap);
\draw[thick,color=red,dashed] (-\s+\gap,-\v) arc (270-30:270+30:\ra-2*\gap);
\draw[thick,color=blue,dashed] (-\s-4*\gap,\v) arc (90-10:270+10:\v);
\draw[thick,color=blue,dashed] (\s+4*\gap,\v) arc (90+10:-90-10:\v);
\draw[thick,color=black!50!green,dashed] (-\s-4*\gap,\v+0.15) arc (90-10:270+10:\v+0.15);
\draw[thick,color=black!50!green,dashed] (\s+4*\gap,\v+0.15) arc (90+10:-90-10:\v+0.15);
\draw[thick,color=violet,dashed] (-\s-4*\gap,\v-0.2) arc (90:270:\v-0.2);
\draw[thick,color=violet,dashed] (\s+4*\gap,\v-0.2) arc (90:-90:\v-0.2);

\draw[thick, color=black]
{
(-\s-4*\gap - 0.15,0) node [right] {\small{$\bf \cdots$}} 
(\s+\gap - 0.15,0) node [right] {\small{$\bf \cdots$}} 
(-\s-12*\gap - 0.15,0) node [right] {\small{$\bf \cdots$}} 
(\s+9*\gap - 0.15,0) node [right] {\small{$\bf \cdots$}} 
};
\end{tikzpicture}
\caption{Disconnected diagram of order $\mathcal{O}(N^{2D-1})$ in Pillow channel in rank-$\rank$ tensor model}
\label{fig:disconnected diagram in pillow channel in rank D}
\end{subfigure}\\
\vspace{0.5cm}


\begin{subfigure}[t]{\linewidth}
\centering
\def\c{0} \def\s{2}  \def\v1{1/2} \def\b{2.3}  

\begin{tikzpicture}[scale=1]
\draw[thick,color=red] (-\s+\c,0.1) -- (\s+\c,0.1);
\draw[thick,color=red] (-\s+\c,-\b-0.1) -- (\s+\c,-\b-0.1);
\draw[thick,color=red,dashed] (-\s+\c,0.1) arc (270:90:0.05);
\draw[thick,color=red,dashed] (\s+\c,0.1) arc (-90:90:0.05);
\draw[thick,color=red,dashed] (-\s+\c,-\b-0.1) arc (90:270:0.05);
\draw[thick,color=red,dashed] (\s+\c,-\b-0.1) arc (90:-90:0.05);

\draw[thick,color=red,dashed] (\s+\c,0.1-\b-0.2) arc (90-20:-90+20:0.2);
\draw[thick,color=red,dashed] (-\s+\c,0.1-\b-0.2) arc (90+20:270-20:0.2);
\draw[thick,color=red,dashed] (-\s+\c,-\b-0.1-0.4) arc (-90-10:-90+10:11);
\draw[thick,color=red,dashed] (-\s+\c,0.1) arc (270-20:95:0.2);
\draw[thick,color=red,dashed] (\s+\c,0.1) arc (-90+20:90-5:0.2);
\draw[thick,color=red,dashed] (-\s+\c,0.5) arc (90+10:90-10:11);

\draw[thick,color=black!50!green,dashed] (-\s+\c,0) arc (90:270:\b/2);
\draw[thick,color=black!50!green] (-\s+\c,0) arc  (90:-90:\b/2);
\draw[thick,color=black!50!green,dashed] (\s+\c,0) arc (90:-90:\b/2);
\draw[thick,color=black!50!green] (\s+\c,0) arc (90:270:\b/2);

\draw[thick,color=blue,dashed] (-\s+\c,-0.1) arc (90:270:\b/2-0.1);
\draw[thick,color=blue] (-\s+\c,-0.1) arc (90:-90:\b/2-0.1);
\draw[thick,color=blue,dashed] (\s+\c,-0.1) arc (90:-90:\b/2-0.1);
\draw[thick,color=blue] (\s+\c,-0.1) arc (90:270:\b/2-0.1);

\draw[thick,color=violet,dashed] (-\s+\c,-0.6) arc (90:270:\b/2-0.6);
\draw[thick,color=violet] (-\s+\c,-0.6) arc (90:-90:\b/2-0.6);
\draw[thick,color=violet,dashed] (\s+\c,-0.6) arc (90:-90:\b/2-0.6);
\draw[thick,color=violet] (\s+\c,-0.6) arc (90:270:\b/2-0.6);

\draw[thick, color=black]
{
(\c-0.3,-\b/2) node [right] {\small{$\cdots$}} 
(-\s-\b/2+\c + 0.0,-\b/2) node [right] {\small{$\bf \cdots$}} 
(\s-\b/2-\c - 0.0,-\b/2) node [right] {\small{$\bf \cdots$}}
(-\s+\b/2+\c - 0.7 ,-\b/2) node [right] {\small{$\bf \cdots$}} 
(\s+\b/2+\c- 0.70,-\b/2) node [right] {\small{$\bf \cdots$}} 
};
\end{tikzpicture}
\caption{Unbroken ladder diagram of the same color of order $\mathcal{O}(N^{D+1})$ in Pillow channel in rank-$\rank$ tensor model}
\label{fig:unbroken ladder diagram of the same color in pillow channel in rank D}
\end{subfigure}
\caption{Leading and sub-leading diagrams in Pillow channel in rank-$\rank$ tensor model}
\label{fig:leading and sub-leading diagrams in pillow channel in rank D}
\end{figure}

\begin{figure}
\centering
\begin{subfigure}[t]{\linewidth}
\centering
\def\c{0} \def\s{2}  \def\v1{1/2} \def\b{2.3}  
\begin{tikzpicture}[scale=1]

\draw[thick,color=violet] (-\s+\c,0.1) -- (\s+\c,0.1);
\draw[thick,color=violet] (-\s+\c,-\b-0.1) -- (\s+\c,-\b-0.1);
\draw[thick,color=violet,dashed] (\s+\c,0.1) arc (90:-90:\b/2+0.1);
\draw[thick,color=violet,dashed] (-\s+\c,0.1) arc (90:270:\b/2+0.1);

\draw[thick,color=red,dashed] (-\s+\c,0.0) arc (270:95:0.25);
\draw[thick,color=red,dashed] (\s+\c,0.0) arc (-90:90-5:0.25);
\draw[thick,color=red,dashed] (-\s+\c,0.5) arc (90+10:90-10:11.6);
\draw[thick,color=red,dashed] (\s+\c,-\b) arc (90:-90+5:0.25);
\draw[thick,color=red,dashed] (-\s+\c,-\b) arc (90:270-5:0.25);
\draw[thick,color=red,dashed] (-\s+\c,-\b-0.1-0.4) arc (-90-10:-90+10:11.6);

\draw[thick,color=red] (-\s+\c,0) arc  (90:-90:\b/2);
\draw[thick,color=red] (\s+\c,0) arc (90:270:\b/2);

\draw[thick,color=black!50!green,dashed] (-\s+\c,-0.1) arc (90:270:\b/2-0.1);
\draw[thick,color=black!50!green] (-\s+\c,-0.1) arc (90:-90:\b/2-0.1);
\draw[thick,color=black!50!green,dashed] (\s+\c,-0.1) arc (90:-90:\b/2-0.1);
\draw[thick,color=black!50!green] (\s+\c,-0.1) arc (90:270:\b/2-0.1);

\draw[thick,color=blue,dashed] (-\s+\c,-0.6) arc (90:270:\b/2-0.6);
\draw[thick,color=blue] (-\s+\c,-0.6) arc (90:-90:\b/2-0.6);
\draw[thick,color=blue,dashed] (\s+\c,-0.6) arc (90:-90:\b/2-0.6);
\draw[thick,color=blue] (\s+\c,-0.6) arc (90:270:\b/2-0.6);

\draw[thick, color=black]
{
(\c-0.3,-\b/2) node [right] {\small{$\cdots$}} 
(-\s-\b/2+\c + 0.0,-\b/2) node [right] {\small{$\bf \cdots$}} 
(\s-\b/2-\c - 0.0,-\b/2) node [right] {\small{$\bf \cdots$}}
(-\s+\b/2+\c - 0.7 ,-\b/2) node [right] {\small{$\bf \cdots$}} 
(\s+\b/2+\c- 0.70,-\b/2) node [right] {\small{$\bf \cdots$}} 
};
\end{tikzpicture} 
\caption{Unbroken ladder diagram of different color in Pillow channel in rank-$\rank$ tensor model}
\label{fig:unbroken ladder diagram of different color in pillow channel in rank D}	
\end{subfigure}\\
\vspace{0.5cm}

\begin{subfigure}[t]{\linewidth}
\centering
\def\c{0} \def\s{2}  \def\v1{1/2} \def\b{2.3}  
\begin{tikzpicture}[scale=1]

\draw[thick,color=red] (-\s+\c,0.1) arc (90:-90:\b/2+0.1);
\draw[thick,color=red] (\s+\c,0.1) arc (90:270:\b/2+0.1);
\draw[thick,color=red,dashed] (-\s+\c,0.1) arc (270:95:0.2);
\draw[thick,color=red,dashed] (\s+\c,0.1) arc (-90:90-5:0.2);
\draw[thick,color=red,dashed] (-\s+\c,0.5) arc (90+10:90-10:11.6);
\draw[thick,color=red,dashed] (\s+\c,0.1-\b-0.2) arc (90:-90+5:0.2);
\draw[thick,color=red,dashed] (-\s+\c,0.1-\b-0.2) arc (90:270-5:0.2);
\draw[thick,color=red,dashed] (-\s+\c,-\b-0.1-0.4) arc (-90-10:-90+10:11.6);

\draw[thick,color=black!50!green,dashed] (-\s+\c,0) arc (90:270:\b/2);
\draw[thick,color=black!50!green] (-\s+\c,0) arc  (90:-90:\b/2);
\draw[thick,color=black!50!green,dashed] (\s+\c,0) arc (90:-90:\b/2);
\draw[thick,color=black!50!green] (\s+\c,0) arc (90:270:\b/2);

\draw[thick,color=blue,dashed] (-\s+\c,-0.1) arc (90:270:\b/2-0.1);
\draw[thick,color=blue] (-\s+\c,-0.1) arc (90:-90:\b/2-0.1);
\draw[thick,color=blue,dashed] (\s+\c,-0.1) arc (90:-90:\b/2-0.1);
\draw[thick,color=blue] (\s+\c,-0.1) arc (90:270:\b/2-0.1);

\draw[thick,color=violet,dashed] (-\s+\c,-0.6) arc (90:270:\b/2-0.6);
\draw[thick,color=violet] (-\s+\c,-0.6) arc (90:-90:\b/2-0.6);
\draw[thick,color=violet,dashed] (\s+\c,-0.6) arc (90:-90:\b/2-0.6);
\draw[thick,color=violet] (\s+\c,-0.6) arc (90:270:\b/2-0.6);

\draw[thick, color=black]
{
(\c-0.3,-\b/2) node [right] {\small{$\cdots$}} 
(-\s-\b/2+\c + 0.0,-\b/2) node [right] {\small{$\bf \cdots$}} 
(\s-\b/2-\c - 0.0,-\b/2) node [right] {\small{$\bf \cdots$}}
(-\s+\b/2+\c - 0.7 ,-\b/2) node [right] {\small{$\bf \cdots$}} 
(\s+\b/2+\c- 0.70,-\b/2) node [right] {\small{$\bf \cdots$}} 
};
\end{tikzpicture}	
\caption{Broken ladder diagram in Pillow channel in rank-$\rank$ tensor model}
\label{fig:broken ladder diagram in pillow channel in rank D}
\end{subfigure}
\caption{Unbroken and broken ladder diagram contribution of order $O(N^{D-1})$ in rank-$\rank$ tensor model}
\label{fig:unbroken and broken ladder diagram contribution of order n2 in rank D}
\end{figure}
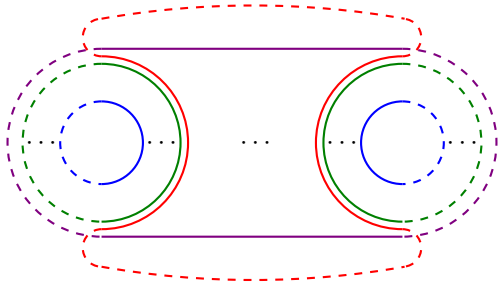
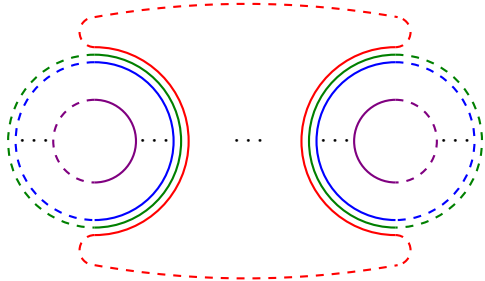

The leading disconnected diagram in Pillow channel is of order $\mathcal{O}(N^{2\rank-1})$ (See Fig.~\ref{fig:disconnected diagram in pillow channel in rank D}). Just like rank-3 case, various ladder diagrams have different $N$ scalings. The ladder diagram in Pillow channel have different scalings. For example, consider Pillow channel of red color. The dominant subleading contribution comes from unbroken ladder diagram which gains $2D$ extra loops via Pillow contraction (See Fig.~\ref{fig:unbroken ladder diagram of the same color in pillow channel in rank D}) to become of order $\mathcal{O}(N^{\rank+1})$. In contrast, unbroken ladder diagram of different color (\eg violet color in Fig.~\ref{fig:broken ladder diagram in pillow channel in rank D}) get $2\rank-2$ extra loops whereas broken ladder diagram gains $2\rank-1$ extra loops (\eg See Fig.~\ref{fig:broken ladder diagram in pillow channel in rank D}). Hence, both of them becomes of order $\mathcal{O}(N^{\rank-1})$.

We will argue below that these $\mathcal{O}(N^{\rank-1})$ diagrams gives the next-to-subleading corrections in ${1\over N}$. For this, we consider $N$ scaling of two other classes of diagrams: non-melonic and melonic ladder diagram of type which we have not considered till now.

The case of non-melonic diagrams was considered in \cite{GurauSchaeffer,Bonzom:2017pqs}. For our purpose, it is sufficient to consider large $N$ scaling of the leading non-melonic diagrams. In \cite{Bonzom:2017pqs}, such a leading non-melonic diagrams (without contraction) are shown to be at most of order $\mathcal{O}(N^{4-2\rank})$. Note that this contribution comes from a new class of diagrams\footnote{For $\rank=3$, non-melonic contribution is of order $\mathcal{O}(N^{-3})$.} which start to give a contribution for $\rank\geqq 5$~\cite{Bonzom:2017pqs}. By Pillow contraction, such diagrams obtain at most $2D$ loops like unbroken ladder diagram in the Pillow channel, and becomes of order $\mathcal{O}(N^{4})$, which does not increase with $\rank$.

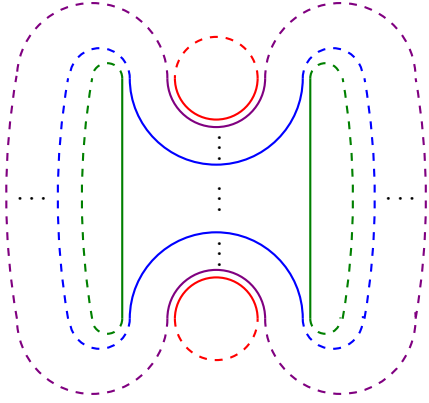
\begin{figure}
\centering
\def\c{0} \def\s{1.6}  \def\v1{1/2} \def\b{2.3}  \def\shi{2.3}
\begin{tikzpicture}[scale=1]

\draw[thick,color=blue] (\shi,-\s+\c) arc  (0:180:\b/2);
\draw[thick,color=violet] (-0.5+\shi,-\s+\c) arc (0:180:\b/2-0.5);
\draw[thick,color=red] (-0.6+\shi,-\s+\c) arc (0:180:\b/2-0.6);

\draw[thick,color=blue] (0,\s+\c) arc (180:360:\b/2);
\draw[thick,color=violet] (0.5,\s+\c) arc (180:360:\b/2-0.5);
\draw[thick,color=red] (0.6,\s+\c) arc (180:360:\b/2-0.6);

\draw[thick,color=red,dashed] (0.6,-\s+\c) arc (180:360:\b/2-0.6);
\draw[thick,color=red,dashed] (\b/2+0.55,\s+\c) arc (0:180:\b/2-0.6);

\draw[thick,color=black!50!green] (-0.1,-\s+\c) -- (-0.1,\s+\c);
\draw[thick,color=black!50!green] (0.1+\shi,-\s+\c) -- (0.1+\shi,\s+\c);

\draw[thick,color=black!50!green,dashed] (-0.1,\s+\c) arc (0:180-5:0.2);
\draw[thick,color=black!50!green,dashed,dashed] (-0.5,-\s+\c) arc (180+9.8:180-9.8:9.4);
\draw[thick,color=black!50!green,dashed] (-0.1,-\s+\c) arc (0:-180+5:0.2);

\draw[thick,color=blue,dashed] (0,\s+\c) arc (0:180-5:0.4);
\draw[thick,color=blue,dashed,dashed] (-0.82,-\s+\c) arc (180+9.8:180-9.8:9.4);
\draw[thick,color=blue,dashed] (0,-\s+\c) arc (0:-180+5:0.4);

\draw[thick,color=violet,dashed] (0.5,\s+\c) arc (0:180-5:1);
\draw[thick,color=violet,dashed,dashed] (-1.5,-\s+\c) arc (180+10:180-9.8:9.4);
\draw[thick,color=violet,dashed] (0.5,-\s+\c) arc (0:-180+5:1);

\draw[thick,color=black!50!green,dashed] (\shi+0.1,\s+\c) arc (180:0+5:0.2);
\draw[thick,color=black!50!green,dashed,dashed] (\shi+0.53,-\s+\c) arc (0-9.8:0+9.8:9.4);
\draw[thick,color=black!50!green,dashed] (\shi+0.1,-\s+\c) arc (180:360+5:0.2);

\draw[thick,color=blue,dashed] (\shi,\s+\c) arc (180:0+5:0.4);
\draw[thick,color=blue,dashed,dashed] (\shi+0.82,-\s+\c) arc (0-9.8:0+9.8:9.4);
\draw[thick,color=blue,dashed] (\shi,-\s+\c) arc (180:360+5:0.4);

\draw[thick,color=violet,dashed] (\shi-0.5,\s+\c) arc (180:0+5:1);
\draw[thick,color=violet,dashed,dashed] (\shi+1.5,-\s+\c) arc (0-9.8:0+9.8:9.4);
\draw[thick,color=violet,dashed] (\shi-0.5,-\s+\c) arc (180:360+5:1);

\draw[thick, color=black]
{
(\shi/2-0.15,0.1) node [right] {\small{$\vdots$}} 
(\shi/2-0.15,-\s+\b/2 - 0.18) node [right] {\small{$\bf \vdots$}} 
(\shi/2-0.15, \s-\b/2 + 0.33) node [right] {\small{$\bf \vdots$}} 
(\shi+0.95, -\s+\b/2 + 0.44) node [right] {\small{$\bf \cdots$}} 
(-\shi+0.65, -\s+\b/2 + 0.44) node [right] {\small{$\bf \cdots$}} 
};
\end{tikzpicture}	
\caption{Example: Other contribution from unbroken (melonic) ladder diagram of order~$O(N^{3})$ in rank-$\rank$ tensor model. The unbroken ladder diagram (of green color) of order $\mathcal{O}(N^{-D+1})$ gains $(D+2)$ loops in the Pillow channel of red color.}
\label{fig:Example: Other contribution from unbroken}
\end{figure}

In addition, one also has to consider other class of ladder diagrams which we have ignored so far. For example, a vertical ladder in Fig.~\ref{fig:Example: Other contribution from unbroken}, which is of lower order than the leading ladder diagrams. Indeed, for $\rank=3$, the vertical ladder diagram is of the same order $\mathcal{O}(N^3)$ as the non-melonic diagrams. However, unlike the usual (horizontal) ladder diagrams, the $N$ scaling of the (vertical) ladder diagram does not increase with the rank $\rank$. Hence, we can also ignore them in the higher rank (\ie $\rank>3$).

To summarize, for $\rank>5$ the next-to-subleading contribution is dominated by broken ladder diagrams and unbroken ladder diagrams of different color.

Now, let us evaluate Pillow channel of color $\col_a$ ($a=1,2,\cdots D)$:
\begin{align}
&F^{P,\col_a} (\tau_1,\tau_2,\tau_3,\tau_4)\equiv  \langle \psi_{i_1 \cdots  i_a \cdots i_\rank}(\tau_1) \psi_{i_1 \cdots  j_a \cdots i_\rank}(\tau_2) \psi_{j_1 \cdots  i_a \cdots j_\rank}(\tau_3) \psi_{j_1 \cdots  j_a \cdots j_\rank}(\tau_4) \rangle \cr
=& N^{2D-1} G(\tau_{12})G(\tau_{34}) + N^{\rank+1} \mathcal{F}^{P,\col_a}_{(0)} (\tau_1,\tau_2,\tau_3,\tau_4) + N^{\rank-1} \mathcal{F}^{P,\col_a}_{(1)} (\tau_1,\tau_2,\tau_3,\tau_4) 
\end{align}
where the leading disconnected diagram is of order $\mathcal{O}(N^{2\rank-1})$ (See Fig.~\ref{fig:disconnected diagram in pillow channel in rank D}). The second term is unbroken ladder diagrams of the color $\col_a$ while the last term corresponds to all unbroken/broken ladder diagram except for the unbroken ladder diagram of the color $\col_a$. Here we consider only $\rank>5$ so that there is no non-melonic contribution up to $\mathcal{O}(N^{D-1})$. Hence, one can write the corresponding geometric series
\begin{align}
\mathcal{F}^{P,\col}_{(0)}=& \sum_{n=0}^\infty (\dipole^\col)^n \mathcal{F}_0\label{eq:geometric series leading Pillow rank d}\\
\mathcal{F}^{P,\col}_{(1)}=&\sum_{n=0}^\infty \left[\sum_{j=1}^\rank\dipole^{\col_j}\right]^n \mathcal{F}_0 -  \sum_{n=0}^\infty (\dipole^\col)^n \mathcal{F}_0\label{eq:geometric series subleading Pillow rank d}
\end{align}
and, their common ratios become
\begin{align}
\dipole^\col(\tau_1,\tau_2,\tau_3,\tau_4)=&{1\over \rank}\mathcal{K}(\tau_1,\tau_2,\tau_3,\tau_4)\label{eq:one color dipole in rank d} \\
\sum_{j=1}^\rank\dipole^{\col_j}(\tau_1,\tau_2,\tau_3,\tau_4)=&\mathcal{K}(\tau_1,\tau_2,\tau_3,\tau_4) 
\end{align}
Like rank-3 tensor model, the leading ladder diagram $\mathcal{F}^{P,\col}_{(0)}$ does not grow exponentially. To show this non-chaotic behavior of $\mathcal{F}^{P,\col}_{(0)}$, we also repeat the same calculations for \eqref{eq:geometric series leading Pillow rank d} as in \cite{Maldacena:2016hyu} with additional $1\over D$ factor to get
\begin{align}
\mathcal{F}^{P,\col}_{(0)}(\chi)\equiv&{\mathcal{F}_{(0)}^{P,\col}(\tau_1,\tau_2,\tau_3,\tau_4)\over G(\tau_{12})G(\tau_{34})}=\underset{h=h_\ast }{\text{Res}}\left[{h-{1\over 2}\over \pi \tan{\pi h\over 2}}{k_R(1-h)\over 1- {1\over \rank} k_R(1-h)}\Psi_h(\chi) \right]
\cr
&\hspace{1cm}+{4\pi \over 3}\int_{-\infty}^\infty {ds\over 2\pi} {h-{1\over 2}\over \pi \tan{\pi h\over 2}}\left[{k_c(h)\over 1-{1\over \rank} k_c(h)}-{k_R(1-h)\over 1- {1\over \rank} k_R(1-h)}\right]\Psi_h(\chi)
\end{align}
where $k_R(1-h)$ is defined by
\begin{equation}
k_R(1-h)={\Gamma\left(3-{2\over \rank+1}\right)\Gamma\left(h-1+ {2\over \rank+1}\right)\over \Gamma\left(1+{2\over \rank+1}\right)\Gamma\left(h+1-{2\over \rank+1}\right)}
\end{equation}
The $h_\ast$ in the residue is a solution of an equation\footnote{Strictly speaking, simple poles of the function inside of residue which are greater than ${1\over 2}$ and are not even integer.}
\begin{equation}
1-{1\over \rank} k_{R}(1-h)=0
\end{equation}
This equation has only one solution
\begin{equation}
h_\ast=1
\end{equation}
In the same way as $\xi={1\over 3}$ case ($\rank=3$) in Appendix~\ref{app:large time behavior}, one can see that $\mathcal{F}^{P,\col}_{(0)}(\chi)$ does not grow exponentially. 

One the other hand, one can easily see that $\mathcal{F}^{P,\col}_{(1)}(\chi)$ exponentially grows with time (\ie $\mathcal{F}^{P,\col}_{(1)}(\chi)\sim~e^{{2\pi \over \beta}t}$~) because the first and second term in \eqref{eq:geometric series subleading Pillow rank d} is the same as \eqref{eq:geometric series leading Cooper rank d} and \eqref{eq:geometric series leading Pillow rank d}, respectively: The former shows maximal chaos while the latter does not exponentially grow. Although it would be tempting to read off the Lyapunov exponent and conclude that it saturates the chaos bound from this computation, this is not correct. The reason is that we have not considered ${1\over \beta \local}$ correction to $\mathcal{F}^{P,\col}_{(0)}(\chi)$ arising from perturbation of Schwinger-Dyson equation for two point function (\eg \eqref{eq:SD eq for two point function}). But, it is difficult to evaluate $\mathcal{F}^{P,\col}_{(0)}(\chi)$ at all order in ${1\over \beta\local}$ which we leave for future work. It is interesting to note that if $\mathcal{F}^{P,\col}_{(0)}(\chi)$ does not grow exponentially at all orders in $\beta \local$, the Lyapunov exponent would indeed be maximal and the scrambling time $t_\ast$ is the same as that of Cooper channe~\eqref{eq:scrambling time cooper rank d}.

After calculating various four point functions, one can now ask what are the implications on possible low energy effective action for these tensor models.\footnote{We thank S. Wadia for raising this issue.} In particular, we will see below that the $\beta \effcoupling$ divergence (in the strong coupling limit) in $\mathcal{O}(N^{\rank-1})$ term of the Pillow channel is consistent with a simple effective action. To do this, first note that as in the SYK model, the SYK-like tensor models also have emergent reparametrization symmetry in the strong coupling limit, which is broken explicitly by kinetic term in the action as well as spontaneously by classical solution of two point function. This (explicitly and spontaneously) broken symmetry leads to the low energy effective action for the zero mode of the reparametrization symmetry~\cite{Maldacena:2016hyu,Jevicki:2016bwu,Jevicki:2016ito} (The analogous broken symmetry for $AdS_2$ bulk was discussed in~\cite{Maldacena:2016upp,Mandal:2017thl}). From the Cooper channel which is exactly the same as the four point function of SYK model, one may expect that the effective action would be
\begin{equation}
S\sim {N^\rank \over \beta \effcoupling } \int d\tau \{ f(\tau),\tau\}\label{eq:schwarzian action}
\end{equation}
where $\{ f(\tau),\tau\}$ is the Schwarzian derivative. Now, let us consider the contribution of this zero mode to the Pillow channel. For this, we perform infinitesimal conformal transformation $\tau\;\longrightarrow \; \tau+\epsilon$ of the Pillow channel\footnote{Strictly speaking, this is a conformal transformation of one point function of a non-local (quartic) gauge invariant operator.}, and this leads to the contribution of zero mode to the Pillow channel. From \eqref{eq:schwarzian action}, the leading contribution in $N$ from the zero mode can be evaluated to be~\cite{Maldacena:2016hyu}
\begin{equation}
\delta_\epsilon F^{P,\col}= N^{2\rank-1}\sum_n \delta_n G \delta_n G\;\; \langle \epsilon_n \epsilon_{-n}\rangle\sim N^{\rank-1} \beta\effcoupling e^{ {2\pi\over \beta } t }
\end{equation}
which corresponds to the $h=2$ mode contribution to the Pillow channel in order $\mathcal{O}(N^{\rank-1})$.

\section{Conclusion}
\label{sec:conclusion}

In this work, we studied tensor models on lattice, and introduced general techniques to compute the four point functions in large $N$ and strong coupling. As a concrete example, we worked out the KT chain model, and evaluated not only the four point function in Cooper channel (the analogous one in KT model was shown to saturate the chaos bound~\cite{Klebanov:2016xxf} like SYK model) but also other channels~(Pillow and Tetrahedron channels). The Pillow channels exhibited more interesting results. In fact, we find that the leading connected diagram does not even have chaotic behavior, and one has to look at subleading terms to discern chaos. Moreover, in the Pillow channel, we found new spectrum in addition to that observed in KT or SYK model. As an aside, note that we can read off\footnote{The lattice translational-invariant mode~({\small $p=0$}) of KT chain model gives the corresponding mode of KT model.} the Tetrahedron and Pillow channel four point function in KT model which are actually new results.

We pointed out that our techniques are not restricted to KT models, but can be applied to broad classes of tensor models which satisfies certain property which we refer to as ULF property in Section~\ref{sec:models}. In this context, we analyzed the GW model and a generalized GW model which we proposed in this work, and again obtained the four point functions in various channels. Especially, the generalized GW model possesses an interesting feature that it interpolates between KT model and GW model, and some of the channels contain a varying chaos exponent depending on the details of interaction term.

We also generalized our techniques to the rank-$\rank$ tensor model. First, we pointed out that ``uncoloring'' process enables us to study the GW and KT models of rank-$\rank$ in the same framework. Then, we worked out the KT model of rank-$\rank$ because the lattice generalizations thereof are straightforward. We found that the four point functions show the analogous behavior with $\rank=3$ case. Namely, the Cooper channel saturates chaos bound, and the Pillow channel does not grow exponentially up to certain order in $N$. We also showed that in Pillow channel for $\rank>5$, specific ladder diagrams suppress other ladder diagrams as well as non-melonic diagrams in large $N$. This enabled us to consider the next-to-subleading contributions in $N$, which turned out to be the same behavior as Cooper Channel. Furthermore, we show that this contribution to the Pillow channel is consistent with the effective action for the Goldstone boson associated with the symmetry breaking.

Here are some future directions that are possible:
\begin{itemize}
\item Rank-$\rank$ tensor theories : It is interesting to work out the large $N$ diagrammatics for rank-$\rank$ theories. The structure of four point function is more intricate now since there are more possible external gauge contractions. It is also interesting if there is an appropriate large $\rank$ limit in which the class of diagrams simplify and look for possible simplifications in the Schwinger-Dyson equation for four point functions. 

\item It is also interesting to explore the generalization of tensor models to other type of lattice interactions~\cite{Berkooz:2016cvq,Berkooz:2017efq}. 
\end{itemize}
We hope to report on some of these future directions soon. 

\acknowledgments
We thank Spenta Wadia, Antal Jevicki, Aaditya Salgarkar, Soumyadeep Chaudhuri, Anosh Joseph, Victor Ivan Giraldo Rivera and especially R. Loganayagam for extensive discussions. JY thanks the Galileo Galilei Institute for Theoretical Physics (GGI) for the hospitality and INFN for partial support during the completion of this work, within the program ``New Developments in AdS3/CFT2 Holography''. JY also thanks the International Centre for Theoretical Physics (ICTP) for the hospitality and APCTP, Simons Foundation for partial support during the completion of this work, within the program ``Spring School on Superstring Theory and Related Topics''. We gratefully acknowledge support from International Centre for Theoretical Sciences (ICTS), Tata institute of fundamental research, Bengaluru. We would also like to acknowledge our debt to the people of India for their steady and generous support to research in the basic sciences.

\appendix

\section{Appendix: Chaos from four point function} 
\label{app:temp appendix}

In this Appendix, following \cite{Maldacena:2016hyu,Gu:2016oyy}, we will study the large time behavior of the out of time ordered correlator and also the spectrum of four point functions. The connected part of the (Euclidean) four point function we found for all the models studied in this work is of the form
\begin{align}
\mathcal{F}_\xi(\chi)\equiv& {\mathcal{F}_\xi(\tau_1,\tau_2,\tau_3,\tau_4)\over G(\tau_{12})G(\tau_{34})}={4\pi \over 3}\left[\int_{-\infty}^\infty {ds\over 2\pi} \left.{h-{1\over 2}\over \pi \tan{\pi h\over 2}}{k_c(h)\over 1- \xi k_c(h)}\Psi_h(\chi)\right|_{h={1\over 2}+is }\right.\cr
&\hspace{4cm}\left.+\left.\sum_{n=1}^\infty \left({2h-1\over \pi^2 }{k_c(h)\over 1- \xi k_c(h)}\Psi_h(\chi) \right)\right|_{h=2n}\right]\label{app eq:four point function}
\end{align}
where $k_c(h)$ is given by
\begin{equation}
k_c(h)=-{3\over 2} {\tan {\pi\over 2}( h-{1\over 2})\over h-{1\over 2}}\ ,
\end{equation}
and, $\xi$ depends on the details of the model and channels (\ie Cooper, Pillow) of four point function that we are interested in.

\subsection{Large Time Behavior} 
\label{app:large time behavior}

The out of time ordered correlator is obtained by taking an appropriate analytic continuation of the Euclidean correlator given in \eqref{app eq:four point function}. The analytic continuation which takes the Euclidean correlator to the out of time ordered correlator is given by
\begin{equation}
\chi = {2 \over 1  - i \sinh({2 \pi t \over \beta})} \quad\xrightarrow{t \to \infty}\quad  4 i e^{- {2 \pi  \over \beta}t}
\label{def:conformal transformation}
\end{equation}
Before we perform this analytic continuation,  we first massage the expression eq(\ref{app eq:four point function}) following the method of \cite{Maldacena:2016hyu}. First, note that for $h\in 2\mathbb{Z}$, we have
\begin{equation}
k_c(2n)={3\over 4n-1}\hspace{1cm} (n\in \mathbb{Z})
\end{equation}
Hence, the summation in \eqref{app eq:four point function} can diverge if
\begin{equation}\label{xispecial}
\xi={4n-1\over 3} \hspace{1cm} (n=1,2,3,\cdots)
\end{equation}
For those cases, one has to carefully treat the divergence from $h={1\over 2}(3\xi+1)$ by doing a ${1 \over \beta \effcoupling}$ perturbation theory where $\effcoupling$ is the effective coupling in the two point function for the model being studied. Such an analysis for $\xi=1$ (\ie $h=2$) case was done in \cite{Maldacena:2016hyu} where they found that $h=2$ term leads to maximal Lyapunov exponent $\lambda_L={2\pi \over \beta}$. The other finite terms in the summation gives a ${1\over\beta \effcoupling}$~correction to the Lyapunov exponent.

For values of $\xi$ corresponding to $n \geqq 2$ in \eqref{xispecial}, one has to do in principle a more careful analysis, but since we never encounter such values for $\xi$'s in the models we study (in fact, we always find $\xi\leqq 1$), we will not pursue this further.\footnote{
As we mention later, we do find that any value $\xi>1$ other than ${4n-1\over 3}$ $(n=2,3,\cdots)$ violates the chaos bound. Therefore it is unlikely that $\xi = {4n-1\over 3}$ for $(n=2,3,\cdots)$ will arise in any reasonable theories.}

If there is no divergence in \eqref{app eq:four point function}, one can rewrite it as
\begin{align}
\mathcal{F}_\xi(\chi)\equiv&{\mathcal{F}_\xi(\tau_1,\tau_2,\tau_3,\tau_4)\over G(\tau_{12})G(\tau_{34})}={4\pi \over 3}\left[\int_{-\infty}^\infty {ds\over 2\pi} \left.{h-{1\over 2}\over \pi \tan{\pi h\over 2}}{k_c(h)\over 1- \xi k_c(h)}\Psi_h(\chi)\right|_{h={1\over 2}+is }\right.\cr
&\hspace{4cm}\left.+\sum_{n=1}^\infty \underset{h=2n}{\text{Res}}\left({h-{1\over 2}\over \pi \tan{\pi h\over 2}}{k_c(h)\over 1- \xi k_c(h)}\Psi_h(\chi) \right)\right]\label{app eq:four point function2}
\end{align}
As in \cite{Maldacena:2016hyu}, we define a function
\begin{equation}
k_R(1-h)\equiv{\cos {\pi\over 2} (h-{1\over 2}) \over \cos {\pi\over 2} (h + {1\over 2})  }k_c(h)={3\over 2h-1}
\end{equation}
Note that for $h=2n$ ($n\in\mathbb{Z}$):
\begin{equation}
k_R(1-2n)=k_c(2n)\ ,
\end{equation}
and therefore, one can replace $k_c(h)$ with $k_R(1-h)$ in the residue of \eqref{app eq:four point function2}. Then, one can pull the small contours around $h=2,4,6,\cdots$ to the contour $h={1\over 2} +is$ ($s\in \mathbb{R}$). In this procedure, one picks up a pole at
\begin{equation}
h_\ast={1\over 2}+{3\over 2}\xi
\end{equation}
from the denominator $1- \xi k_R(h)$ unless $\xi<0$. It is convenient to perform the analysis seperately for different regimes of $\xi$.
\begin{itemize}
\item \textbf{$\xi<0$}: In this case, the contour will not pass the pole at $h_\ast$, we have  
\begin{align}
\mathcal{F}_\xi(\chi)=&{4\pi \over 3}\int_{-\infty}^\infty {ds\over 2\pi} {h-{1\over 2}\over \pi \tan{\pi h\over 2}}\left[{k_c(h)\over 1-\xi k_c(h)}-{k_R(1-h)\over 1- \xi k_R(1-h)}\right]\Psi_h(\chi)
\end{align}
Note that with the analytic continuation $\chi\sim e^{-{2\pi \over \beta}t}\rightarrow 0$ as given in eq(\ref{def:conformal transformation}), the above term does not grow exponentially. 
\item \textbf{$\xi>0$}: On the other hand, for $\xi>0$, we pick up the pole at $h_\ast$. Consequently, we get
\begin{align}
\mathcal{F}_\xi(\chi)=&{4\pi \over 3}\int_{-\infty}^\infty {ds\over 2\pi} {h-{1\over 2}\over \pi \tan{\pi h\over 2}}\left[{k_c(h)\over 1-\xi k_c(h)}-{k_R(1-h)\over 1- \xi k_R(1-h)}\right]\Psi_h(\chi)\cr
&\hspace{3cm}+ \underset{h=h_\ast }{\text{Res}}\left[{h-{1\over 2}\over \pi \tan{\pi h\over 2}}{k_R(1-h)\over 1- \xi k_R(1-h)}\Psi_h(\chi) \right]\label{eq:app four point calculation with xi}
\end{align}
The first integral does not grow exponentially as before. From the behavior of hypergeometric function $\Psi_{h_\ast}(\chi)$ (and together with $\xi>0$), one has
\begin{equation}
\mathcal{F}_\xi(\chi)\sim e^{(h_\ast-1){2\pi \over \beta} t }
\end{equation}
Therefore, Lyapunov exponent $\lambda_L$ for general $\xi$ is given by
\begin{equation}
\lambda_L={3\xi-1\over 2} {2\pi \over \beta}\label{eq:lyapunov exponent xi}
\end{equation}
%

\end{itemize}

We summarize long time behavior of four point functions.
\begin{equation}
 \mathcal{F}_\xi(\chi)\sim e^{(h_\ast -1) {2\pi \over \beta}t}
\end{equation}
\begin{itemize}

\item $\xi>1$ : Violation of chaos bound
\begin{equation}
\lambda_L>{2\pi \over \beta}
\end{equation}

\item $\xi=1$ : Maximally chaotic
\begin{equation}
\lambda_L={2\pi \over \beta}
\end{equation}

\item ${1\over 3} <\xi<1$ : Non-maximal chaotic
\begin{equation}
0< \lambda_L<{2\pi \over \beta}
\end{equation}

\item $\xi \leqq {1\over 3}$ : Non-chaotic
\begin{equation}
\lambda_L=0
\end{equation}

\end{itemize}

\subsection{Spectrum} 
\label{app:spectrum}

\begin{figure}[t]
\centering
\includegraphics[width=0.8\textwidth]{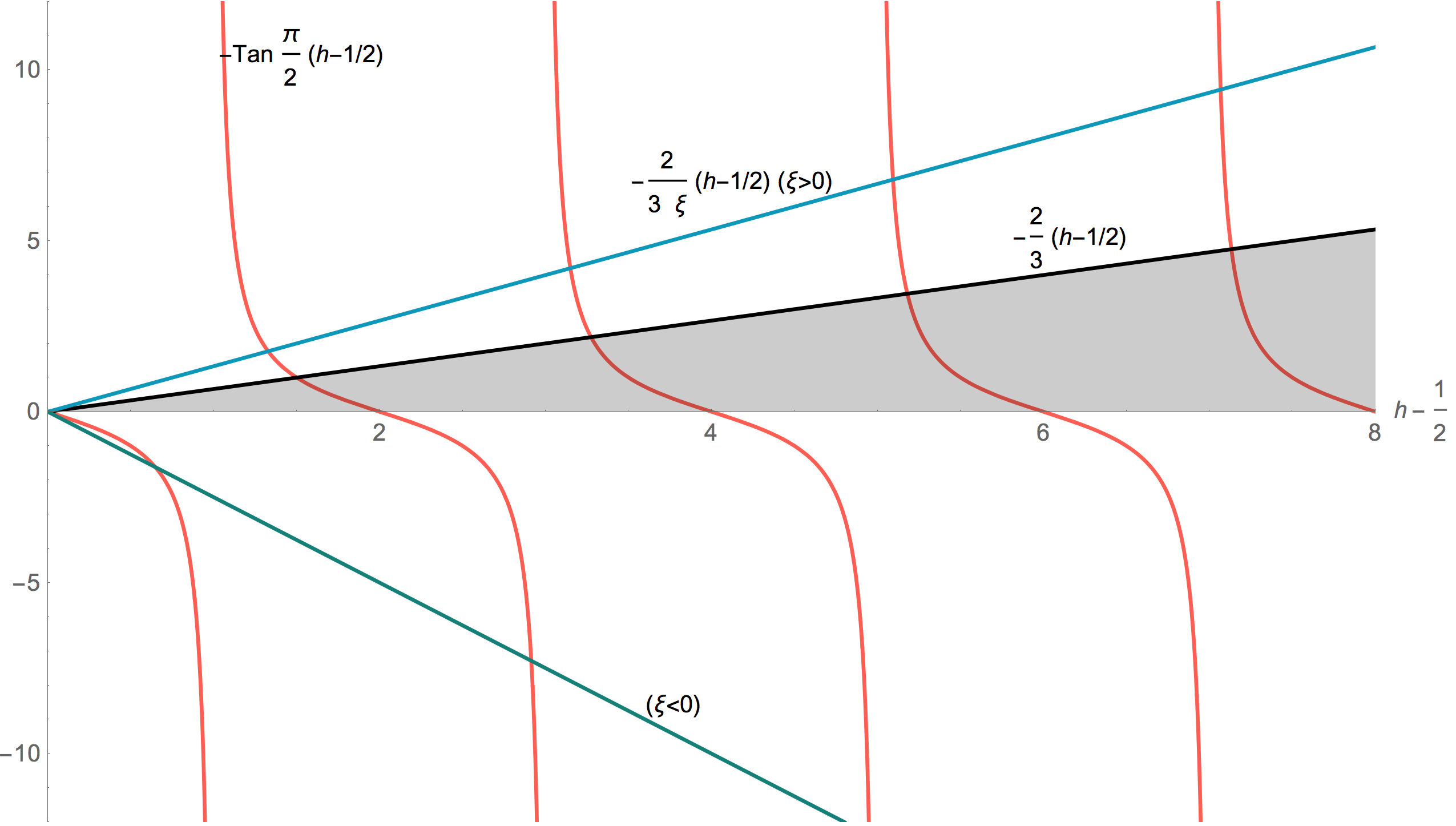}
\caption{The intersections of red curve and blue line are simple poles corresponding to conformal dimension $h_n$ ($n=0,1,2,\cdots$). The black line represents $\xi=1$ case which exhibits maximal chaos, and the shaded region leads to violation of chaos bound.}
\label{fig:spectrum}
\end{figure}

Again, it is enough to consider only $\xi<1$ case because the spectrum for $\xi=1$ case was already found in~\cite{Maldacena:2016hyu}. Following the method in~\cite{Maldacena:2016hyu}, we move the contour in the first term in~\eqref{app eq:four point function2} (\ie ${1\over 2}+is $, $\; s\in \mathbb{R}$) to the right. Then, this contour will cancel the residue in the second term in~\eqref{app eq:four point function2}. At the same time, it will pick up poles at
\begin{equation}
h_n> {1\over 2} \hspace{1cm} \mbox{where}\quad 1-\xi k_c(h_n)=0\hspace{1cm}(n=0,1,2,\cdots) 
\end{equation}
which were not included in the second term of \eqref{app eq:four point function2} (See Fig.~\ref{fig:spectrum}). For large $n$, $h_n$ asymptotes to
\begin{equation}
h_n\simeq 2n+{3\over 2} +{3\xi\over 2\pi n}\hspace{1cm} (n\gg1)
\end{equation}
Note that $h_0$ is a simple pole for $\xi<1$ unlike $\xi=1$ case where it is a double pole. For $\xi=-{4\over 3\pi} $, $h_0$ is located at ${1\over 2}$ which the contour ${1\over 2} + is$ passes through. Since the contribution from this pole becomes ambiguous for $\xi=-{4\over 3\pi} $, we consider a case where
\begin{equation}
-{4\over 3\pi} <\xi<1
\end{equation}
and, all models in this paper lies in this range. From \cite{Maldacena:2016hyu}, we found 
\begin{equation}
\mathcal{F}_\xi(\chi)= \sum_{n=0}^\infty c_n^2 \chi^{h_n} {}_2 F_1(h_n, h_n; 2h_n;\chi) 
\end{equation}
where (the square of) OPE coefficient is given by
\begin{equation}
c_n^2 = {4\pi  \over 3 \xi^2  } {h_n-{1\over 2} \over  (-\pi  \tan ({\pi h_n \over 2}))}{[ \Gamma(h_n)]^2\over \Gamma(2h_n)} \left(-{1\over k'_c(h_n)}\right)
\end{equation}
Especially, for $\xi=0$, one can find the conformal dimension $h_n$ exactly. 
\begin{equation}
h_n=2n+{3\over 2}
\end{equation}
and the OPE coefficient is
\begin{equation}
c_n^2=  {4[\Gamma(2n+{3\over 2})]^2\over \pi \Gamma(4n+3)} 
\end{equation}

%
%
%
%
 %

\bibliographystyle{JHEP}
\bibliography{paper}

\end{document}